\documentclass[PAPER, atlasdraft=false, UKenglish, cernpreprint, texmf, orcidlogo]{atlasdoc}
\usepackage{atlaspackage}
\usepackage{atlasbiblatex}

\usepackage{atlasphysics}
\usepackage{atlasjetetmiss}

\graphicspath{{logos/}{figures/}}

\usepackage{ANA-IDTR-2023-01-PAPER-defs}

\AtlasTitle{Expected Tracking Performance of the ATLAS Inner Tracker at the High-Luminosity LHC}

\AtlasAbstract{%
The high-luminosity phase of LHC operations (HL-LHC),
will feature a large increase in
simultaneous proton--proton interactions per bunch crossing up to 200,
compared with a typical leveling target of 64 in Run 3. Such an increase
will create a very challenging environment in which to perform charged
particle trajectory reconstruction, a task crucial for the
success of the ATLAS physics program, and will exceed the
capabilities of the current ATLAS Inner Detector (ID).
A new all-silicon Inner Tracker (ITk) will replace
the current ID in time for the start of the HL-LHC\@.
To ensure successful use of the ITk capabilities in Run 4 and
beyond, the ATLAS tracking software has been successfully adapted to
achieve state-of-the-art track reconstruction in challenging
high-luminosity conditions with the ITk detector. This paper presents
the expected tracking performance of the ATLAS ITk based on the latest available developments
since the ITk technical design reports.
}

\AtlasRefCode{IDTR-2023-01}

\PreprintIdNumber{CERN-EP-2024-342}

\AtlasJournal{JINST}
\AtlasJournalRef{JINST 20 (2025) P02018}
\AtlasDOI{10.1088/1748-0221/20/02/P02018}

\hypersetup{pdftitle={ATLAS document},pdfauthor={The ATLAS Collaboration}}

\usepackage{pdflscape}

\begin{document}

\maketitle

\tableofcontents

\section{Introduction}
%

The Large Hadron Collider (LHC)~\cite{Evans:2008zzb} has enabled detailed
exploration of the energy frontier since 2009 with a successful program of
proton--proton and heavy-ion collisions. It is slated to be upgraded in time for its
fourth period of operation (Run 4) to deliver a
nominal luminosity of $5.0\times 10^{34} \mathrm{cm^{-2}s^{-1}}$ to the general-purpose detectors
ATLAS~\cite{PERF-2007-01} and CMS~\cite{CMS-CMS-00-001}, in a
configuration known as high-luminosity LHC (HL-LHC)~\cite{Aberle:2749422}.
Such an increase in instantaneous luminosity creates a very challenging
environment for particle detectors since it is achieved at the cost of a
drastic increase in the average number of proton--proton
interactions in the same bunch crossing ($\langle\mu\rangle$, also known as ``\pileup''), expected to rise up to as much as 140 in
the ATLAS detector during Run 4, as compared with a typical leveling target of 64
during Run~3.
For subsequent runs, further luminosity increases up to
$7.5\times 10^{34} \mathrm{cm^{-2}s^{-1}}$ are under discussion, which would bring the \pileup up to 200,
extreme conditions under which the reconstruction will have to maintain robust performances.
Moreover, the performance of the ATLAS Inner Detector subsystem~\cite{ATLAS-TDR-04,ATLAS-TDR-05}
is already affected by radiation
damage~\cite{IDET-2017-10-fix,IDET-2020-01}, which degrades its operational capabilities.
Furthermore, the subsystem will face challenges in addressing the increased fluence and occupancy resulting from HL-LHC collisions, with bandwidth limitations and the need to minimize the material budget likely exacerbating these issues.
Therefore, the ATLAS Collaboration will install a new
all-silicon tracking detector, the Inner Tracker
(ITk)~\cite{ATLAS-TDR-25,ATLAS-TDR-30}, as a part of its Phase-II detector upgrade
program~\cite{CERN-LHCC-2012-022,ATLAS-TDR-PhaseII} in preparation for HL-LHC data-taking.
Following the publication of the technical design reports, further refinements were made to the detector design, which are detailed in this document.

The reconstructed trajectories (tracks) of charged particles produced in collisions are
a key input to many other reconstruction algorithms~\cite{PERF-2015-09, FTAG-2019-07,ATL-PHYS-PUB-2022-027,EGAM-2021-01,ATL-PHYS-PUB-2022-044, MUON-2018-03},
and they are also used directly by a
number of physics analyses (for example Refs.~\cite{HION-2018-03-fix,SUSY-2018-19,STDM-2019-22}).
To achieve high tracking efficiency, the ATLAS tracking
chain~\cite{Cornelissen:2008zzc,IDTR-2022-04-fix} is designed around a combinatorial
variant of the well-known Kalman filter
algorithm~\cite{Fruhwirth:1987fm}, augmented by an ambiguity solving
stage designed to keep the low-quality and duplicate track creation rates at a
minimum.  This paper presents the expected performance of a version of
this track reconstruction chain tuned specifically for the challenging
environment of HL-LHC collisions and using the ITk detector.

The ATLAS Collaboration has
produced notes detailing the expected performance of
flavor-tagging~\cite{ATL-PHYS-PUB-2020-005,ATL-PHYS-PUB-2022-047} as well
as the integrated performance of the whole Run~4
detector~\cite{ATL-PHYS-PUB-2019-005,ATL-PHYS-PUB-2021-023-fix}.  Reports
about the concomitant upgrade to the CMS detector can be found in
Refs.~\cite{CMS-TDR-15-02,CMS-TDR-014}. Extensive information about the
physics prospects at the HL-LHC for both the ATLAS and CMS collaborations
can be found in Refs.~\cite{ATL-PHYS-PUB-2019-006-fix,ATL-PHYS-PUB-2022-018-fix}.

The rest of this paper is structured as follows: the ITk detector layout is
detailed in Section~\ref{sec:atlas_itk}; the simulation framework and samples, and the
track reconstruction strategy are described in Section~\ref{sec:sim} and \ref{sec:reco}, respectively;
the expected ITk tracking and vertexing performance is reported in
Section~\ref{sec:perf}; and concluding remarks can be found in
Section~\ref{sec:conclusion}.


%
\section{The ATLAS Inner Tracker}
\label{sec:atlas_itk}
%

\subsection{Overview}

The ITk is designed to be a silicon-based detector that comprises a pixel~\cite{ATLAS-TDR-30} and a strip subsystem~\cite{ATLAS-TDR-25}.
The pixel subsystem covers a pseudorapidity\footnote{
\begin{minipage}[t]{\textwidth}
ATLAS uses a right-handed coordinate system with its origin at the nominal interaction point (IP)
in the center of the detector and the \(z\)-axis along the beam pipe.
The \(x\)-axis points from the IP to the center of the LHC ring,
and the \(y\)-axis points upwards.
Polar coordinates \((r,\phi)\) are used in the transverse plane,
\(\phi\) being the azimuthal angle around the \(z\)-axis.
The pseudorapidity is defined in terms of the polar angle \(\theta\) as \(\eta = -\ln \tan(\theta/2)\) and is equal to the rapidity
\( y = \frac{1}{2} \ln \left( \frac{E + p_z}{E - p_z} \right) \) in the relativistic limit.
Angular distance is measured in units of \(\Delta R \equiv \sqrt{(\Delta y)^{2} + (\Delta\phi)^{2}}\).
\end{minipage}} range of $|\eta|<4.0$ and consists of five flat barrel layers and five layers of inclined or vertical rings for forward region coverage. The strip subsystem spans $|\eta|<2.7$ and includes four strip layers in the barrel region and six disks in the endcaps, all using double-sided modules.
The active area of the detector will be shielded from the neutron flux coming from the calorimeters by a neutron moderator surrounding the strip detector.

The ITk will operate within a solenoidal magnetic field of 2~T, which is aligned with the beam axis.
This magnetic field plays a crucial role by bending the trajectories of charged particles, allowing to estimate their transverse momentum, \pT.
The overall ITk configuration aims to achieve a minimum of nine measurements per track across the entire expected beam spot size, assuming a Gaussian distribution with a longitudinal width of 50 mm, and aims to reconstruct tracks left by charged particles with $\pT>1~\GeV$ passing through the detector in the $|\eta| < 4.0$ range.

\subsection{ITk layout}

\begin{figure}%
\centering
\subfloat[]{\includegraphics[width=0.8\linewidth]{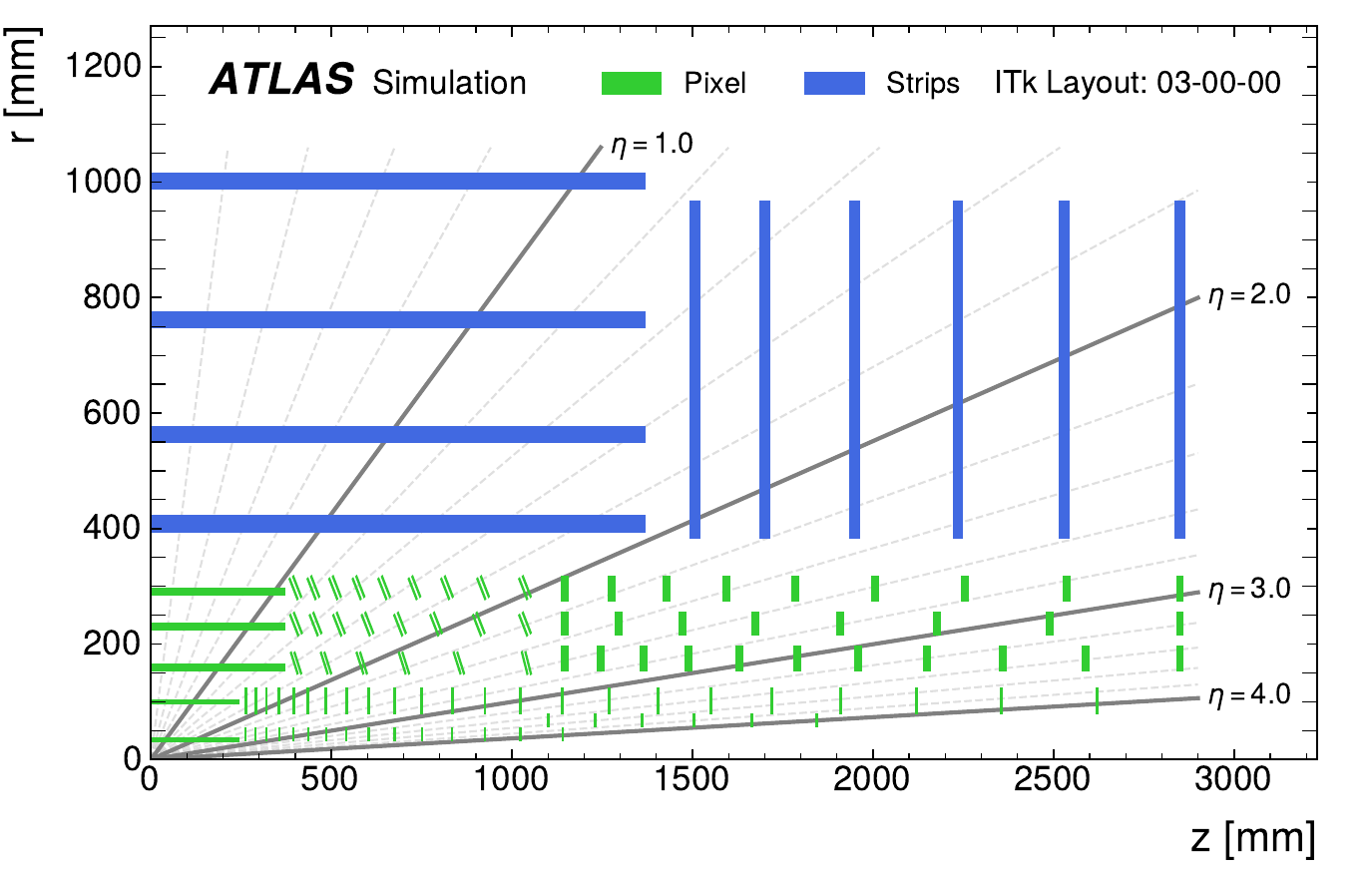}} \\
\subfloat[]{\includegraphics[width=0.8\linewidth]{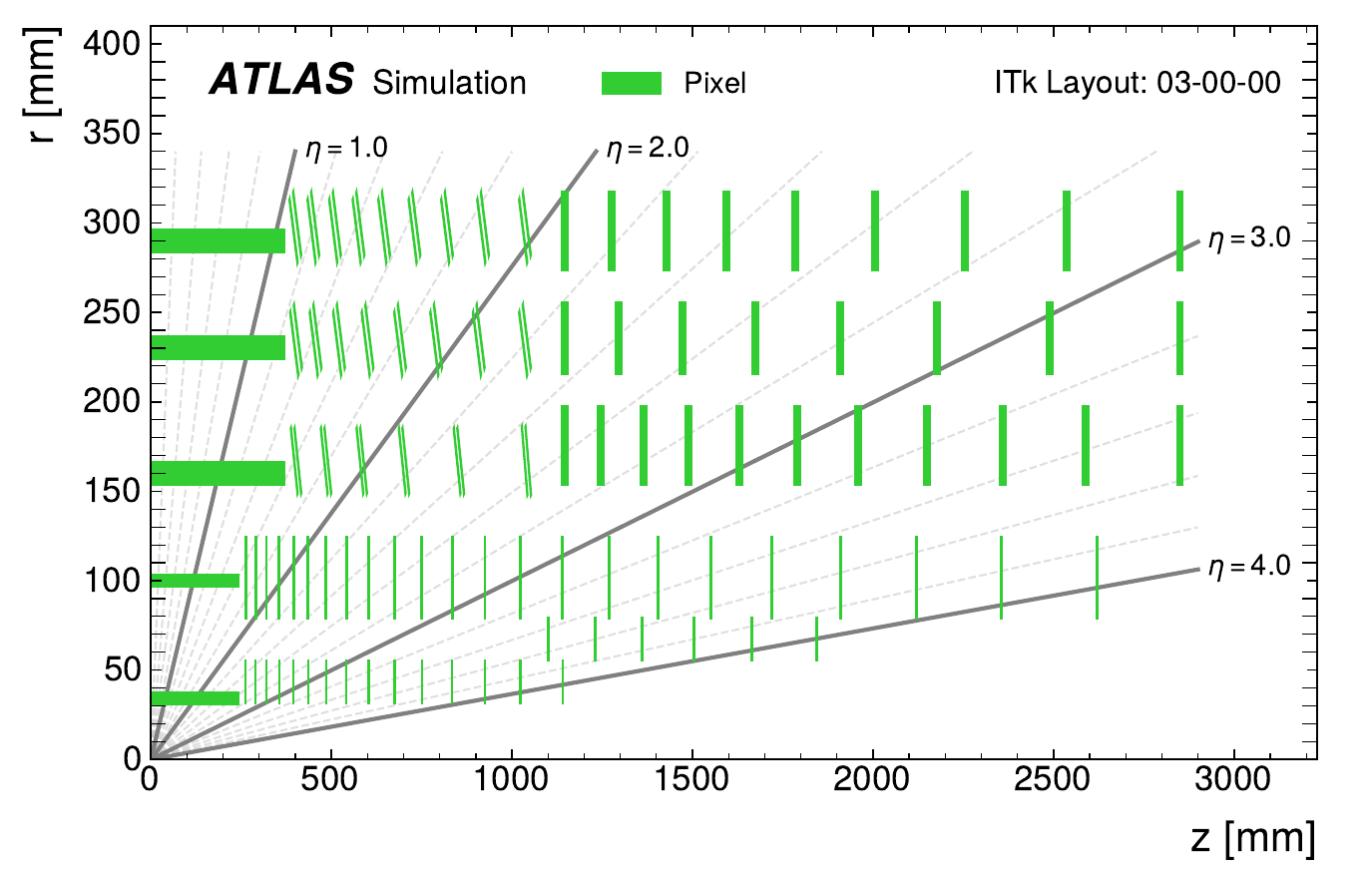}}\\
\caption{
(a) A schematic depiction of the ITk Layout 03-00-00 as presented in this document. (b) A zoomed-in view of the pixel detector. In each case, only one quadrant and only active detector elements are shown. The horizontal axis is along the beam line with zero being the nominal interaction point. The vertical axis is the radius measured from the interaction point. Thicker lines in the flat barrel sections are due to their tilt in the $\phi$-direction, while thicker lines in the endcap disks are due to their staggering along the $z$-direction, where $z$-positions vary between distinct values as a function of $\phi$ location.
\label{fig:ITK_layout}
}
\end{figure}

After the release of the technical design reports~\cite{ATLAS-TDR-25, ATLAS-TDR-30}, the detector design underwent additional refinement, as outlined in the following description. The resulting ITk layout is labeled as 03-00-00 and is presented in Figure~\ref{fig:ITK_layout}.

Updates were made to the tilt angles in the azimuthal direction of the barrel strip modules in the first two layers, and the radius of the first layer was modified. These changes were implemented to ensure sufficient space during installation, while still maintaining ample overlap between the modules of adjacent staves for alignment purposes and hermeticity\footnote{Hermeticity is the ability of the detector to provide complete coverage of the particle trajectories in all directions, without any gaps or regions where particles could potentially escape detection.}, as detailed in Table~\ref{Tab:StripRadiiTilts}.
Following the decision to incorporate the high-granularity timing detector (HGTD)~\cite{ATLAS-TDR-31} into the Phase-II upgrade of ATLAS, necessary adjustments were also made to the overall spatial dimensions of the ITk to accommodate the HGTD\@.
Table~\ref{Tab:StripDiskPos} provides a summary of the resulting disk locations.
The innermost pixel layer uses smaller modules based on 3D pixel sensor technology.
These modules consist of a single active element with front-end (FE) read-out chips bump-bonded to the sensors and grouped in triplets.
In contrast, all other layers use planar ``quad'' modules, where four FE chips are bump-bonded to the silicon sensor.

\begin{table}[htb]
\centering
\caption{Radius, extent in the $z$ direction, tilt angle and strip length for the strip barrel in the ITk Layout 03-00-00. The number of staves reported are for the full barrel. \label{Tab:StripRadiiTilts}}

\begin{tabular}{|c|c|c|c|c|c|} \hline
Barrel & Number of & Radius [mm] & |z| [mm] & Tilt                     & Strip  \\
layer & staves       &                       &              & angle [degrees] & length [cm]  \\ \hline
0  & 56 & 399 &  0 -- 1372   & 13 & 2.5 \\
1  & 80 & 562 &  0 -- 1372  & 12 & 2.5  \\
2  & 112 & 762 &   0 -- 1372 & 12 & 5 \\
3  & 144 & 1000 &   0 -- 1372   & 11 & 5  \\ \hline
\end{tabular}

\end{table}

\begin{table}[htb]
\centering
\caption{Position of disks of the strip endcap in the ITk Layout 03-00-00. \label{Tab:StripDiskPos}}

\begin{tabular}{|c|c|c|c|c|c|c|} \hline
Disk & Radius [mm] & Position [mm] \\ \hline
0  &  385  &  1512 \\
1  &  385  &  1702 \\
2  &  385  &  1952 \\
3  &  385  &  2237 \\
4  &  385  &  2532 \\
5  &  385  &  2850 \\ \hline
\end{tabular}

\end{table}

A redesign of the pixel endcap ring system took into consideration the altered envelope along the $z$-axis and the need to provide space for service routing in the radial direction.
To meet the latter requirement, adjustments were made to the radial extent of the support tube in the inner pixel barrel.
The redesign of the inner two layers of the pixel system involved incorporating staves for the traditional barrel region, featuring ``flat''  modules parallel to the beam line, and replacing the former inclined section with coupled rings sharing the same support structure for the two layers.
This change has the advantage of allowing the services to be routed to modules in the innermost pixel layer outward in the radial direction, thereby minimizing the material in the innermost part of the detector.
The transition to coupled rings required a re-optimization of the ring locations in the previously inclined section to ensure hermeticity and an adequate number of predicted hits within the constraints of the new support structure.

A substantial redesign was also implemented in the three outermost layers of the barrel pixel system to help the incorporation of quad modules, each consisting of $2 \times 2$ readout chips, within the inclined section.
These modules in the inclined sections are mounted on rings supported by a shell structure.
The transition from the initially planned dual chip ($2 \times 1$) to quad chip modules results in a 50\% reduction in the total number of rings.
This adjustment allows the entire ITk to be equipped with both single- and quad-chip modules, eliminating the necessity for dual-chip modules.
Additionally, it contributes to a reduction in the length of the inclined outer barrel.
The inclination angles of the modules in the $r-z$ plane were re-optimized in accordance with the new layout.
Furthermore, the length of the flat section in the three outermost pixel barrel layers was shortened to accommodate nine quad modules.

One of the most recent and significant modifications pertains to the innermost pixel layer.
Following an extensive technical review, the ATLAS Collaboration decided in early 2020 to adjust the ITk layout, bringing this layer closer to the beam line to enhance the tracking performance.
Consequently, the radius of the barrel section for this pixel layer was reduced from 39 to 34 mm, and the innermost point of the endcap rings was shifted inward from 36 to 33.2 mm.
The reduction in the radius of the barrel section allowed a decrease in the number of staves from 16 to 12, and a re-optimization of the number and position of endcap rings in the innermost pixel layers, maintaining hit coverage equivalent to the previous layout.
Additionally, the pixel sensor dimensions for the innermost pixel layer in the barrel were adjusted to 25~$\mu\mathrm{m}$ in the transverse plane and 100~$\mu\mathrm{m}$ in the longitudinal plane, while the rest of the pixel detector uses the original $50 \times 50$ $\mu\mathrm{m}^2$ pixels.
While the innermost layer will suffer higher radiation damage with this smaller radius, the two innermost pixel layers are designed to be replaceable before the end of the HL-LHC data-taking.

\begin{figure}[ht!]
\centering
\includegraphics[width=1.\linewidth]{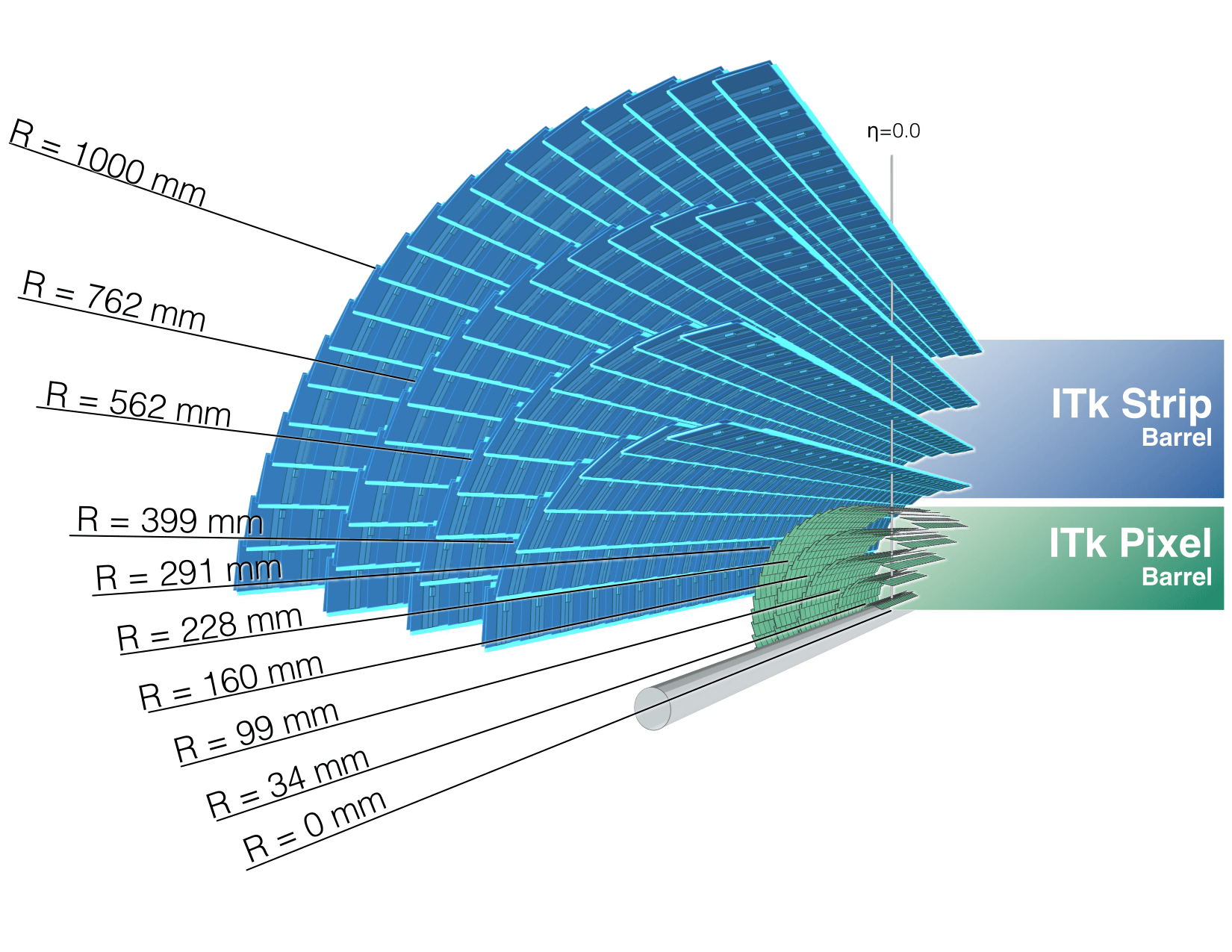}
\caption{
Transverse view of the Inner Tracker Layout 03-00-00 presented in this document.
\label{fig:ITk_display_barrel}
}
\end{figure}

\begin{figure}[ht!]
\centering
\includegraphics[width=1.\linewidth]{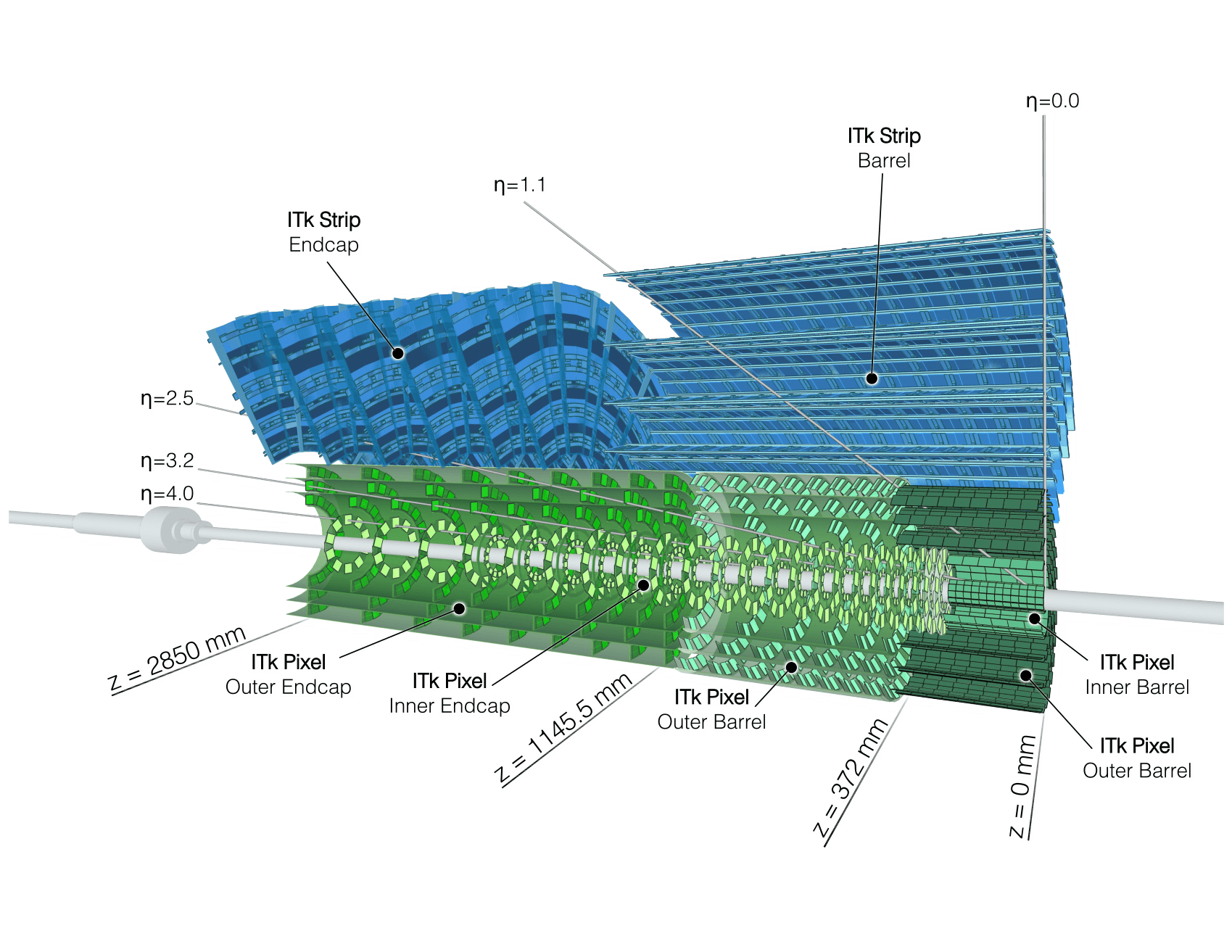}
\caption{
Longitudinal view of the Inner Tracker Layout 03-00-00 presented in this document.
\label{fig:ITk_display_full}
}
\end{figure}

The updated layout of the pixel barrel and endcaps is summarized in Tables~\ref{Tab:pixel_flat} and \ref{Tab:pixel_endcaps}. A display of the full ITk layout presented in this document is shown in Figures~\ref{fig:ITk_display_barrel} and \ref{fig:ITk_display_full}.

\begin{table}[htb]

\centering
\caption{Parameters for the pixel flat barrel and inclined rings in the ITk Layout 03-00-00. The number of sensors per row refers to a complete stave in the central, flat part of the barrel where sensors are placed parallel to the beam line. The number of inclined rings refers to both sides of the detector.  ``Triplets'' consist of three connected read-out chips, each associated with a $2\times2~\mathrm{cm}^2$ sensor, while ``quad'' modules are made of four connected read-out chips associated with a single $4\times4~\mathrm{cm}^2$ sensor.}

\begin{tabular}{|c|c|c|c|c|c|c|c|c|} \hline
Barrel & Radius   & Rows of & Flat barrel   & Flat sensors  & Incl. rings               &         Incl.           & Module & Sensor \\
layer   & [mm]      & sensors  & $|z|$ [mm]  & per row          & $|z|$ [mm]              &        rings           & type    & dim. [$\mu\mathrm{m}^{2}$] \\ \hline
0      &        34   &        12    &   0 -- 245    &          24         &            --                 &               --         & triplets  & $25 \times 100$ \\
1      &        99   &        20    &   0 -- 245    &          12         &            --                 &               --         & quads  & $50 \times 50$ \\
2      &       160  &        32    &   0 -- 372    &          18         &     380 --1035         &      $2\times 6$  & quads  & $50 \times 50 $ \\
3      &       228  &        44    &   0 -- 372    &          18         &     380 -- 1035        &     $2\times 8$   & quads  & $50 \times 50$ \\
4      &       291  &        56    &   0 -- 372    &          18         &      380 --1035        &     $2\times 9$   & quads  & $50 \times 50$ \\ \hline
\end{tabular}

\label{Tab:pixel_flat}

\end{table}

\begin{table}[htb]

\centering
\caption{Parameters for the pixel endcaps in the ITk Layout 03-00-00. The radii refer to the innermost point of the sensors on a ring. The number of rings refers to both sides of the detector. ``Triplets'' consist of three connected read-out chips, each associated with a $2\times2~\mathrm{cm}^2$ sensor, while ``quad'' modules are made of four connected read-out chips associated with a single $4\times4~\mathrm{cm}^2$ sensor.}

\begin{tabular}{|c|c|c|c|c|c|c|} \hline
Ring  &    Radius   &   $|z|$ [mm]     &         Rings          &         Sensors       &  Module & Sensor \\
layer  &     [mm]      &                         &                            &        per ring          &    type   &  dim. [$\mu\mathrm{m}^{2}$]   \\ \hline
0     &     33.20     &    263 -- 1142   &   $2\times 15$    &           18               & triplets  & $50 \times 50$  \\
0.5    &     58.70     &  1103 -- 1846   &   $2\times 6$      &           30               & triplets  & $50 \times 50$  \\
1     &     80.00     &    263 -- 2621   &   $2\times 23$    &           20               & quads  & $50 \times 50$    \\
2     &    154.50    & 1145.5 -- 2850 &   $2\times 11$    &           32               & quads  & $50 \times 50$    \\
3     &    214.55    & 1145.5 -- 2850 &   $2\times 8$      &           44               & quads  & $50 \times 50$      \\
4     &    274.60    & 1145.5 -- 2850 &   $2\times 9$      &           52               & quads  & $50 \times 50$     \\ \hline
\end{tabular}

\label{Tab:pixel_endcaps}

\end{table}

\begin{figure}%
\centering
\includegraphics[width=0.8\linewidth]{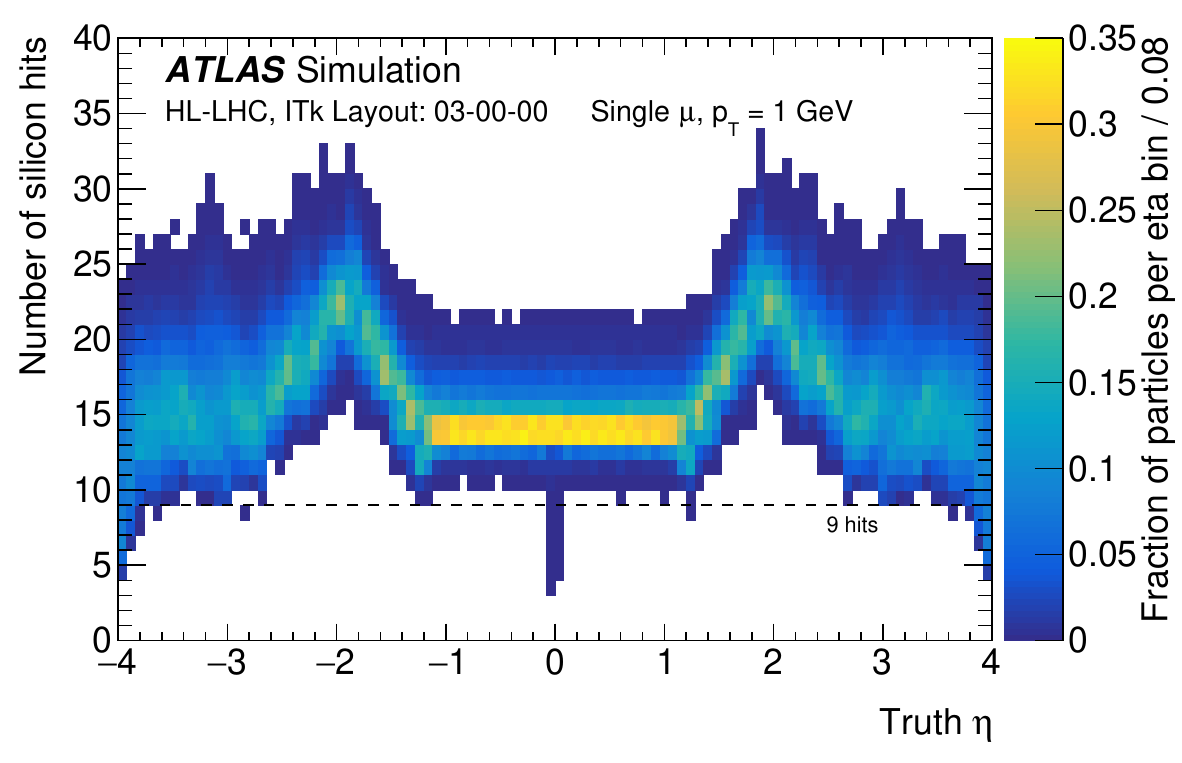}\\
\caption{Number of combined potential strip and pixel measurements along a particle trajectory as a function of the truth particle pseudorapidity for the ITk Layout 03-00-00. A sample of single-muon events with $\pT=1~\GeV$ is used. The muons are produced with a uniform distribution between 0~to~2~mm in transverse distance to the beam line and at fixed values of $z =$ -15 cm, 0 cm, and 15 cm, in equal amounts.
\label{fig:nhits}
}
\end{figure}

Figure~\ref{fig:nhits} depicts the overall count of expected pixel and strip measurements along a track depending on its pseudorapidity. The data are derived from simulated single-muon events with a transverse momentum of 1~\GeV. Across the entire detector acceptance, a minimum of nine measurements is maintained, with rare exceptions occurring only in cases where tracks pass through the gaps between pixel or strip barrel modules, in particular for particles produced at $z=0$ cm and $\eta \sim 0$. This feature is enhanced in Figure~\ref{fig:nhits} given the generation of particles at $z=0$ cm.

%
\section{Simulation}
\label{sec:sim}
%

\subsection{Geometry description and material}
\label{sec:geom}

A major effort has been put over the last years in a new implementation of the simulation of the ITk Pixel and Strip detectors using the ATLAS software suite~\cite{SOFT-2022-02}. This new implementation makes use of the GeoModel tool suite~\cite{GeoModel} and relies on atomic XML configuration files assembled to create the full detector. It benefits from a significantly improved clarity and long-term maintainability, the latter guaranteed by the commitment of the ATLAS Collaboration to use the GeoModel library as the basis for its detector simulation for Run 4 and beyond~\cite{CERN-LHCC-2020-015}.

\begin{figure}[t]
\centering
\includegraphics[width=.8\linewidth]{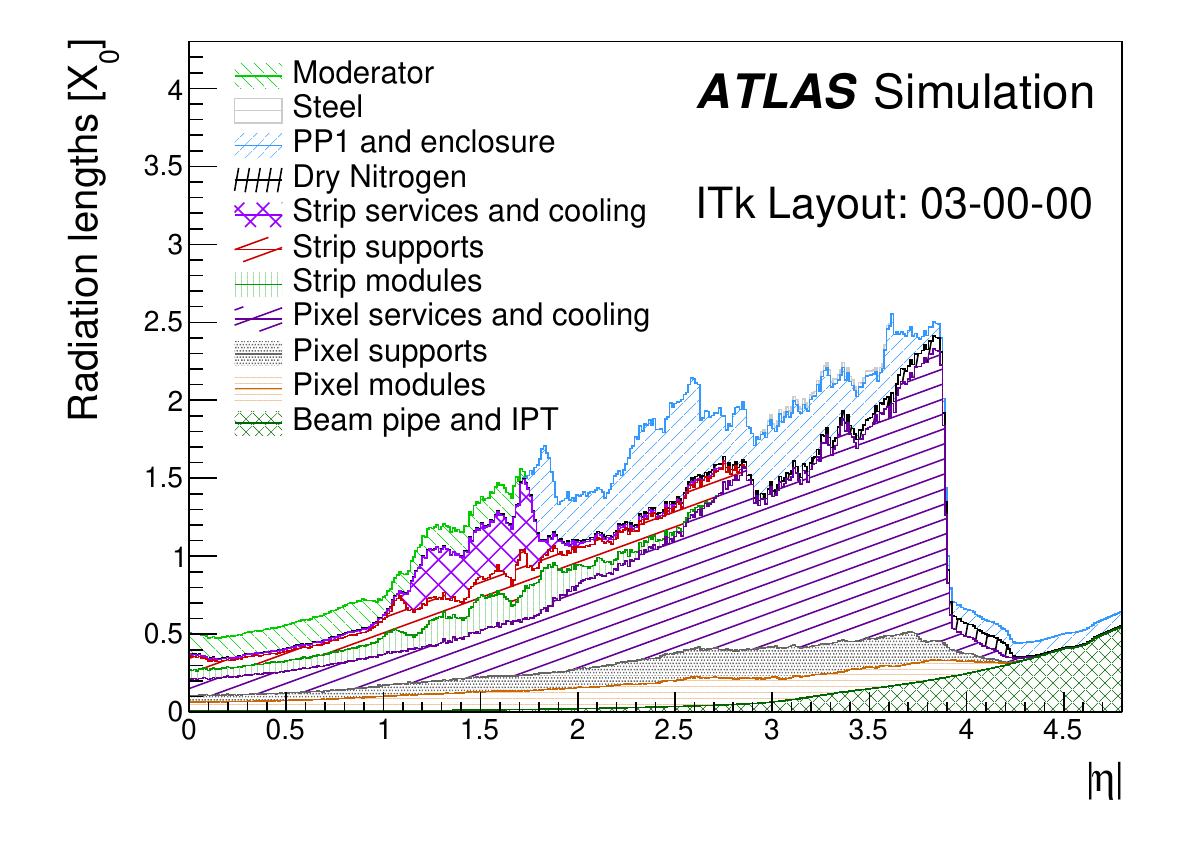}\\
\caption{
Integrated radiation length ($\mathrm{X}_0$) traversed by a straight track as a function of the absolute pseudorapidity $|\eta|$ at the exit of the ITk volume for the ITk Layout 03-00-00, broken down by sub-system and material category. The Inner Positioning Tube (IPT) is a support carbon-fibre cylinder just outside the beam pipe. The Patch Panel~1 (PP1) is an interface located in the endcaps that facilitates power distribution, signal transmission, and optical conversion between the detector modules and external systems. The moderator is located beyond the active detector area.
\label{fig:material_itk_X0_eta}
}
\end{figure}

\begin{figure}[t]
\centering
\includegraphics[width=.8\linewidth]{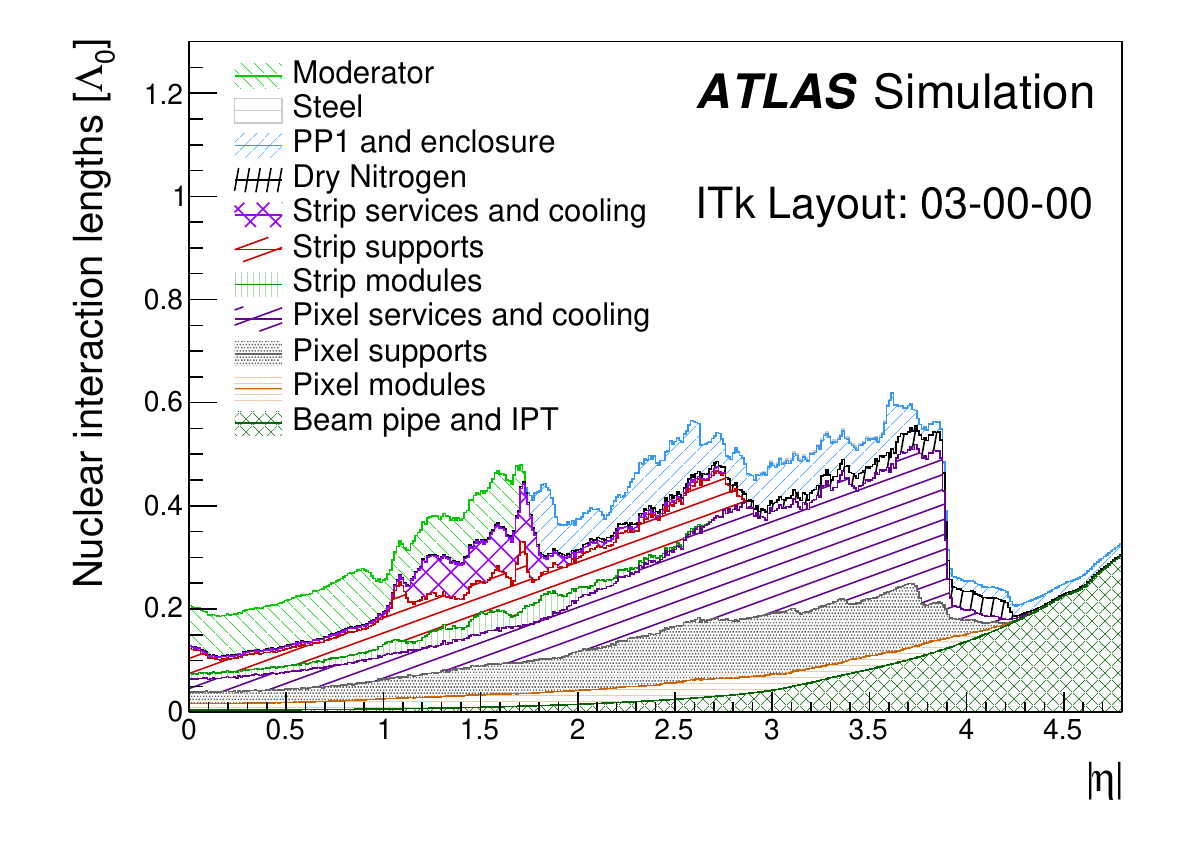}\\
\caption{
Integrated nuclear interaction length ($\Lambda_0$) traversed by a straight track as a function of the absolute pseudorapidity $|\eta|$ at the exit of the ITk volume for the ITk Layout 03-00-00, broken down by sub-system and material category. The Inner Positioning Tube (IPT) is a support carbon-fibre cylinder just outside the beam pipe. The Patch Panel~1 (PP1) is an interface located in the endcaps that facilitates power distribution, signal transmission, and optical conversion between the detector modules and external systems. The moderator is located beyond the active detector area.
\label{fig:material_itk_L0_eta}
}
\end{figure}

\begin{figure}[t]
\centering
\subfloat[]{\includegraphics[width=.49\linewidth]{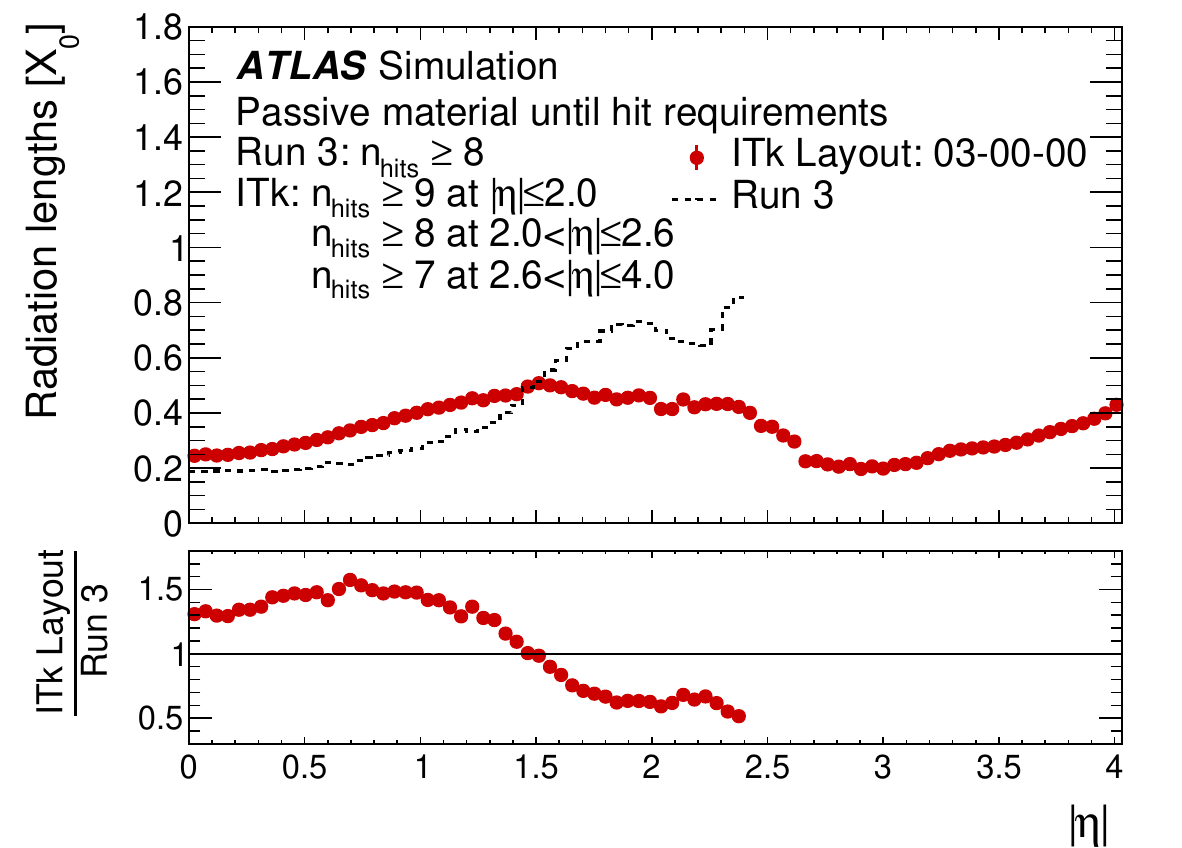}}
\subfloat[]{\includegraphics[width=.49\linewidth]{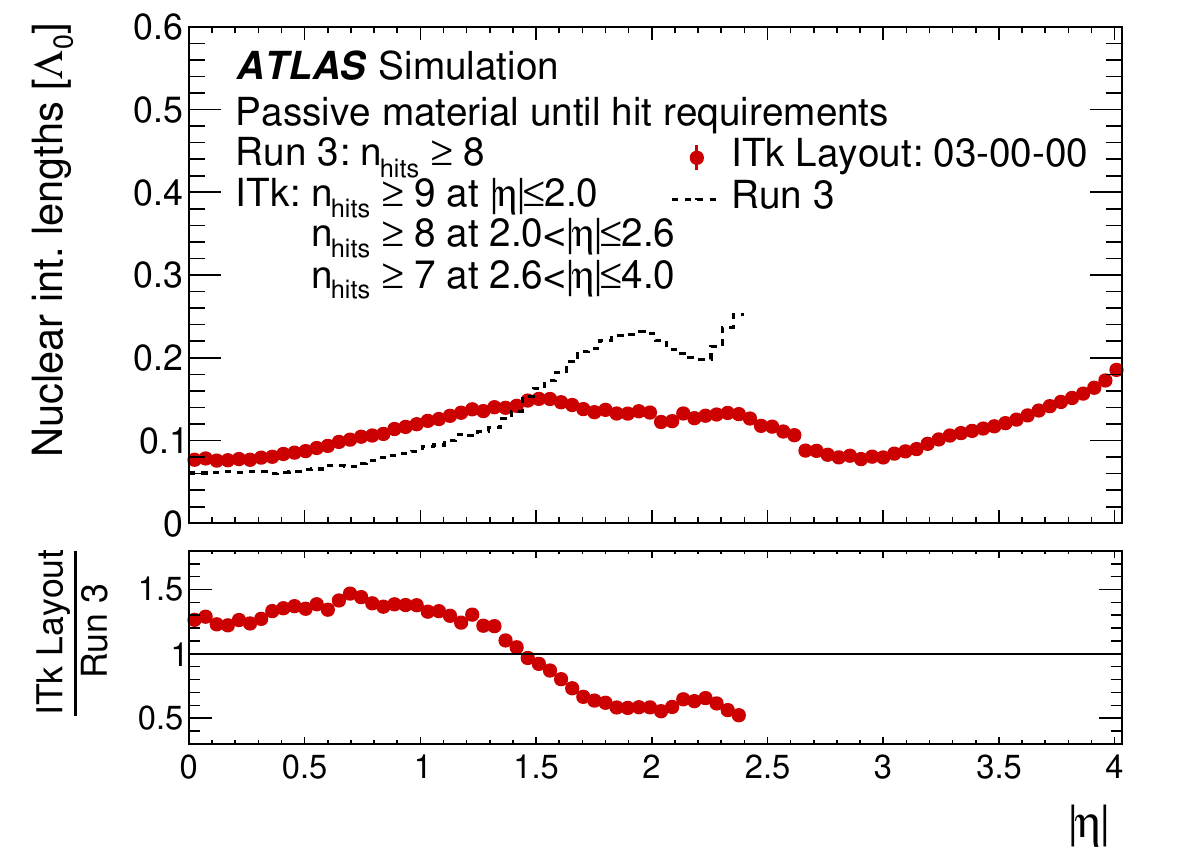}}\\
\caption{Material thickness in (a)~radiation lengths ($X_{0}$) and (b) nuclear interaction lengths ($\Lambda_{0}$) seen by particles until reaching the minimum number of hits required for track reconstruction. The ITk detector is compared with the Run 3 detector.}
\label{fig:materialhits}
\end{figure}

\begin{figure}[h]
\centering
\includegraphics[width=.75\linewidth]{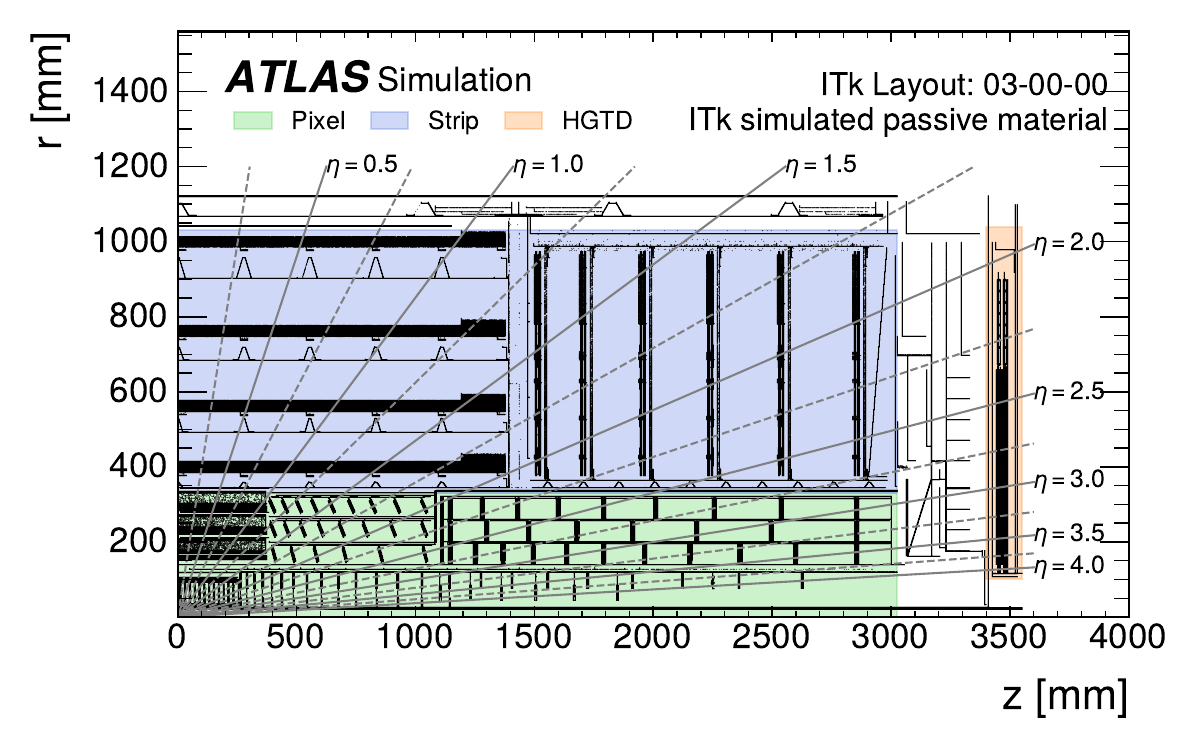}\\
\caption{
Location of the material for one quadrant of the ITk Layout 03-00-00.
\label{fig:material_itk}
}
\end{figure}

Figures~\ref{fig:material_itk_X0_eta} and \ref{fig:material_itk_L0_eta} present the integrated radiation length ($\mathrm{X}_0$) and nuclear interaction length ($\Lambda_0$)\footnote{The radiation length is defined as the average distance after which an electron's energy is decreased to 1/e of its initial value by material interactions, while the nuclear interaction length is defined as the mean distance between nuclear interactions sustained by a hadron.} traversed by a straight track as a function of the absolute pseudorapidity at the exit of the ITk volume.
The position of the origin of those tracks follows the expected HL-LHC beam spot distribution.
The mean radiation and interaction lengths traversed by particles before reaching the minimum number of hits required for track reconstruction (see Section~\ref{sec:Track_reco}) is shown in Figure~\ref{fig:materialhits}.
The material budget in the central region $|\eta|<1.5$ demonstrates an increase of up to 50\% relative to the Run 3 Inner Detector, while significant reductions in the material are present in the more forward region.
The material location in the $r-z$ frame simulated for the studies presented here is illustrated in Figure~\ref{fig:material_itk}.

\FloatBarrier

\subsection{Simulation samples}

To evaluate the performance of the generic track reconstruction, single-particle (muon, electron and charged pion) samples are simulated with a uniform $\eta$ distribution for $|\eta|<4.0$ and a fixed transverse momentum of 2, 10, or 100~\GeV.
To replicate the expected HL-LHC beam spot dimensions and positions, the particle origin positions are
generated based on a Gaussian distribution with a width of 50~mm centered at $z=0$ in the longitudinal direction, while a width of 12~$\mu$m centered at $r=0$ is used for the transverse direction.

In addition, a $t\bar{t}$ sample at $\sqrt{s}=14$~\TeV is simulated to assess the performance of the track and vertex reconstruction, with different average \pileup ranges from 50 to 70, 130 to 150, and 190 to 210, labeled respectively as $\left\langle \mu \right\rangle=60$, 140 and 200.
For every simulated event, the number of \pileup interactions is randomly sampled from a Poisson distribution with a mean corresponding to the average \pileup.
For comparison with the Run 3 detector performance, a $t\bar{t}$ sample at $\sqrt{s}=13.6$~\TeV produced with the Run 3 detector geometry and a uniform \pileup profile between 0~and 80 is used.
The production of \ttbar events is modeled using the \POWHEGBOX[v2]~\cite{Frixione:2007nw,Nason:2004rx,Frixione:2007vw,Alioli:2010xd} generator at next-to-leading-order in QCD with the \NNPDF[3.0nlo]~\cite{Ball:2014uwa} parton distribution functions (PDF) and the \hdamp parameter\footnote{The \hdamp parameter is a resummation damping factor and one of the parameters that controls the matching of \POWHEG matrix elements to the parton shower and thus effectively regulates the high-\pT radiation against which the \ttbar system recoils.} fixed to 1.5\,\mtop~\cite{ATL-PHYS-PUB-2016-020}, with $\mtop = 172.5~\GeV$ used for the top-quark mass.
The events are interfaced to \PYTHIA[8.230]~\cite{Sjostrand:2014zea} to model the parton shower, hadronization, and underlying event, with the A14 set of tuned parameters~\cite{ATL-PHYS-PUB-2014-021} and using the \NNPDF[2.3lo] set of PDFs~\cite{Ball:2012cx}. The decays of bottom and charm hadrons are performed by \EVTGEN[1.6.0]~\cite{Lange:2001uf}.
Only \ttbar events with a single leptonic $W$ boson decay are considered.
Minimum bias simulation samples are used in addition for specific studies, in which the average \pileup ranges are extended to cover the ranges 0~to 105, 95 to 175 and 165 to 210.

Finally, a sample of hypothetical $Z'$ particles with a mass of $m_{Z'} = 4~\TeV$ decaying with roughly equal probabilities into $b$-, $c$-, and light-quark jets is used to study the tracking performance in the core of high-$\pT$ jets. The sample is constructed so that the resulting jet $\pT$ spectrum is roughly uniform up to 5~\TeV.
The $Z'$ events are produced with the \PYTHIA[8.307]~\cite{Bierlich:2022pfr} Monte Carlo generator, using the A14 set of tuned parameters and the \NNPDF[2.3lo] PDF set. The decay of heavy-flavor hadrons is handled by the \EVTGEN[2.1.1]~\cite{Lange:2001uf} package. No additional proton--proton interactions are included in this sample.

The simulated events are processed through the full ATLAS detector simulation~\cite{SOFT-2010-01} based on the \GEANT toolkit~\cite{Agostinelli:2002hh}.

\subsection{Simulation of digitized readout signals}

The energy depositions created by \GEANT undergo a digitization process to mimic the behavior of detector electronics and produce the readout signals~\cite{IDET-2017-10-fix}.
This involves utilizing the energy deposited during each \GEANT step within the active silicon volume to calculate both the free charge and the drift time to the readout surface.
These calculations take into account various parameters such as sensor thickness, carrier mobility, depletion and bias voltages, and effects due to the magnetic field such as the directional drift of electrons, known as the Lorentz shift.
Additionally, the simulation accounts for contributions from noise and capacitive coupling to neighboring channels.
The algorithm then assesses the amount of charge collected in each channel, and if it surpasses a predetermined threshold, the corresponding channel is labeled as having fired.

The Bichsel straggling function~\cite{bib:bichsel} is used to simulate realistic charge depositions for thin pixel sensors.
Modules in the innermost layer of the barrel, equipped with pixels with a pitch of $25 \times 100~\mathrm{\mu m}^2$, use a discriminator threshold of $900~e$, while all other modules use a threshold of $600~e$.
Those thresholds are motivated by the expected behavior of the ITkPixV2 front-end chip to be used for the ITk Pixel detector~\cite{RD53}.
When a pixel registers a charge exceeding the discriminator threshold, a 4-bit measurement of time-over-threshold (ToT) is emulated through a calibration function. This emulation results in an average ToT of seven bunch crossings for a charge of $10000~e$.
All hits with a charge surpassing the threshold are attributed to the specific bunch crossing associated with the particle that contributed the most to the charge.
Noise is added to the charge from ionizing particles with a standard deviation of $75~e$ ($100~e$ for the innermost barrel layer); however, random/thermal noise is not yet included in the modeling.
The resolution of the charge measurement is in any case expected to be primarily influenced by the limited number of bits in the ToT measurement rather than the noise.
In the innermost layer, 3D pixel sensors are approximated as planar sensors, with Lorentz effects disabled to mimic the 3D sensor designs.


%
\section{Reconstruction}
\label{sec:reco}
%

\subsection{Track reconstruction and selection}
\label{sec:Track_reco}

\begin{figure}[t]
\centering
\includegraphics[width=.8\linewidth]{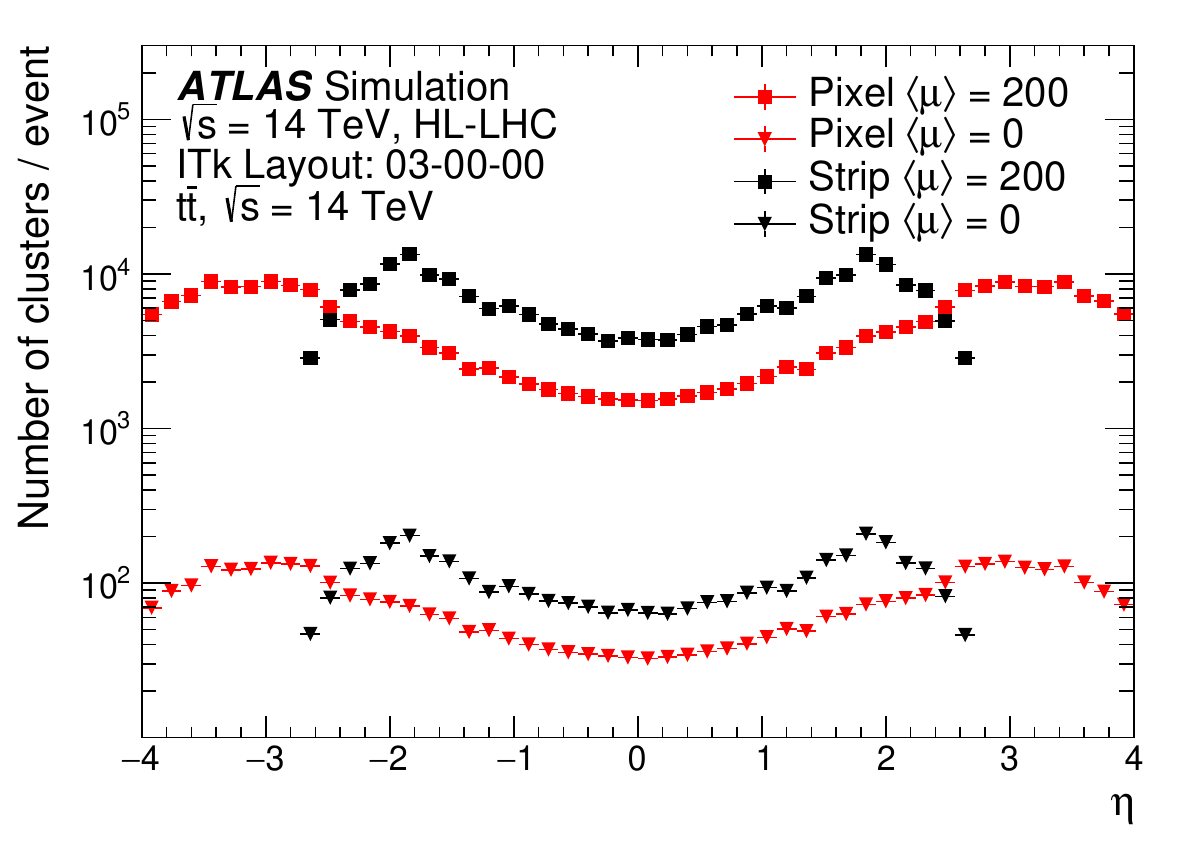}\\
\caption{
Distribution of the mean number of pixel and strip clusters vs $\eta$, at $\langle\mu\rangle=0$ and $\langle\mu\rangle=200$.
\label{fig:cluster_occupancy}
}
\end{figure}

The track reconstruction process\footnote{This section describes the main tracking reconstruction pass, aiming to reconstruct prompt particles originating from the interaction point. In addition, there exist some specialized passes, for instance aiming to reconstruct tracks produced away from the interaction point (i.e., large-radius tracking), or a calorimeter-seeded back-tracking pass aiming to recover tracks from photon conversions. Describing such passes is outside the scope of this work.} begins by creating clusters from the individual channels within the strip and pixel subdetectors.
In the case of the pixel detector, ToT information from each channel is accessible and is transformed into a representative charge measurement.
The ToT-based calibration exploits the charge distribution within the cluster, enhancing the precision of the cluster position determination and its estimated uncertainty~\cite{CERN-EP-2001-032}.
Additionally, the position estimate incorporates an assessment of the local incident angle of the track to further refine the cluster measurements.
The distribution of pixel and strip clusters can be seen in Figure~\ref{fig:cluster_occupancy}, showing a total occupancy of approximately $5\times 10^{5}$ clusters per event at $\langle\mu\rangle=200$.

Pixel and strip clusters are transformed into three-dimensional representations, referred to as space-points. Strip space-points are constructed combining clusters on opposite module sides, incorporating the relative stereo angle between the strips on the two sides and assuming that the track trajectory points to the interaction region.

Track finding starts with the track seeding stage~\cite{ATL-PHYS-PUB-2023-034}, during which seeds are built from triplets of pixel or strip space-points compatible with a helical track model.
These seeds define a search path, and a combinatorial Kalman filter (CKF)~\cite{Fruhwirth:1987fm, Lund_2009} identifies track candidates compatible with the initial seed parameters, extending the seeds with additional space-points.
In the CKF, the material interactions and noise contributions are approximated based on the number of detector layers crossed, rather than relying on detailed material and magnetic field description, which are used only in the final track fit.
Seeds are skipped if they are constructed from clusters already associated with a track candidate.
Two iterations of track finding are carried out sequentially, starting with strip seeds followed by pixel seeds, to enhance the overall track reconstruction efficiency.
Strip seeding is performed first, as cluster density is lower in the outermost regions of the detector. This approach allows pixel clusters associated with a strip seed to be removed, reducing the number of possible pixel cluster combinations in the subsequent pixel seeding, which is computationally advantageous.
Initially, a pion hypothesis is adopted to model the energy loss resulting from particle interactions with the detector material. If a track seed cannot be extended to form a complete track but aligns with an electromagnetic cluster detected in the calorimeter, the track finding algorithm is applied under an electron hypothesis.
This electron hypothesis accommodates energy losses of up to 30\% due to bremsstrahlung during each interaction with a detector material layer.

In the final stage of track reconstruction, an ambiguity-solving algorithm is executed to determine the definitive assignment of clusters to competing tracks.
During this stage, a precision fit is performed using a global $\chi^2$-minimization technique~\cite{Cornelissen:2008zza}, incorporating precise data regarding the material model and magnetic field.
Holes, defined as missing measurements on a track where active sensors should have registered hits, are assessed based on the track trajectory and information about the detector geometry and status.
The tracks are ranked according to their hit count and the quality of their fit.
Ambiguities between multiple track candidates are resolved by comparing their respective scores, and the track with the highest score is retained~\cite{ATL-SOFT-PUB-2007-007,IDTR-2022-04-fix}.

In a high track density environment, such as the core of high-$\pT$ jets, the separation between different charged particles is close to the detector granularity.
In such cases, charge depositions from different particles can overlap and be reconstructed as single merged clusters.
Tracks that share clusters are penalized in the ambiguity solver stage and they may consequently fail to satisfy the quality criteria and be rejected.
The identification and special treatment of merged clusters is thus crucial for ensuring a high track reconstruction performance in dense environments, as it has a strong impact on the track reconstruction efficiency and the precision of the reconstructed track parameters.
To optimize the tracking performance in this regime in Run~1, 2~and~3, ATLAS uses machine-learning-based algorithms dedicated to distinguishing clusters compatible with deposits from a single or multiple charged particles~\cite{PERF-2012-05,PERF-2015-08}.
This identification is performed only when a cluster is used by multiple tracks.
The implementation and optimization of such algorithms for the ITk geometry is ongoing as of this writing, and an emulation of those algorithms based on particle-level information that reproduces the performance of the Run 3 machine-learning based algorithms is conservatively used in the meantime.

The accepted tracks must satisfy the $\eta$-dependent criteria outlined in Table~\ref{Tab:req}.
The criteria are made $\eta$-dependent to accommodate for the non-trivial detector layout, both in terms of number of sensitive elements and of distribution of the material as a function of $\eta$.
They give in particular the possibility to maintain a uniform track reconstruction efficiency through the detector.

\begin{table}[htb]

\centering
\caption{Set of requirements applied during the track reconstruction in different pseudorapidity intervals. A hole is an intersection of the predicted trajectory of the particle with an active sensor element from which no measurement is assigned to the track (inactive sensors are not taken into account). The hole counting does not consider layers before the first and after the last hit. The longitudinal and transverse impact parameters, $z_0$ and $d_0$, are defined relative to the mean position of the beam spot.}

\begin{tabular}{|c|c|c|c|}
\hline
Requirements & \multicolumn{3}{c|}{Pseudorapidity interval}\\
\cline{2-4}
& $|\eta|\leq2.0$ & $2.0 < |\eta|\leq2.6$ & $2.6 < |\eta|\leq 4.0$ \\
\hline
Pixel + strip hits & $\geq 9$ & $\geq 8$ & $\geq 7$ \\
Pixel hits & $\geq 1$ & $\geq 1$ & $\geq 1$ \\
Holes & $\leq 2$ & $\leq 2$ & $\leq 2$ \\
$p_T$ [\MeV] & $> 900$ & $> 400$ & $> 400$ \\
$|d_0|$ [mm] & $< 2.0$ & $< 2.0$ & $< 10.0$ \\
$|z_0|$ [cm] & $< 20.0$ & $< 20.0$ & $< 20.0$  \\
\hline
\end{tabular}
\label{Tab:req}

\end{table}

\FloatBarrier

\subsection{Vertex reconstruction and selection}
\label{sec:Vertex_reco}
To distinguish the different proton--proton collisions in a single bunch crossing, the reconstructed tracks are used to identify the position of the interactions along the beam axis.
These are referred to as primary vertices (PVs).
Typically, only a single $pp$ interaction per recorded bunch crossing generates physics of interest for analysis.
The corresponding PV is known as the \emph{hard-scatter} vertex and is selected among all the reconstructed PVs. The position of the hard-scatter vertex is used for example as a reference for \pileup rejection, or in flavor-tagging algorithms.

Primary vertex reconstruction, both with the Run 3 Inner Detector and ITk, relies on the adaptive multi-vertex finder (AMVF) algorithm~\cite{ATL-PHYS-PUB-2019-015}, which simultaneously reconstructs multiple vertices, allowing tracks to be associated with several vertices at the same time.
The primary vertex finding process incorporates only a subset of the entire set of reconstructed tracks.
In addition to the track selection criteria applied during reconstruction, the selections outlined in Table~\ref{Tab:req_PV} are imposed to ensure reliable impact parameter estimates.

\begin{table}[htb]

\centering
\caption{Set of requirements applied to tracks during the AMVF vertex reconstruction, for the ID Run 3 and ITk Run 4 configurations. The uncertainties associated with $d_0$ and $z_0\sin\theta$, $\sigma(d_0)$ and $\sigma(z_0\sin\theta)$, are estimated from the global $\chi^2$ fit used in the track reconstruction, taking into account individual hit position uncertainties.}

\begin{tabular}{|c|c|c|}
\hline
Requirements & ID Run 3 & ITk Run 4 \\
\hline
Pixel hits & $\geq 1$ & $\geq 3$ \\
Pixel holes & $\leq 1$ & $\leq 1$ \\
SCT / Strip hits & $\geq 4$ & $\geq 0$ \\
Silicon hits & $\geq 6$ & $\geq 7$ \\
$p_T$ [\MeV] & $> 500$ & $> 900$ \\
$|d_0|$ [mm] & $< 4.0$ & $< 1.0$ \\
$\sigma(d_0)$ [mm] & $< 5$ & $< 0.35$ \\
$\sigma(z_0\sin\theta)$ [mm] & $< 10 $ & $< 2.5$ \\
\hline
\end{tabular}
\label{Tab:req_PV}
\end{table}

The AMVF uses a vertex seed finder to identify potential vertex candidates along the beamline in the $z$-direction:
\begin{itemize}
\item Each track is represented by a Gaussian probability distribution $P(r,z)$ centered at its point of closest approach $(d_0, z_0)$. The track densities are summed to identify global maxima in track density, representing the initial seeds used to reconstruct vertices.
\item Through an iterative process, a new vertex candidate, along with all tracks with $z_{0}$ agreeing with the vertex candidate within a 0.5~mm window, is added to the pool of vertex candidates managed by the adaptive Kalman filter multi-vertex fit.
\item All vertices undergo a refitting process using the adaptive Kalman filter multi-vertex fit, where each track may contribute to multiple vertices. Tracks with a low adaptive weight in the fit, computed based on a $\chi^2$ compatibility criterion with the vertex position described in Ref.~\cite{ATL-PHYS-PUB-2019-015}, are eliminated from the fit of a specific vertex.
\item The new vertex candidate is preserved if the vertex is identified with more than two tracks, the cumulative adaptive weights are not excessively low, and the new vertex is more than three standard deviations away from any other candidate.
\item This procedure is reiterated until there are no more vertex candidates or the maximum number of iterations is reached.
\end{itemize}

Among all of the reconstructed vertices in each event, the one with the largest $\Sigma \pT^{2}$ of its associated tracks, referred to as the hard-scatter vertex, is selected as the one of interest.
This choice is motivated by the fact that \pileup events are dominated by soft QCD interactions.


%
%
%

\section{Expected tracking and vertexing performance}
\label{sec:perf}

\subsection{Seeding performance}

The performance of the seeding algorithm is assessed using different figures of merit that rely on associating seeds with stable charged particles produced by Monte Carlo simulation. A seed is considered matched with a particle if more than half of its measurements originate from this particle, and the fraction of particles matched to at least one seed is defined as the \textit{physics} seeding efficiency.

The \textit{technical} seeding efficiency represents the efficiency to find seeds for reconstructable particles, and is defined as the fraction of seed matches among particles providing at least three measurements in the detector. By construction, this efficiency does not depend on the detector material or on the layout hermeticity, allowing the isolation of the algorithmic seeding efficiency.

\begin{figure}[!htb]
\centering
\subfloat[]{\includegraphics[width=.48\linewidth]{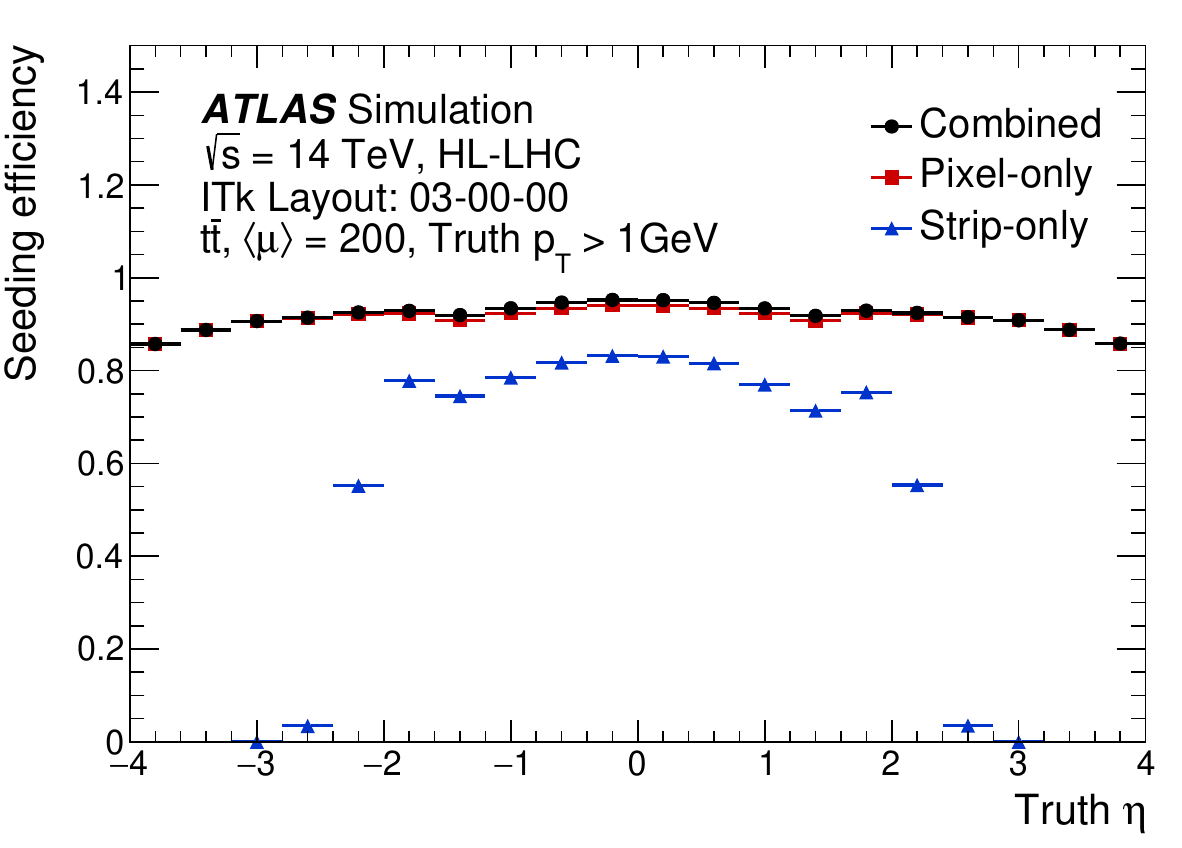}}
\subfloat[]{\includegraphics[width=.48\linewidth]{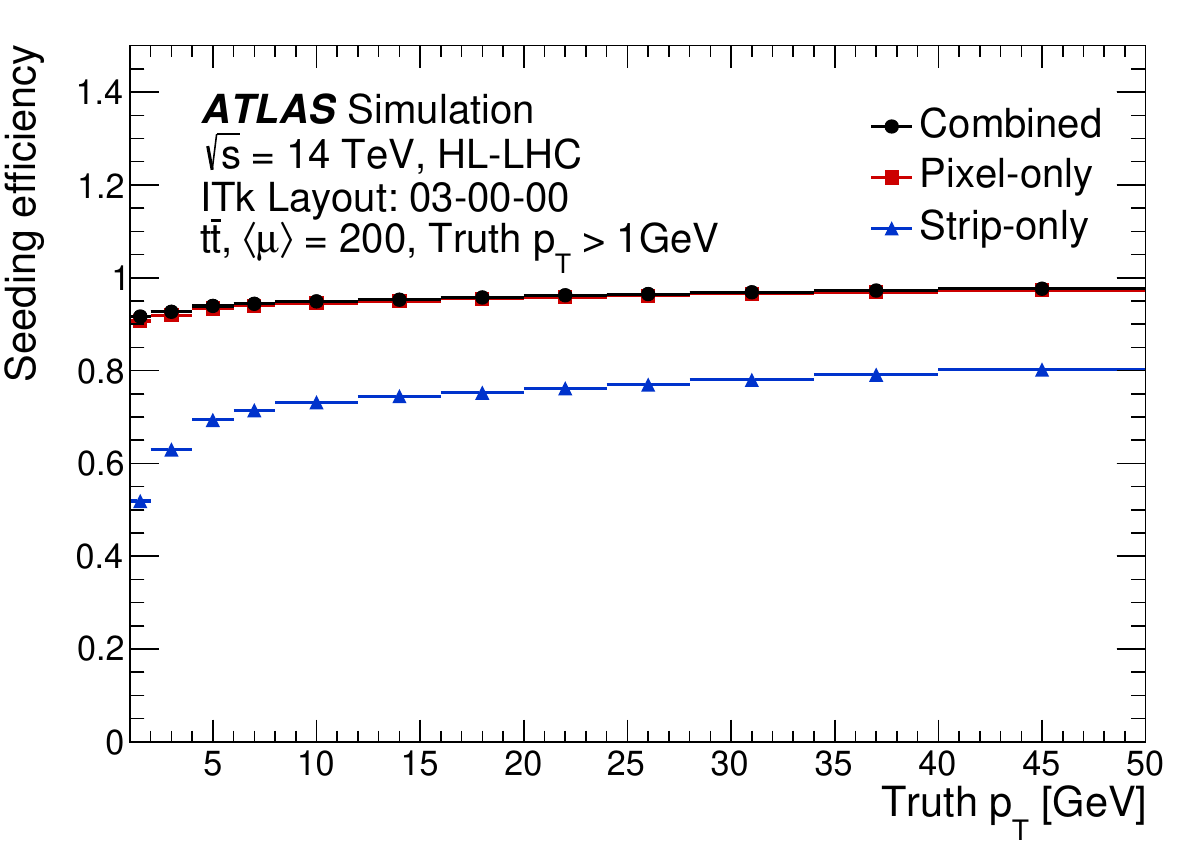}}\\
\caption{Expected physics seeding efficiency as a function of (a) $\eta$ and (b) $\pT$ for $t\bar{t}$ events at $\left\langle\mu\right\rangle$ = 200 for hard-scatter particles with $\pT >1~\GeV$. The seeding efficiency is shown separately for pixel-only, strip-only and both combined.}
\label{fig:seeding_eff}
\end{figure}

\begin{figure}[!htb]
\centering
\subfloat[]{\includegraphics[width=.48\linewidth]{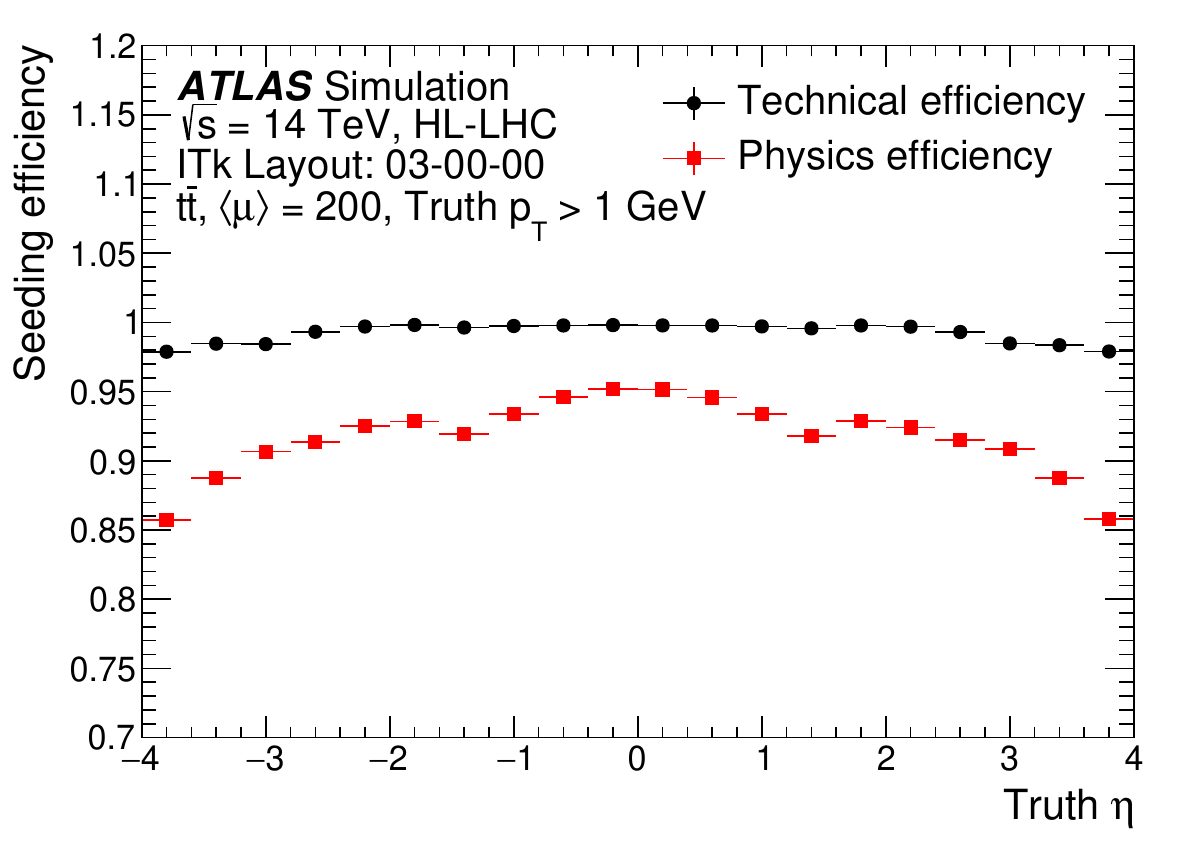}}
\subfloat[]{\includegraphics[width=.48\linewidth]{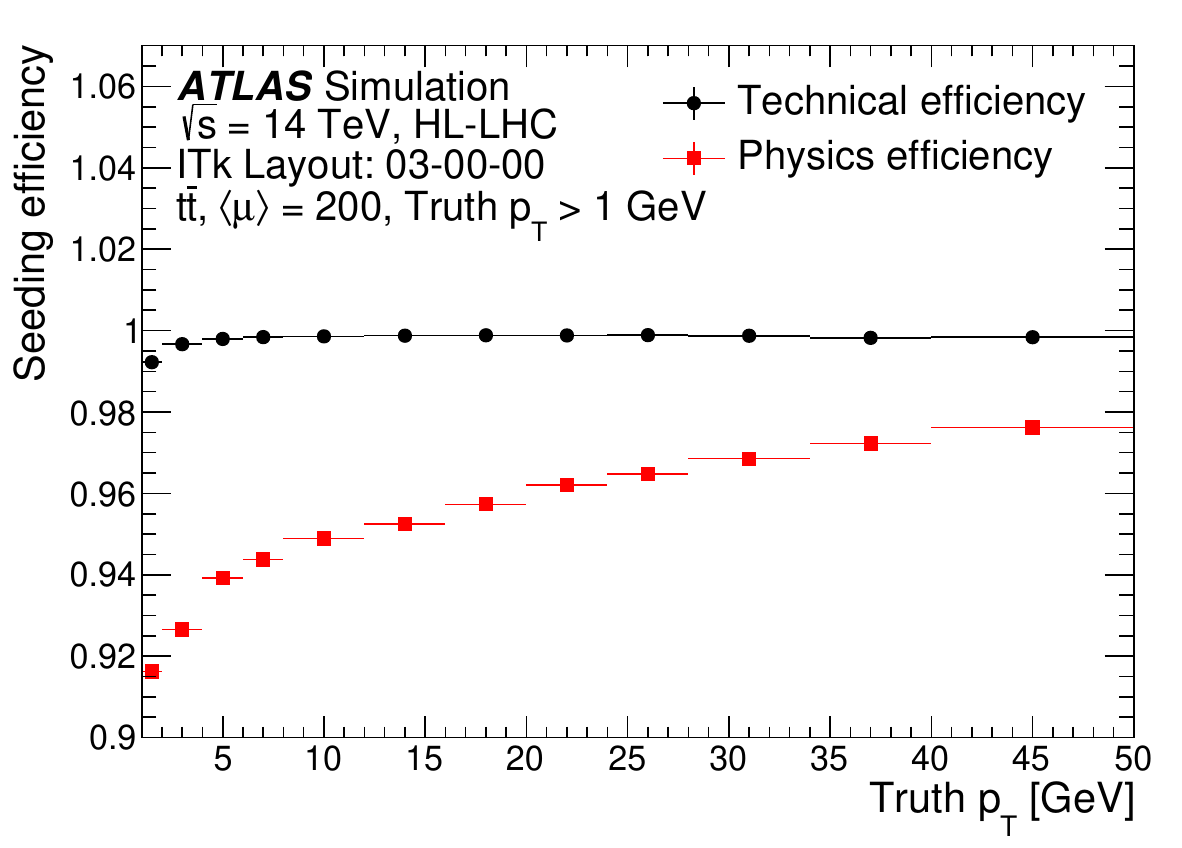}}\\
\caption{Expected technical and physics seeding efficiencies as a function of (a) $\eta$ and (b) $\pT$ for $t\bar{t}$ events at $\left\langle\mu\right\rangle=200$ for hard-scatter particles with $\pT>1~\GeV$. The efficiency is shown for the combined pixel and strip seeding.}
\label{fig:seeding_eff_tech}
\end{figure}

Figure~\ref{fig:seeding_eff} displays the physics seeding efficiency as a function of the particle $\eta$ and $\pT$ (labeled as ``Truth $\eta$'' and ``Truth \pT'') in $t\bar{t}$ events with an average \pileup of 200, for particles from the hard-scatter event with $\pT$ larger than 1~\GeV.
The technical and physics seeding efficiencies combining pixel and strip seeding are both displayed in Figure~\ref{fig:seeding_eff_tech}.
The $\eta$-dependence of the seeding efficiency directly reflects the number of available measurements in the detectors, which can be inferred from the detector layout visible in Figure~\ref{fig:ITK_layout}.
An inclusive physics seeding efficiency larger than 85\% is achieved over the whole phase space and reaches 95\% or more for particles with $\pT$ larger than 10~\GeV.
The lower efficiency at low $\pT$ is related to material interactions, more likely to affect low-$\pT$ particles.
This affects strip seeds more significantly, as they are reconstructed after the particles have crossed a larger amount of material.
The very high technical efficiency confirms that most of the physics inefficiency is indeed associated with material interactions.

\begin{figure}[!htb]
\centering
\subfloat[]{\includegraphics[width=.48\linewidth]{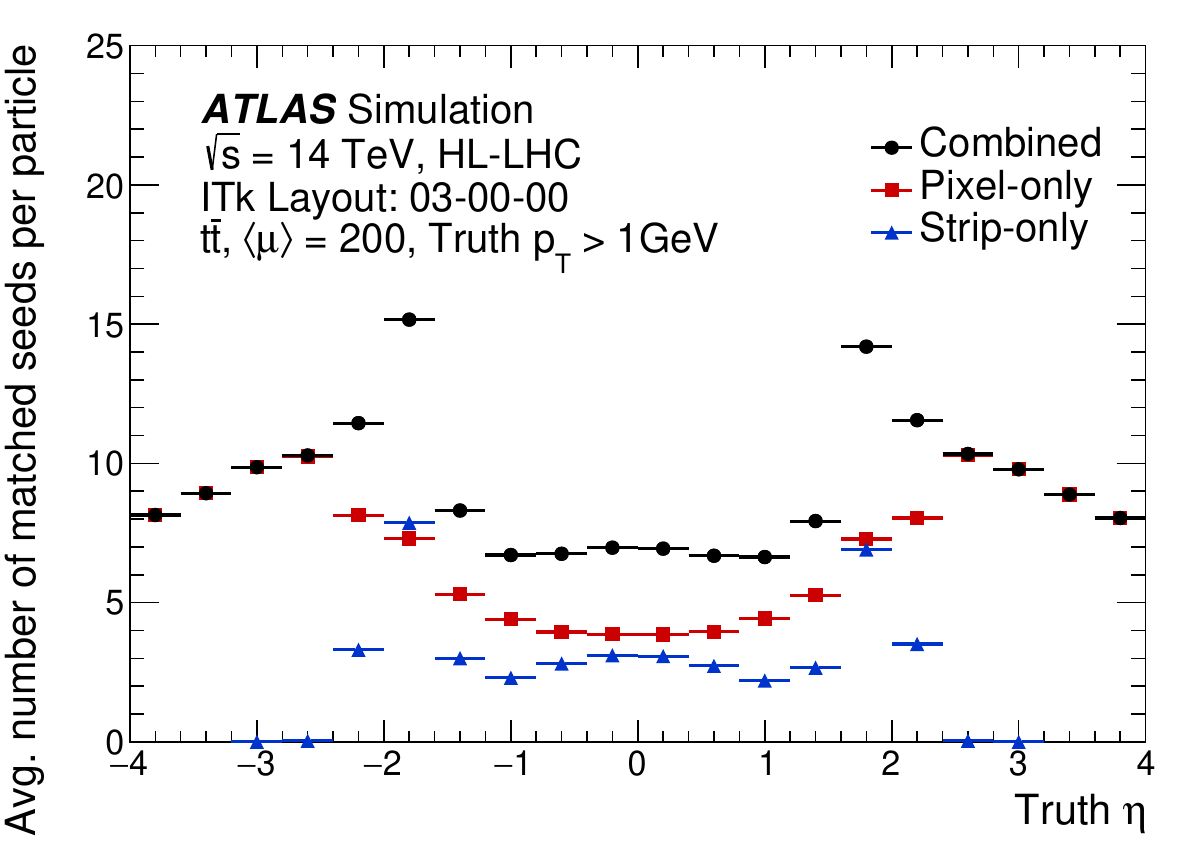}}
\subfloat[]{\includegraphics[width=.48\linewidth]{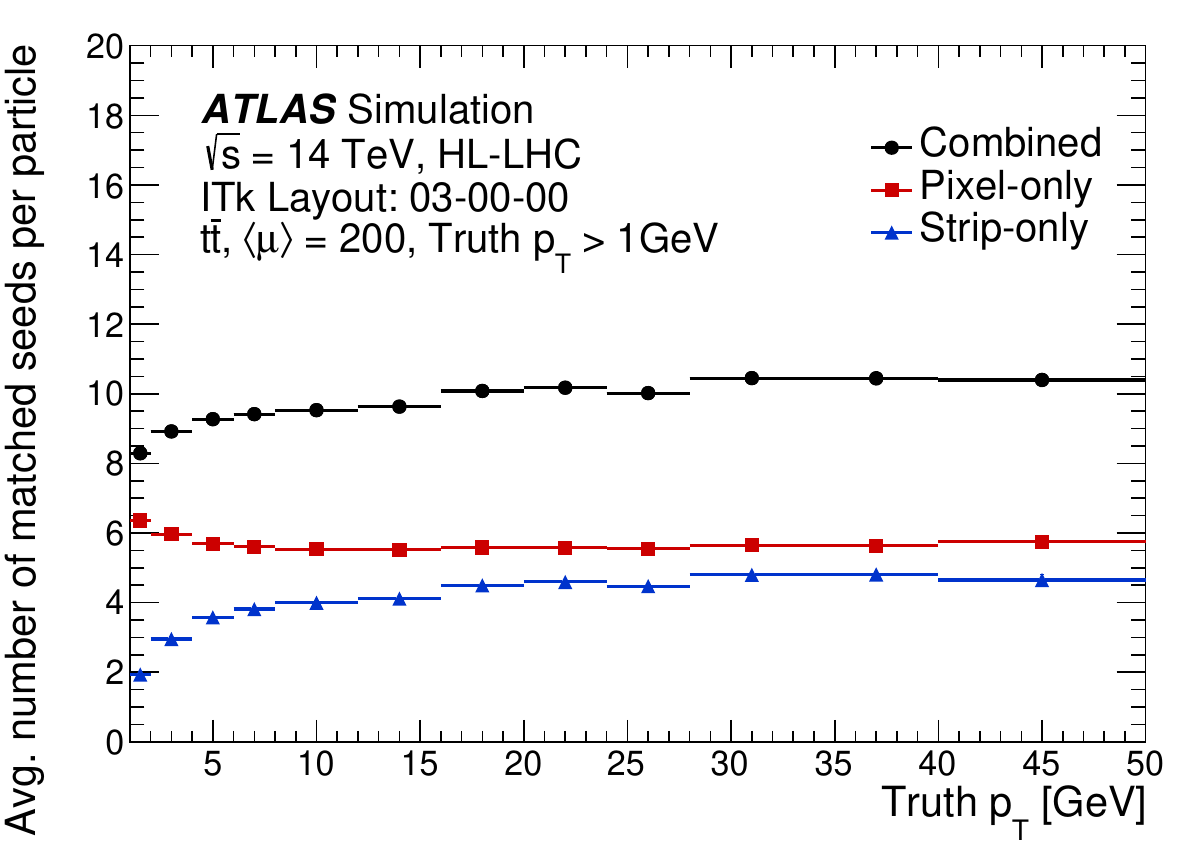}}\\
\caption{Expected number of seeds per particle as a function of (a) $\eta$ and (b) $\pT$ for $t\bar{t}$ events at $\left\langle\mu\right\rangle$ = 200 for hard-scatter particles with $\pT>1~\GeV$, shown separately for pixel-only, strip-only and both combined.}
\label{fig:seeding_redundancy}
\end{figure}

While the pixel physics seeding efficiency is very close to the combined physics seeding efficiency, the strip seeds provide some useful redundancy in case of potential pixel detector defects.
Figure~\ref{fig:seeding_redundancy} presents the number of seeds per particle as a function of $\eta$ and $\pT$.
On average, more than six seeds are found for all particles, as a particle typically leaves many hits in the detector, allowing for the construction of multiple triplets (seeds) from these hits.
This abundance of seeds ensures the robustness of the seeding process against potential detector defects or misalignment.

\FloatBarrier

\subsection{Efficiency}

Similarly to seeds, two different definitions of the tracking efficiency are used to distinguish between pattern recognition effects and detector effects.
The physics tracking efficiency is defined as the fraction of charged particles associated with a reconstructed track.
To do the particle-to-track matching, information about individual particle contributions to a given simulated silicon hit is used.
The matching criterion is based on the weighted fraction of measurements used to reconstruct a track common with the charged particle of interest, defined as

\begin{equation}
R_{\mathrm{match}}=\frac{2 \times N^{\mathrm{pix}}_{\mathrm{common}} + N^{\mathrm{strip}}_{\mathrm{common}}}{2 \times N^{\mathrm{pix}}_{\mathrm{reco}} + N^{\mathrm{strip}}_{\mathrm{reco}}},
\end{equation}

with $N^{\mathrm{pix/strip}}_{\mathrm{reco}}$ ($N^{\mathrm{pix/strip}}_{\mathrm{common}}$) representing the number of clusters from the different subdetectors used in the reconstruction of a given track (associated both to the particle and the reconstructed track).
Since strip sensors are installed in back-to-back stereo pairs, a particle crossing a strip module will usually acquire two silicon hits, as opposed to a single hit per pixel module; the different weights therefore give equal importance to the two measurement types.
A charged particle is considered to be reconstructed if it is matched to a track with $R_{\mathrm{match}}>0.5$.

The \textit{technical} tracking efficiency represents the fraction of track matches among the charged particles providing enough measurements in the detector to satisfy the reconstruction cuts.
It is used to assess the algorithmic efficiency of the tracking reconstruction and decouple it from the detector material or the layout hermeticity.

\begin{figure}[!ht]
\centering
\subfloat[]{\includegraphics[width=.48\linewidth]{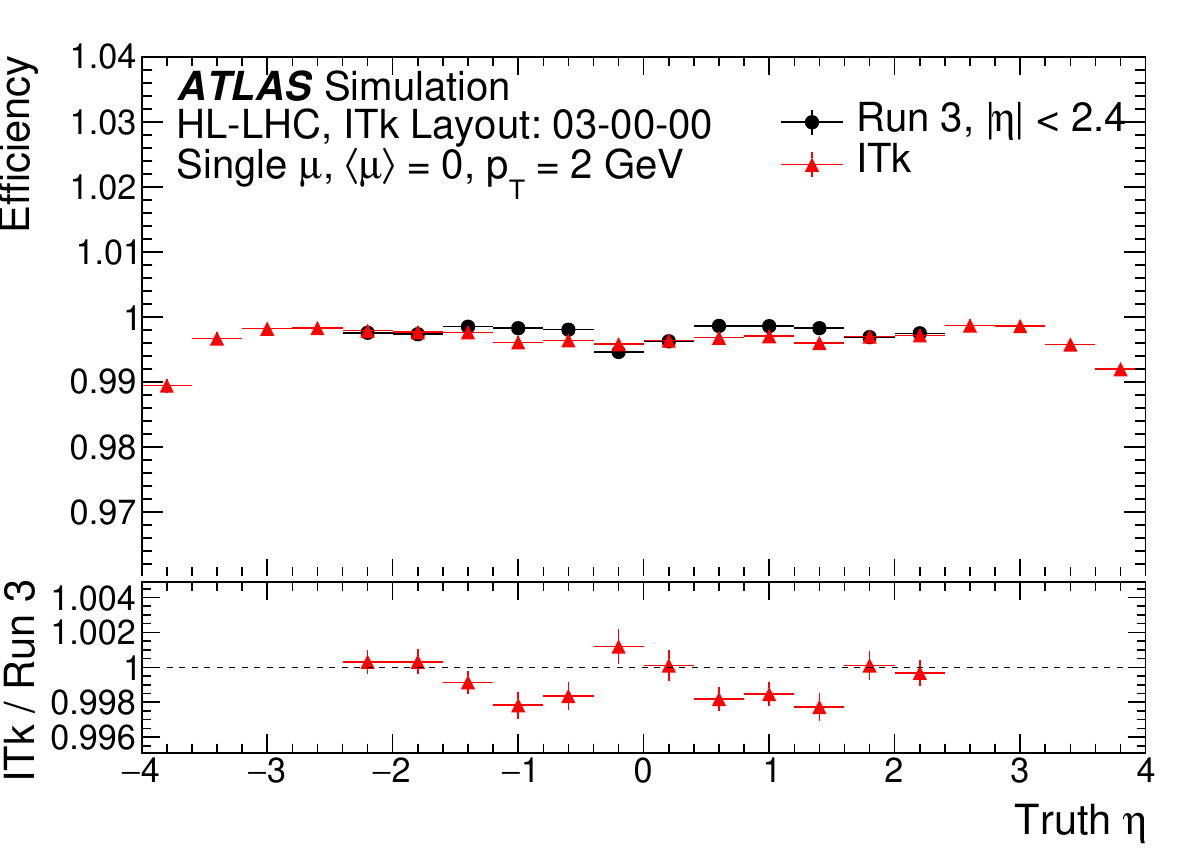}}
\subfloat[]{\includegraphics[width=.48\linewidth]{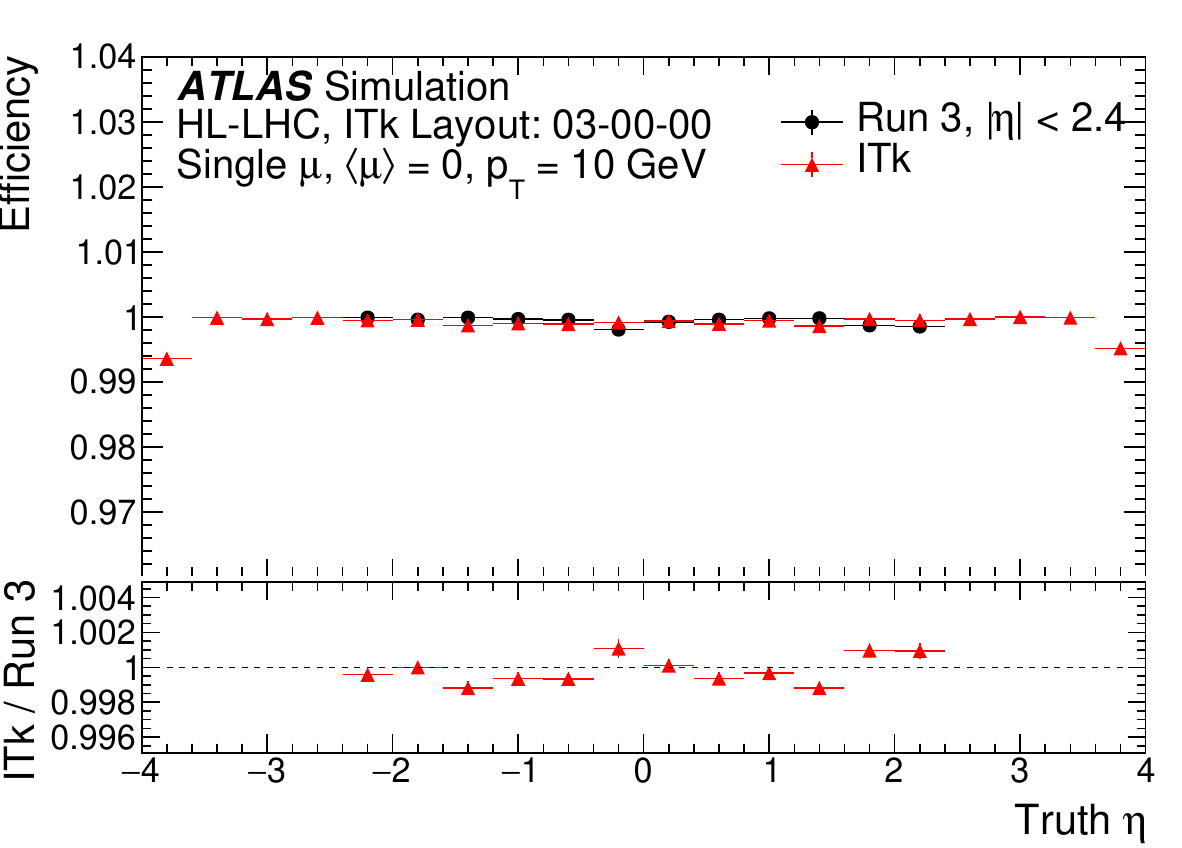}}\\
\subfloat[]{\includegraphics[width=.48\linewidth]{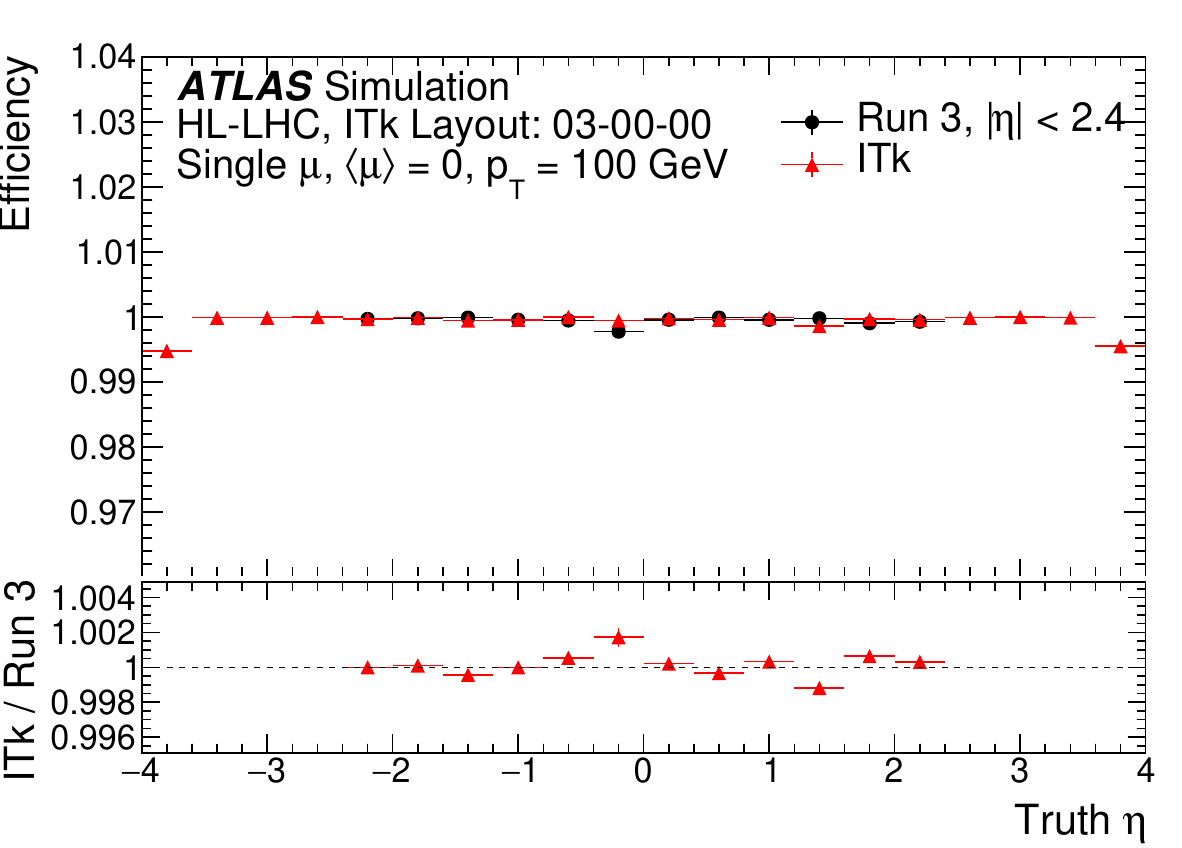}}\\
\caption{
Expected physics tracking efficiency for single muons with (a) $\pT=2$, (b) 10 and (c) 100~\GeV without \pileup. The results are compared for muons between the ITk and the Run 3 detector.
\label{fig:track_eff_sglmu_mu0}
}
\end{figure}

\begin{figure}[!htb]
\centering
\includegraphics[width=0.7\textwidth]{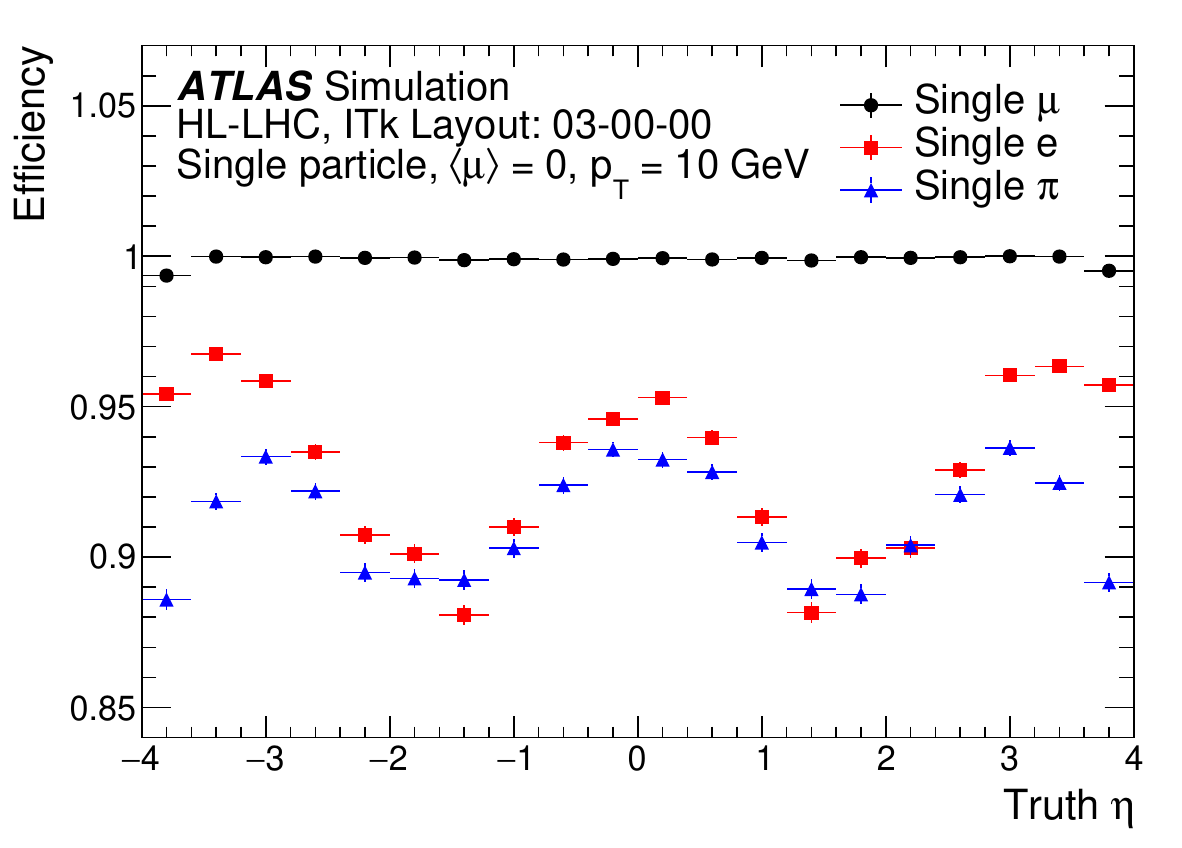}
\caption{Expected physics tracking efficiency for single muons, electrons and pions with $\pT=10~\GeV$ without \pileup.}
\label{fig:track_eff_eta_10gev_single_ptcl}
\end{figure}

The physics tracking efficiency for muons with $\pT=2$, 10 and 100~\GeV without \pileup is presented in Figure~\ref{fig:track_eff_sglmu_mu0}.
The expected efficiency is compatible with the one obtained with the ATLAS Run 3 detector. The efficiency is above 99.5\% for 2~\GeV muons and compatible with 100\% for larger $\pT$ up to $|\eta|=3.6$.
A small drop of efficiency down to 99\% is present in the very forward region of the ITk detector, due to the smaller number of available measurements in that region.
The physics tracking efficiency for 10~\GeV muons, electrons and pions is shown in Figure~\ref{fig:track_eff_eta_10gev_single_ptcl}.
Although the tracking efficiency for electrons and pions is lower due to their higher interaction rate with the detector material, it is expected to remain above 85\% for all types of prompt and stable charged particles.
In the forward region ($|\eta| > 3.0$), Figure~\ref{fig:materialhits} illustrates a steeper increase in nuclear interaction lengths compared with radiation lengths.
Consistent with these material distributions, the tracking efficiency for pions in this region is more significantly reduced than that for electrons.

\begin{figure}[!htb]
\centering
\subfloat[]{\includegraphics[width=.48\linewidth]{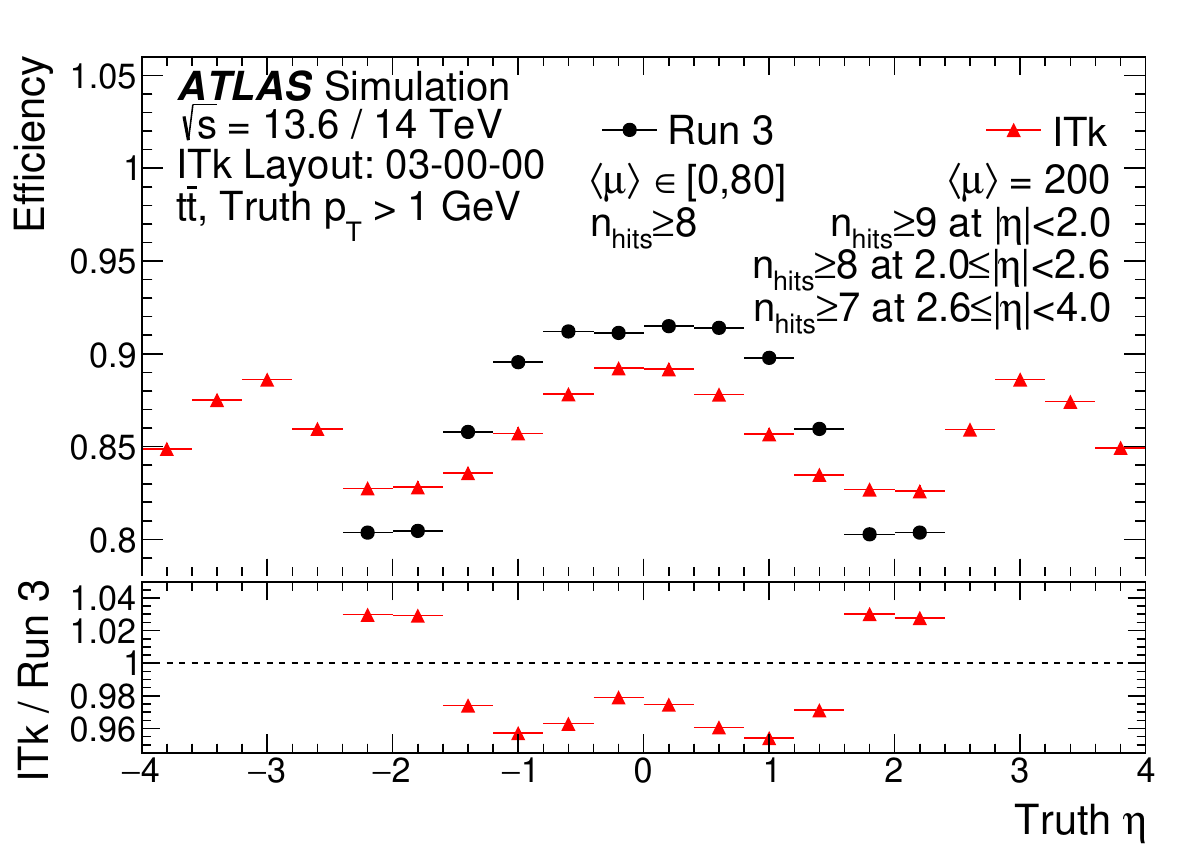}}
\subfloat[]{\includegraphics[width=.48\linewidth]{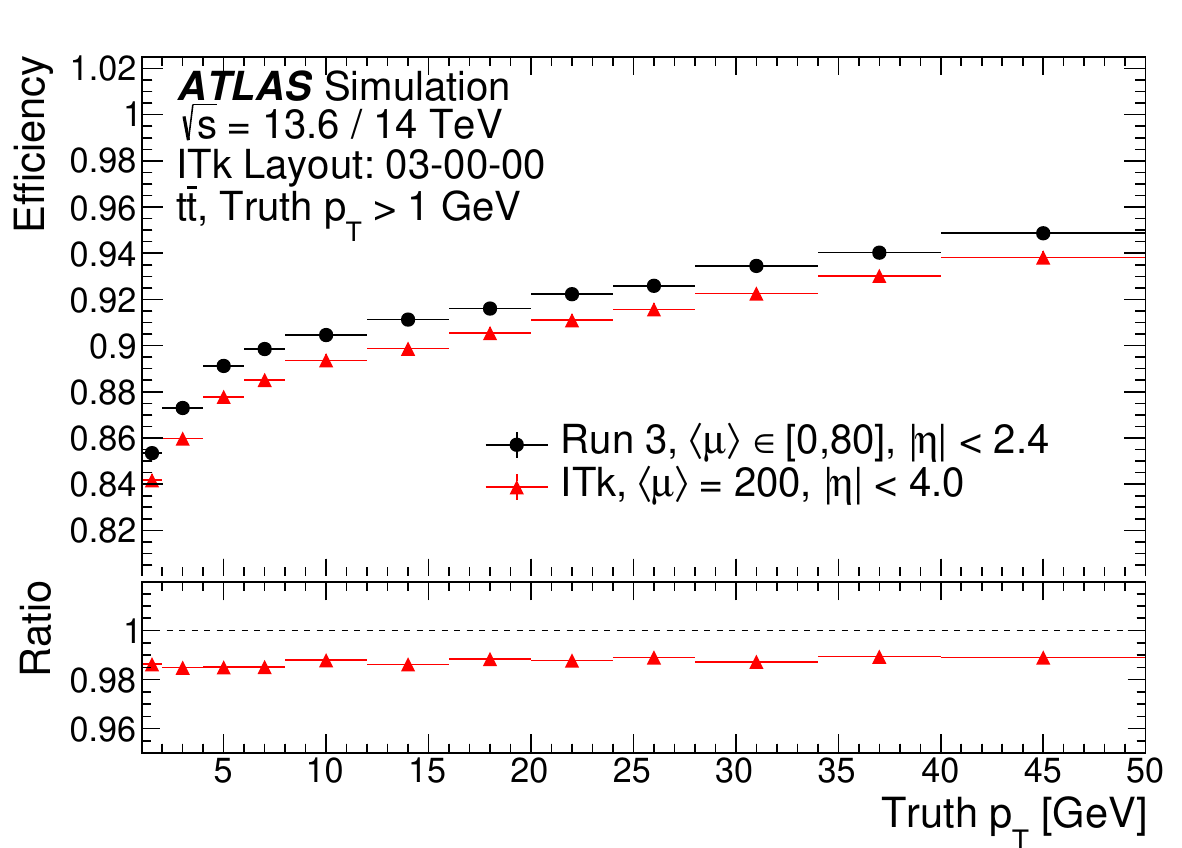}}\\
\caption{Expected physics tracking efficiency as a function of (a) $\eta$ and (b) $\pT$ in $t\bar{t}$ events at $\left\langle\mu\right\rangle$ = 200 for hard-scatter particles with $\pT>1~\GeV$ with the ITk detector compared with the Run 3 detector, in conditions with a uniform $\left\langle\mu\right\rangle$ distribution between 0~and~80.}
\label{fig:track_eff_ttbar}
\end{figure}

Figure~\ref{fig:track_eff_ttbar} emphasizes the expected tracking performance at $\left\langle\mu\right\rangle$ = 200, showcasing the physics efficiency in $t\bar{t}$ events for particles with $\pT > 1~\GeV$ within the detector acceptance, stemming from the hard-scatter interaction.
A comparison is made between this efficiency and that obtained with the Run 3 detector, in conditions with a uniform $\left\langle\mu\right\rangle$ distribution between 0~and~80.
The physics efficiency in the central region of the ITk detector is expected to be maintained within 5\% of that of the Run 3 detector, despite the larger material budget of ITk discussed in Section~\ref{sec:geom}.
The efficiency achieved in the newly accessible forward region with $2.5|<|\eta|<4.0$ is similar to the one achieved in the central region of ITk.

\begin{figure}[!htb]
\centering
\subfloat[]{\includegraphics[width=.48\linewidth]{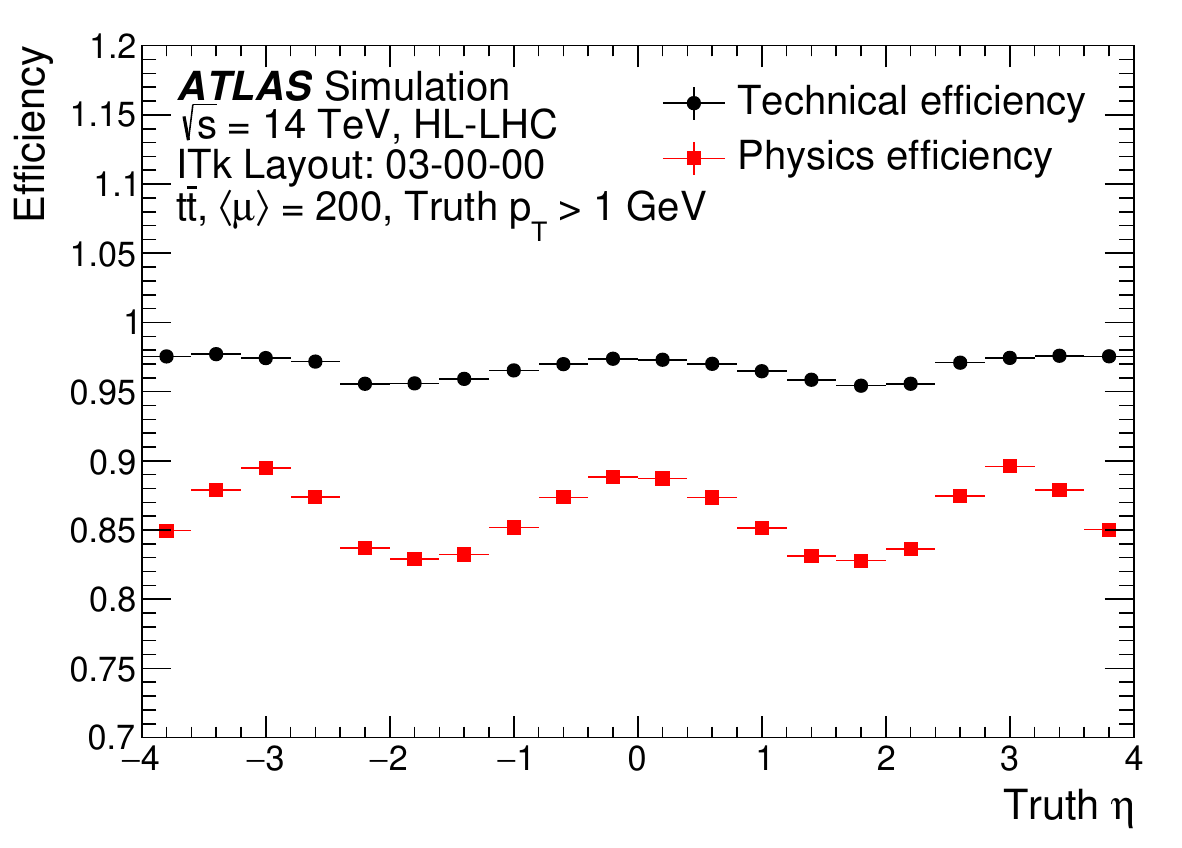}}
\subfloat[]{\includegraphics[width=.48\linewidth]{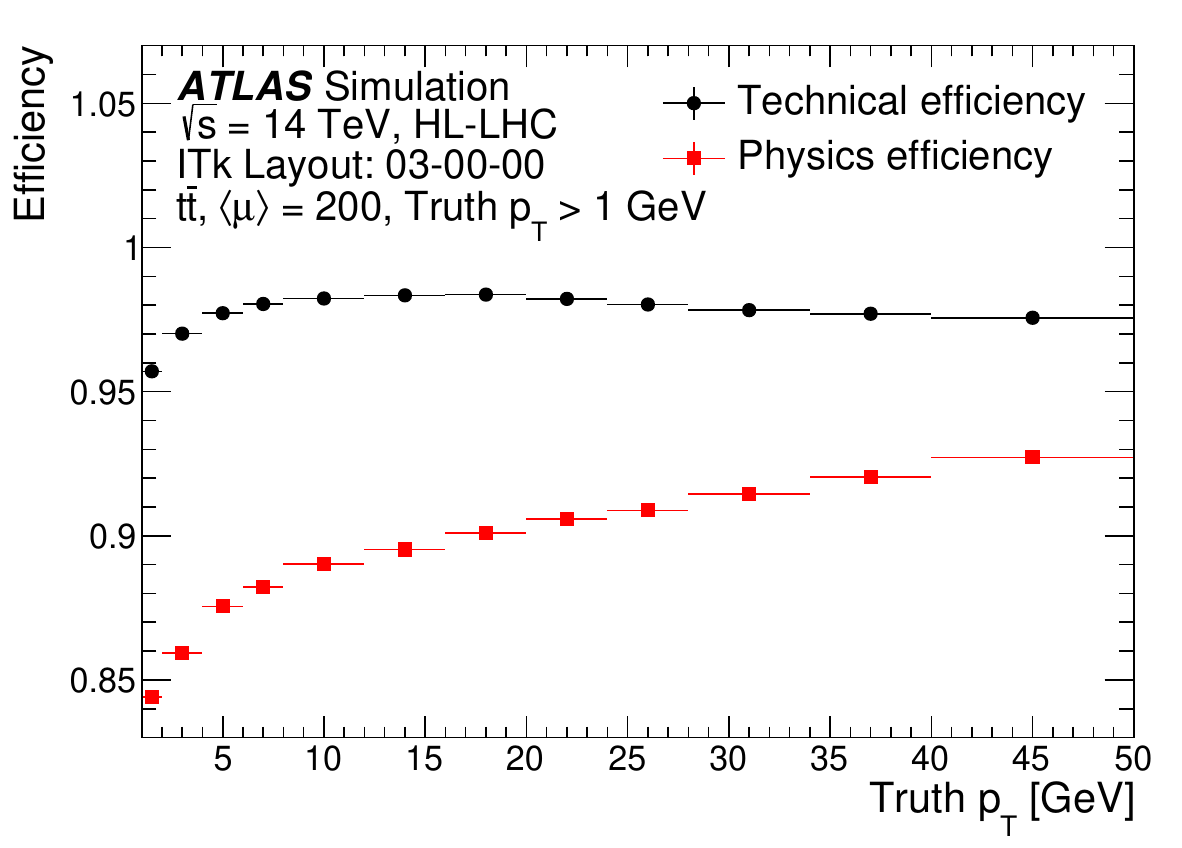}}\\
\caption{Comparison between the expected technical and physics tracking efficiencies as a function of (a) $\eta$ and (b)~$\pT$ in $t\bar{t}$ events at $\left\langle\mu\right\rangle=200$ for hard-scatter particles with $\pT>1~\GeV$ with the ITk detector.}
\label{fig:track_eff_tech_eff_ttbar}
\end{figure}

Figure~\ref{fig:track_eff_tech_eff_ttbar} shows the technical tracking efficiency and the physics tracking efficiency for tracks from the hard-scatter interaction in $t\bar{t}$ events at $\left\langle\mu\right\rangle$ = 200. The difference between the two quantities originates mainly from interactions of charged particles with detector material before reaching the required number of measurements.

\begin{figure}[!htb]
\centering
\includegraphics[width=.8\linewidth]{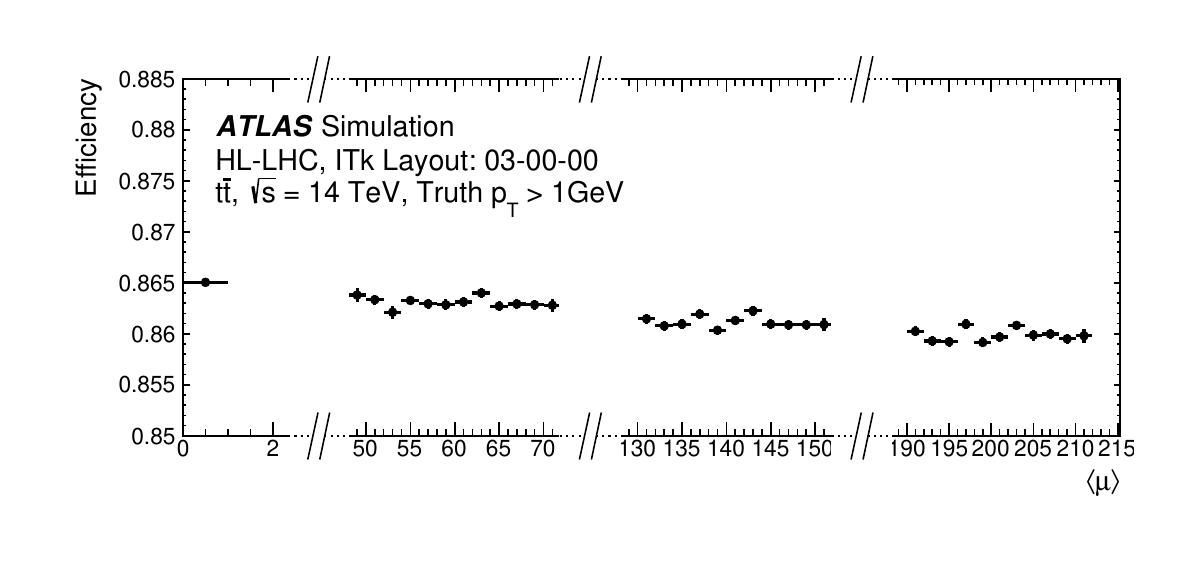}
\caption{Expected physics tracking efficiency as a function of \pileup in $t\bar{t}$ events for hard-scatter particles with $\pT>1~\GeV$ with the ITk detector.}
\label{fig:track_eff_vs_mu}
\end{figure}

\begin{figure}[!htb]
\centering
\includegraphics[width=.48\linewidth]{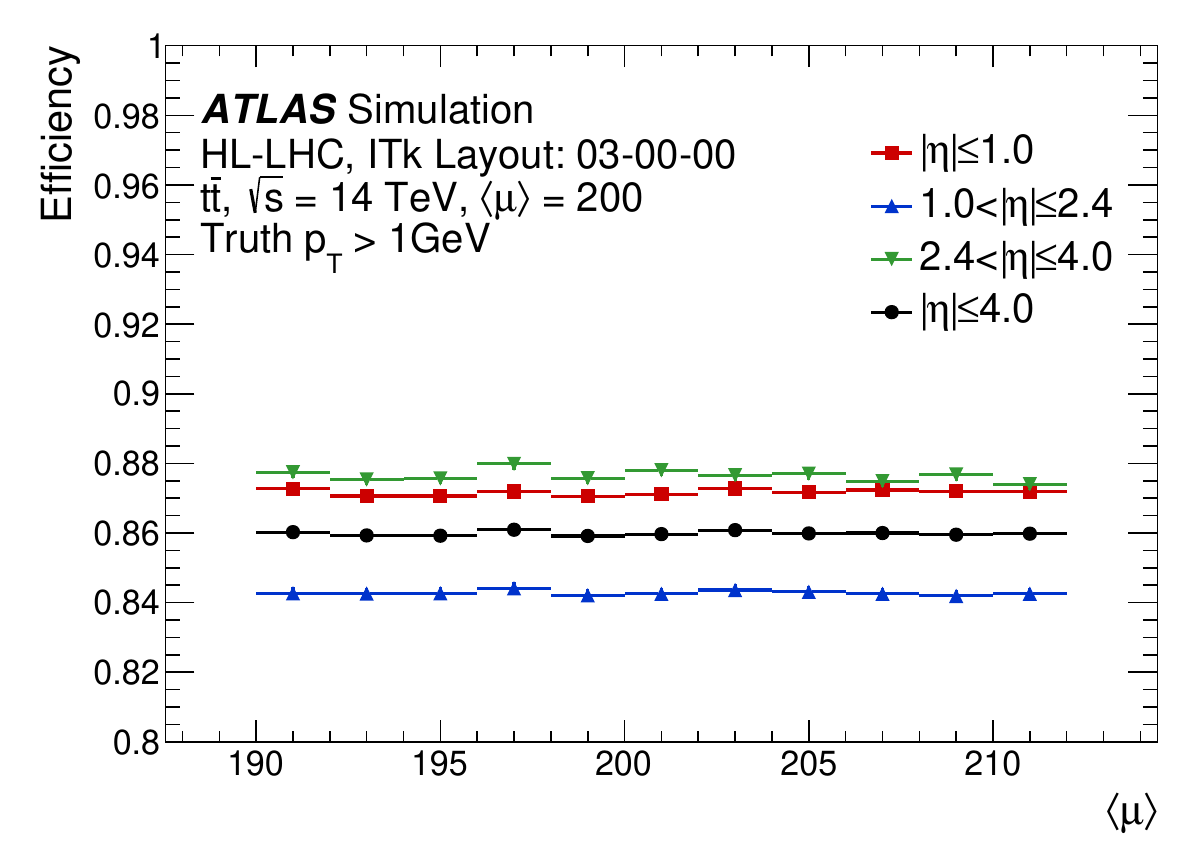}
\caption{Expected physics tracking efficiency as a function of \pileup in $t\bar{t}$ events at $\left\langle\mu\right\rangle$ = 200 for hard-scatter particles with $\pT>1~\GeV$ with the ITk detector in different $|\eta|$ ranges.}
\label{fig:track_eff_vs_mu_eta}
\end{figure}

The stability of the physics tracking efficiency at different \pileup conditions and in different $|\eta|$ ranges is shown in Figures~\ref{fig:track_eff_vs_mu} and~\ref{fig:track_eff_vs_mu_eta}.
The efficiency at $\left\langle\mu\right\rangle$ = 200 is within 0.5\% of the efficiency achieved at $\left\langle\mu\right\rangle = 0$.
A similar \pileup robustness is observed in the different $|\eta|$ ranges considered.

In the following, the performance of track reconstruction within jets is shown for a $Z'(m=4~\TeV)$ Monte Carlo sample, where the rate of cluster merging is much higher than in $t\bar{t}$ events due to the larger fraction of high-$\pT$ jets.
For the studies presented here, jets are reconstructed by clustering energy deposits in the calorimeter with the anti-$\kt$ algorithm~\cite{Cacciari:2008gp} with a radius parameter $R=0.4$, implemented in \Fastjet \cite{Fastjet}.
Tracks are matched to jets based on their angular separation $\Delta R$(track, jet) from the jet axis, required to be lower than 0.4.
To isolate the effects correlated with dense hadronic environments from those due to interactions with the detector material, the efficiency is presented for jets with $|\eta|<1.2$, in the central region where the material budget is the lowest, and for particles with $\pT>10~\GeV$, less impacted by material effects.

\begin{figure}[!htb]
\centering
\subfloat[]{\includegraphics[width=.48\linewidth]{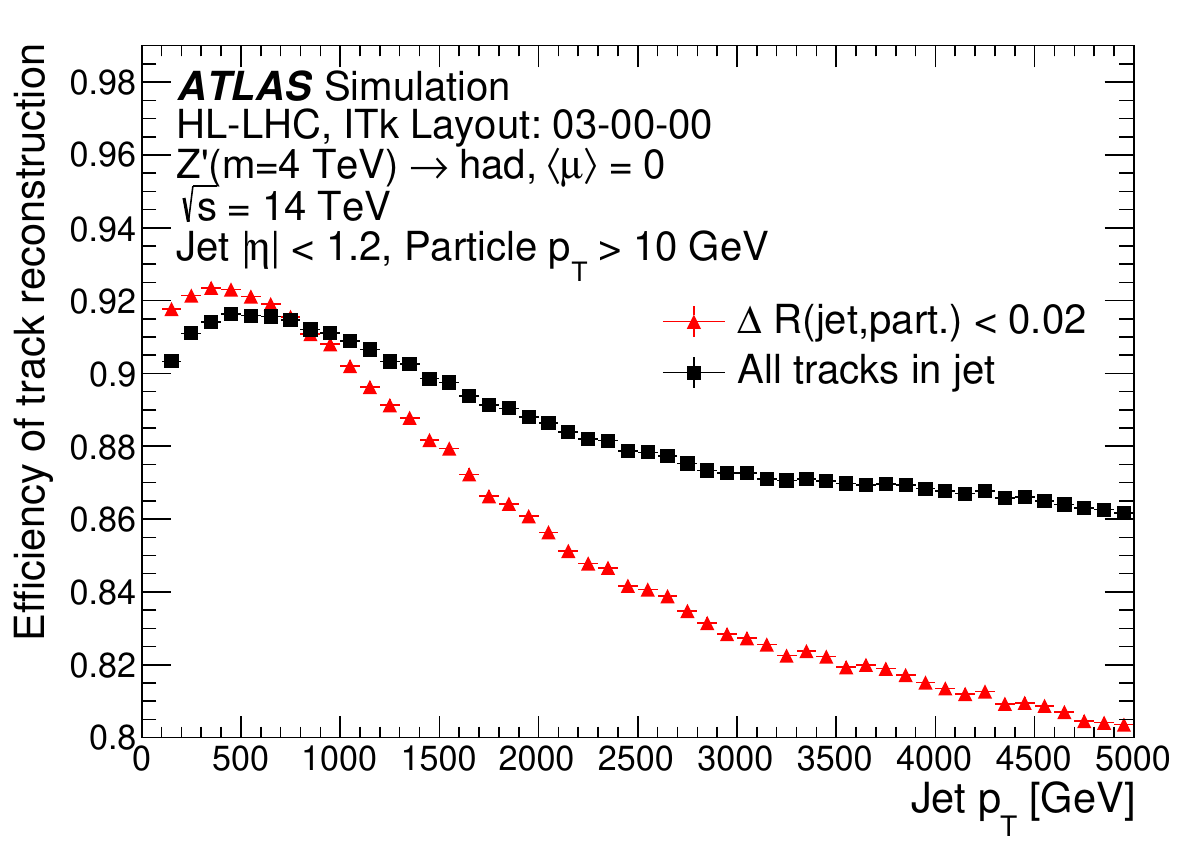} \label{fig:ctide_eff_core}}
\subfloat[]{\includegraphics[width=.48\linewidth]{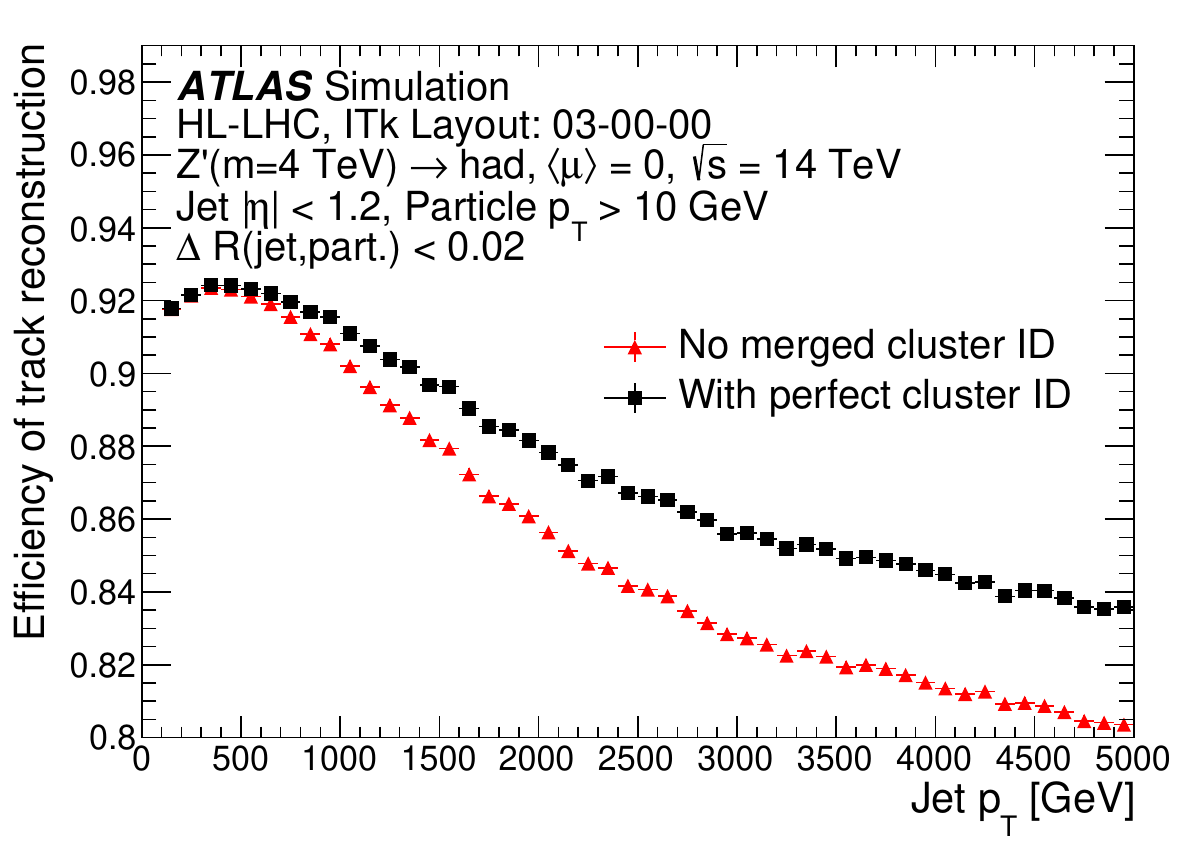} \label{fig:ctide_eff_ID}}\\
\caption{Physics tracking efficiency for tracks in jets as a function of the jet $p_T$ in $Z'$ events in which the $Z'$ decays hadronically, comparing (a) the inclusive efficiency for all tracks matched to the jet ($\Delta R<0.4$) or within the jet core ($\Delta R < 0.02$) and (b)~reconstruction scenarios with perfect, or no, classification of merged clusters in the jet core.}
\end{figure}

Figure~\ref{fig:ctide_eff_core} highlights the physics tracking efficiency expected for charged particles within high-$\pT$ jets when no dedicated merged cluster identification is used.
Tracks located in the core of the jets have on average a larger $\pT$, which explains why the efficiency is higher in the core of jets for a jet $\pT$ up to 700~\GeV.
On the other hand, the particle density increases with the jet $\pT$, inducing a decrease in the physics tracking efficiency.
This effect is more pronounced in the core of the jets where the particle density is highest.
The potential benefit from a dedicated algorithm used for identifying merged clusters is shown in figure~\ref{fig:ctide_eff_ID}, which compares the performance obtained when no merged pixel cluster identification or when a perfect truth-based identification is used.
To optimize the tracking performance in high track-density environments, ATLAS uses a machine-learning-based merged pixel cluster identification algorithm~\cite{PERF-2012-05,PERF-2015-08} that has not yet been adapted for use in Run~4.
The commissioning of dedicated algorithms for the ITk pixel detector will be one of the main focuses of upcoming developments.

\FloatBarrier

\subsection{Number of tracks, mis-reconstructed and fake track rates}

\begin{figure}[t]
\centering
\includegraphics[width=0.7\textwidth]{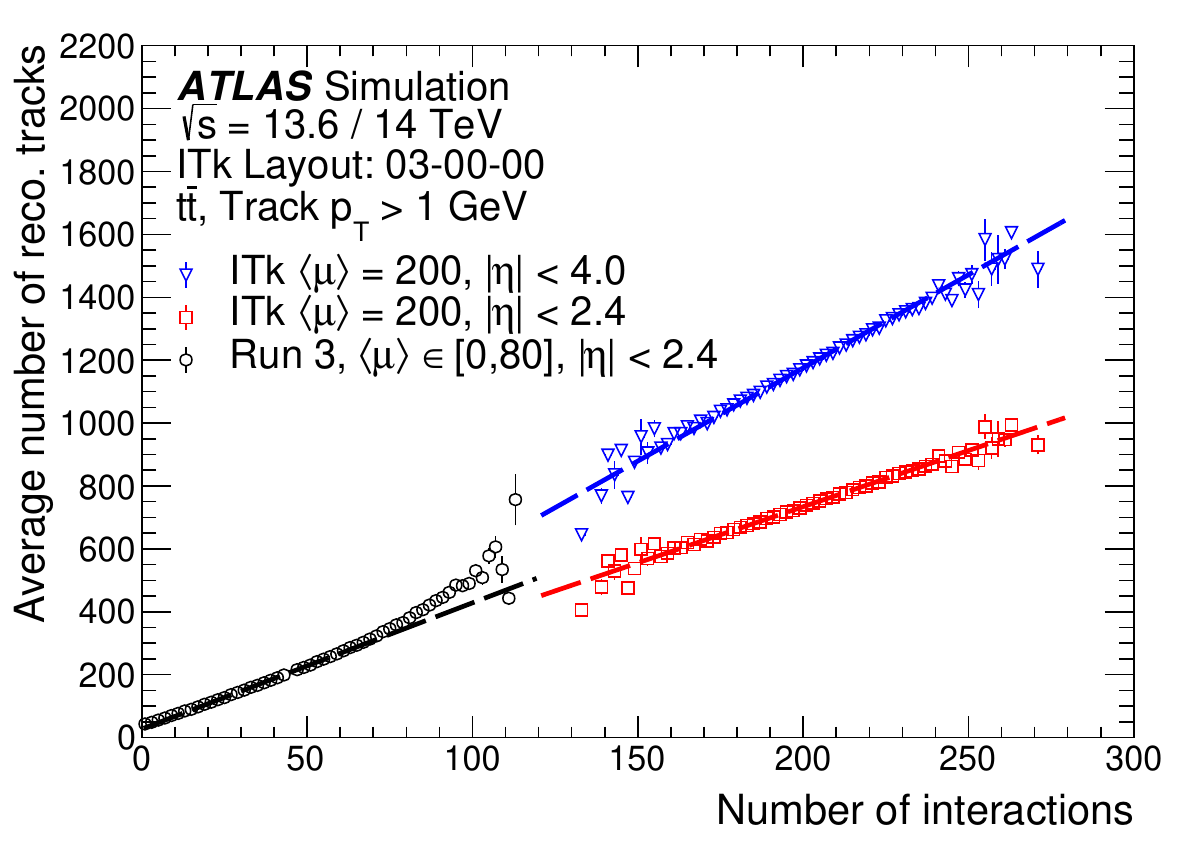}
\caption{Number of reconstructed tracks per event with $\pT> 1~\GeV$ as a function of the number of interactions for $t\bar{t}$ events at $\left\langle\mu\right\rangle$ = 200 with the ITk detector compared with the Run 3 detector, in conditions with a uniform $\left\langle\mu\right\rangle$ distribution between 0~and~80. The dashed lines illustrate the results of linear fits performed over the limited range corresponding to $\langle\mu\rangle$ between 120 and 280 for the ITk and between 20 and 60 (extrapolated to 0--120) for the Run 3 detector to illustrate the \pileup dependence of this quantity.}
\label{fig:ntrack_vs_mu_eta}
\end{figure}

\begin{figure}[!htb]
\centering
\includegraphics[width=.7\linewidth]{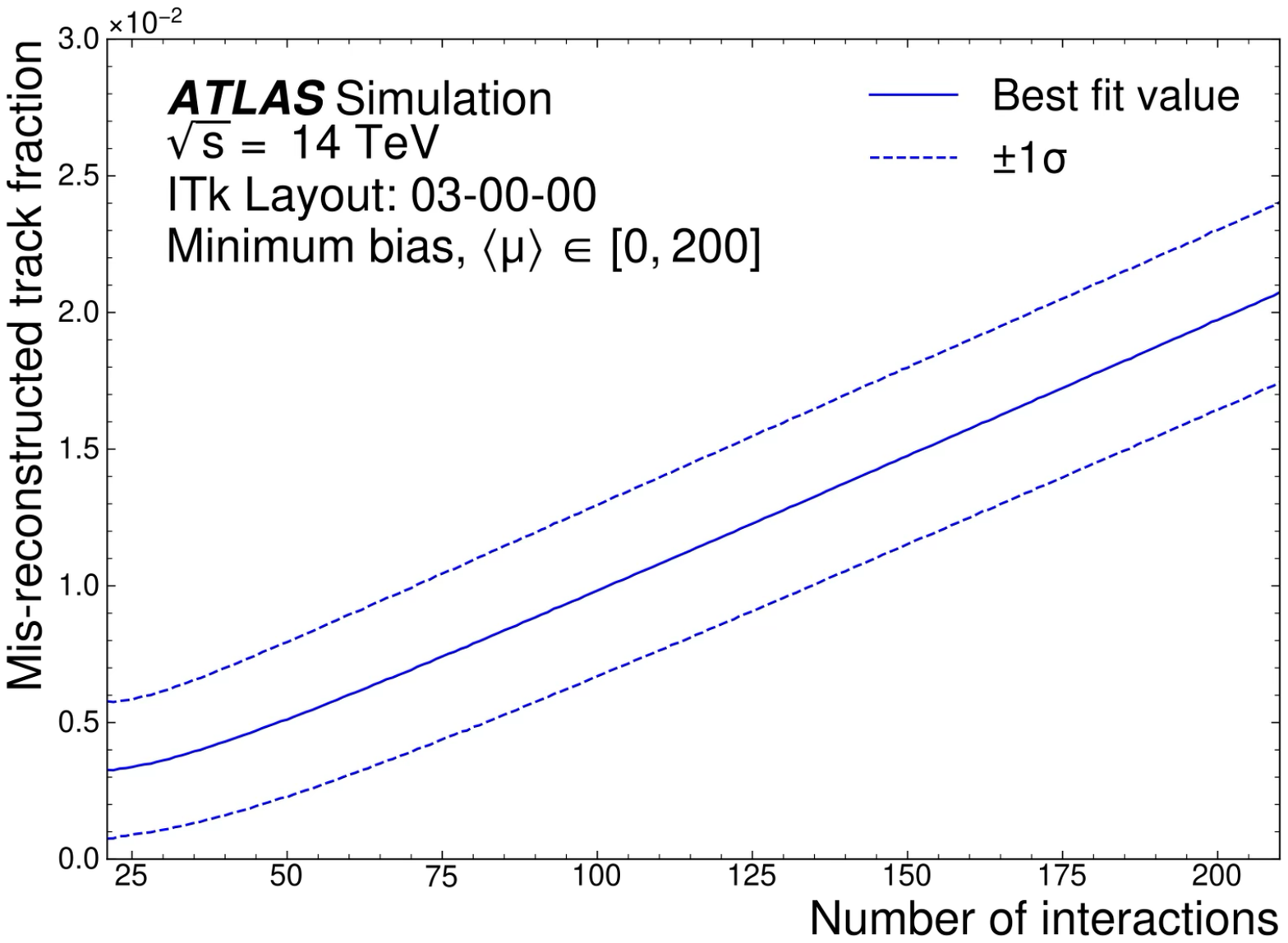}
\caption{Fraction of mis-reconstructed tracks, defined as the
difference between a quadratic fit in the full $\mu$ range and a
linear fit extrapolated from the $\mu < 20$ region. The number of tracks
corresponding to real particles and containing enough hits to
satisfy the reconstruction criteria is expected to scale linearly
with \pileup. The mis-reconstructed track population is composed
of fake tracks not corresponding to any particle and tracks
corresponding to actual particles but that include mis-attributed
hits from \pileup particles, and is expected to increase faster
than linearly with \pileup.}
\label{fig:minbias_misrate}
\end{figure}

\begin{figure}[!htb]
\centering
\includegraphics[width=.7\linewidth]{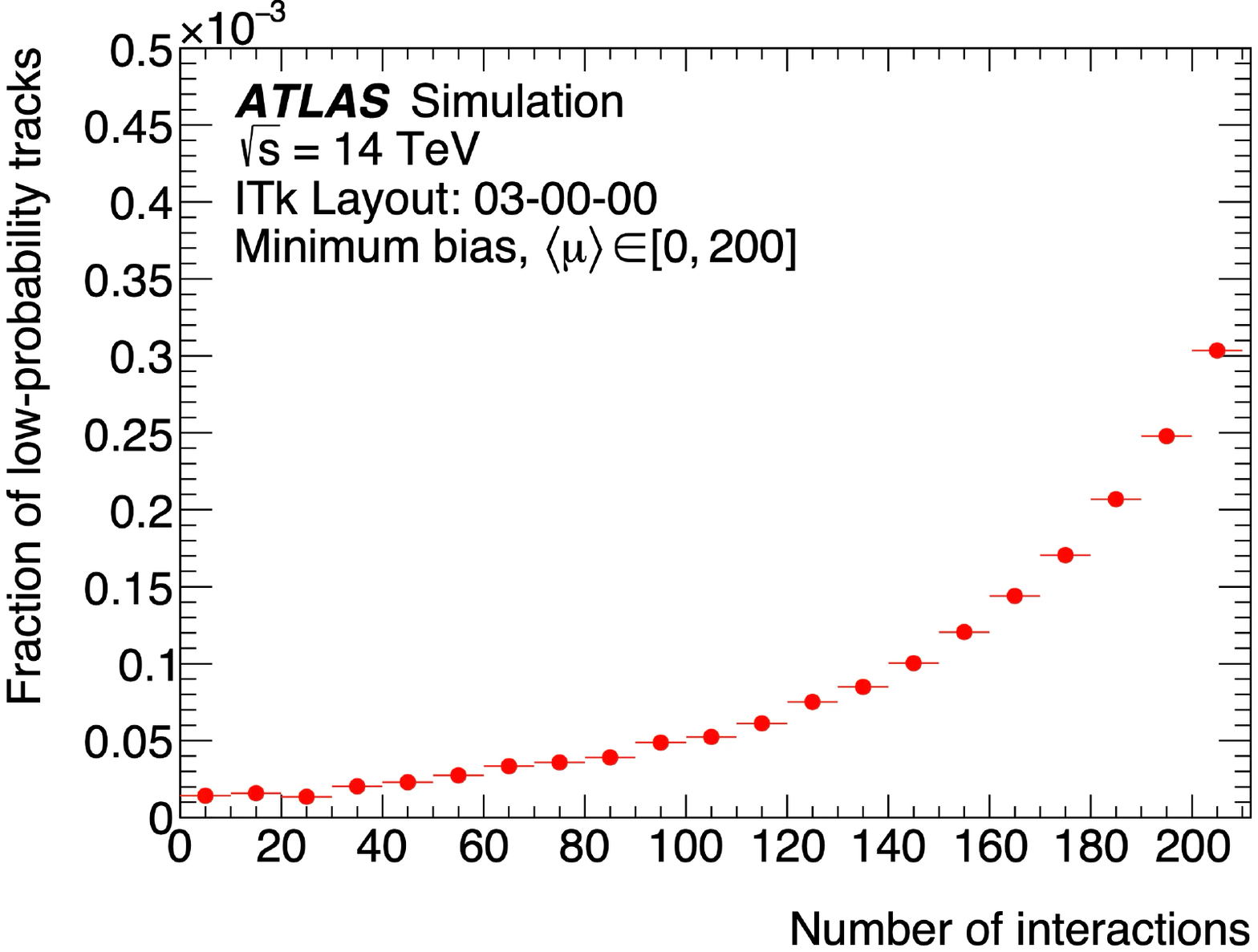}
\caption{Fraction of all tracks with matching probability less than
50\% as computed in Monte Carlo simulation, which estimates the
fraction of tracks not closely corresponding to a
charged particle, also known as the fake track creation rate.}
\label{fig:minbias_fakerate}
\end{figure}

Figure~\ref{fig:ntrack_vs_mu_eta} illustrates the count of reconstructed tracks with $\pT> 1~\GeV$ at $\left\langle\mu\right\rangle$ = 200 in $t\bar{t}$ events.
In the ideal case where all tracks are well-reconstructed and the tracking efficiency is not reduced with \pileup, a linear increase in the number of tracks as a function of the number of interactions\footnote{The number of interactions follows a Poisson distribution with mean $\left\langle\mu\right\rangle$.} is expected.
In reality, a small quadratic component is observed in the number of reconstructed tracks as a function of the number of interactions, caused for instance by
a certain fraction of particles that would not be reconstructed due to the minimum number of hits requirement (c.f. Table~\ref{Tab:req}) but which are assigned to tracks also containing mis-attributed hits from \pileup particles, allowing them to satisfy the reconstruction criteria.
There also exist a small fraction of tracks arising from combinations of hits not closely corresponding to any charged particle, known as fake tracks.
Tracks from either source are collectively referred to as mis-reconstructed tracks, and their number is expected to scale super-linearly with the number of interactions.
The number of well-reconstructed tracks as a function of the number of interactions can be estimated by fitting a linear function to the number of tracks distribution in the low \pileup regime, where the number of fake and mis-reconstructed tracks is expected to be negligible.
The slope of this linear fit is then reduced linearly as a function of $\mu$, to take into account the relative efficiency reduction of up to 0.7\% at $\langle \mu \rangle$ = 200.
On the other hand, the total number of reconstructed tracks as a function of the number of interactions follows a quadratic function; therefore the fraction of mis-reconstructed tracks can be estimated from the difference between a quadratic fit to the number of tracks distribution and this linear fit, extrapolated to the full \pileup range.
This fraction, shown in Figure~\ref{fig:minbias_misrate} for a minimum bias sample, reaches at most 2\% at $\left\langle\mu\right\rangle=200$.
As a comparison, in Run~3 the mis-reconstructed track rate at the typical Run 3 leveling target of  $\left\langle\mu\right\rangle=64$ is approximately 6.5\%.
This reduction in the rate of mis-reconstructed tracks is primarily due to the optimized layout of the ITk detector, which provides in particular a greater number of silicon hits. These additional hits enhance the track reconstruction precision by enabling tighter selection criteria.
The expected fake track creation rate can be estimated from Monte Carlo simulation, using the same matching probability distribution used to calculate the tracking efficiency.
Figure~\ref{fig:minbias_fakerate} shows the expected fraction of tracks with a matching probability lower than 50\% and features an expected fake track creation rate of approximately $3 \times 10^{-4}$ at $\mu = 200$, showcasing the excellent fake track suppression capabilities of the ITk. Moreover, a comparison with Figure~\ref{fig:minbias_misrate} shows that fake tracks are not expected to be a dominant source of mis-reconstructed tracks.

\FloatBarrier

\subsection{Track parameter resolution}

The resolution on track parameters is estimated from reconstructed tracks associated with charged particles.
The distribution of the difference between the reconstructed track parameter and the corresponding quantity for the simulated particle is iteratively truncated from its outliers, which lie more than three standard deviations away from the mean of the distributions, until the distribution is left invariant.
Up to 5\% (1\%) of the tracks associated with muons with $\pT = 2~\GeV$ (100~\GeV) are found to be outliers in the process.
The standard deviation of the core distribution is defined as the intrinsic measurement resolution of the track parameters.

\begin{figure}[b]
\centering
\subfloat[]{\includegraphics[width=.48\linewidth]{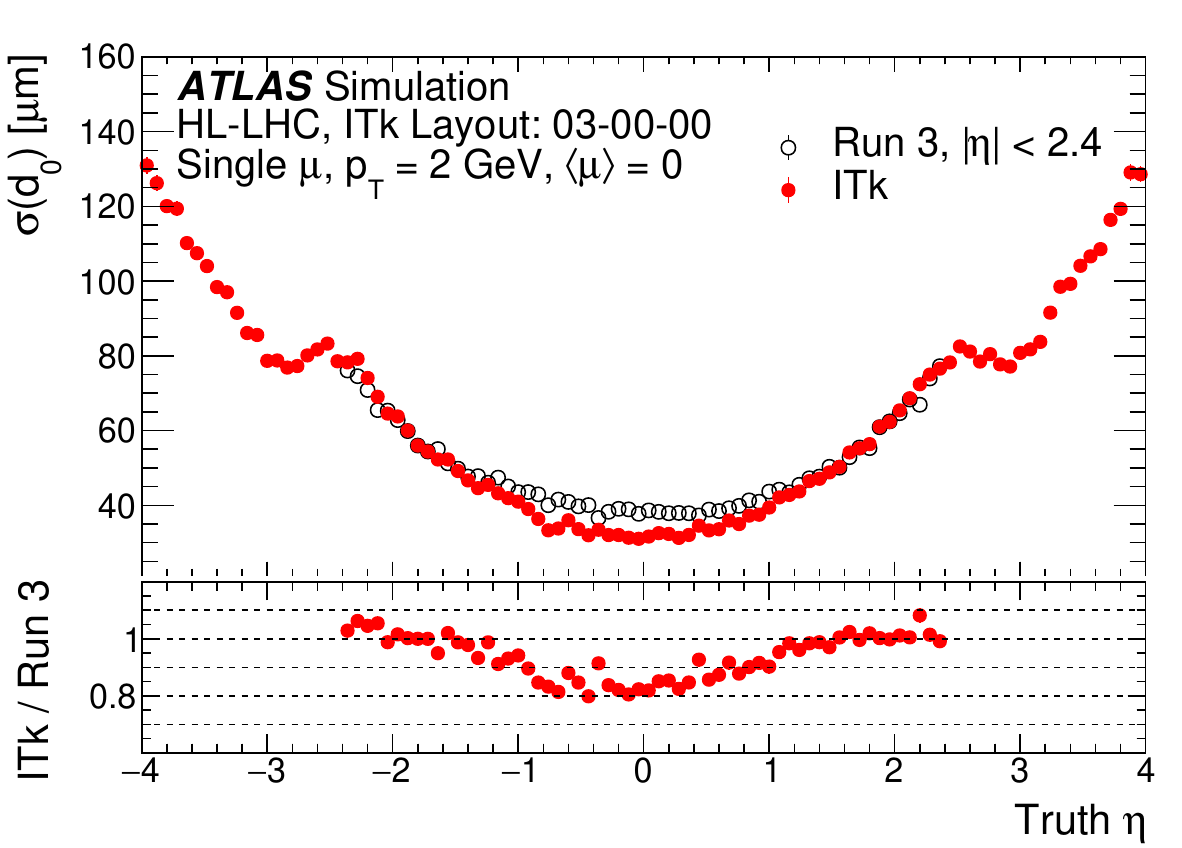}}
\subfloat[]{\includegraphics[width=.48\linewidth]{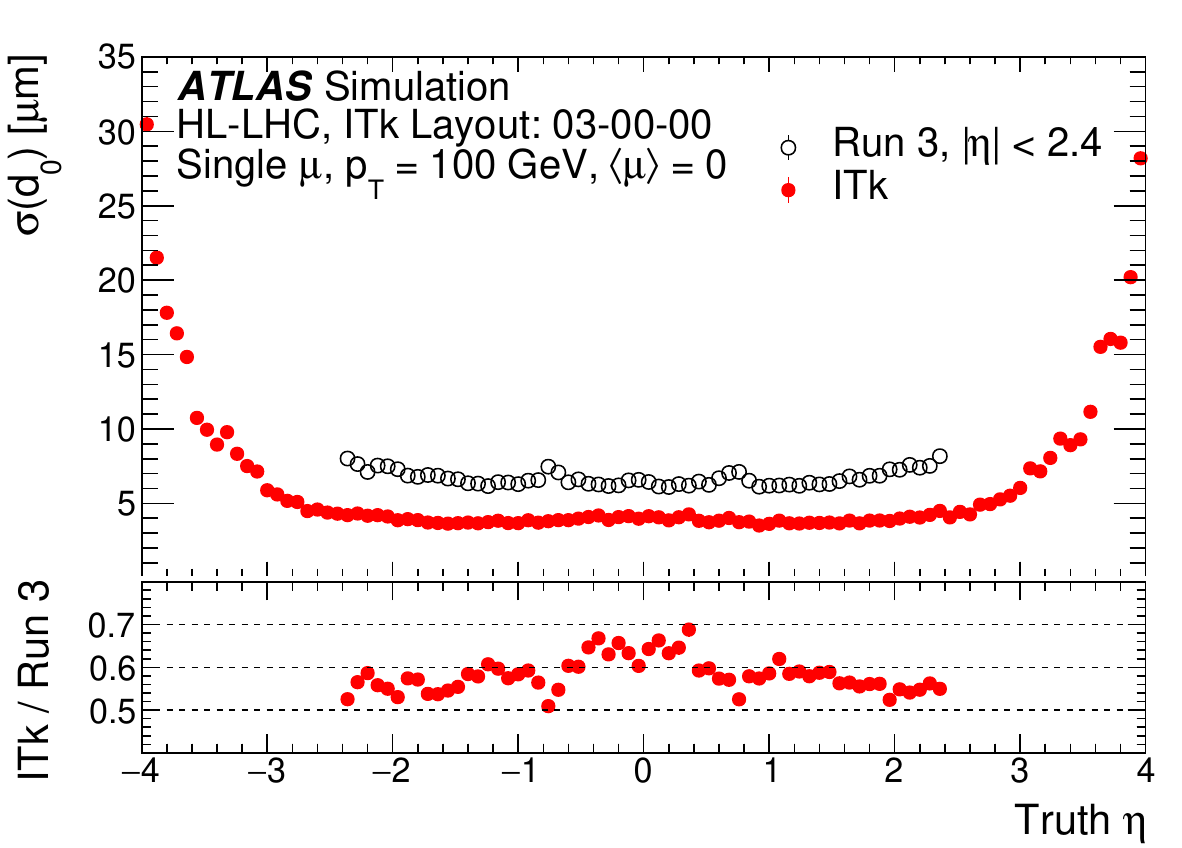}}\\
\caption{Transverse impact parameter ($d_0$) resolution as a function of $\eta$ for (a) 2~\GeV and (b) 100~\GeV muons without \pileup, compared between the ITk and the Run 3 detector.}
\label{fig:d0_res_vs_eta_mu}
\end{figure}

\begin{figure}[t]
\centering
\subfloat[]{\includegraphics[width=.48\linewidth]{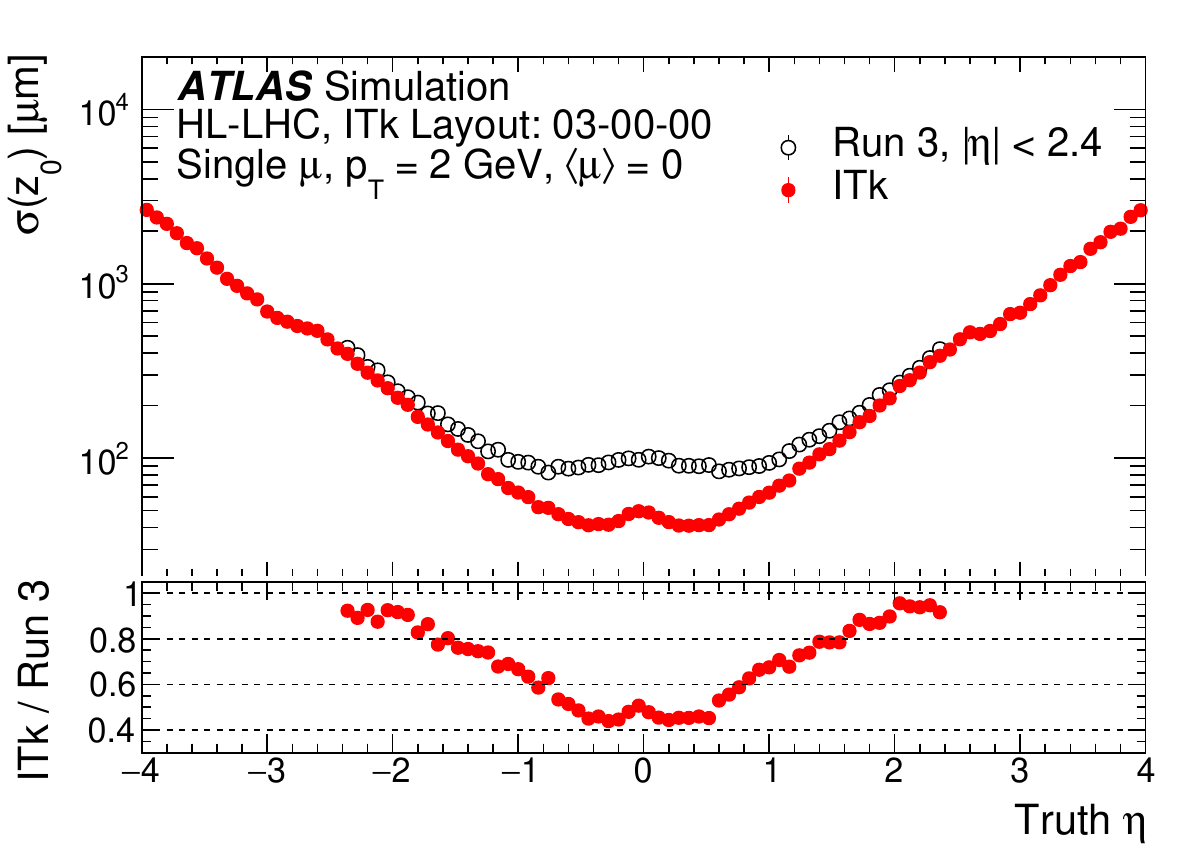}}
\subfloat[]{\includegraphics[width=.48\linewidth]{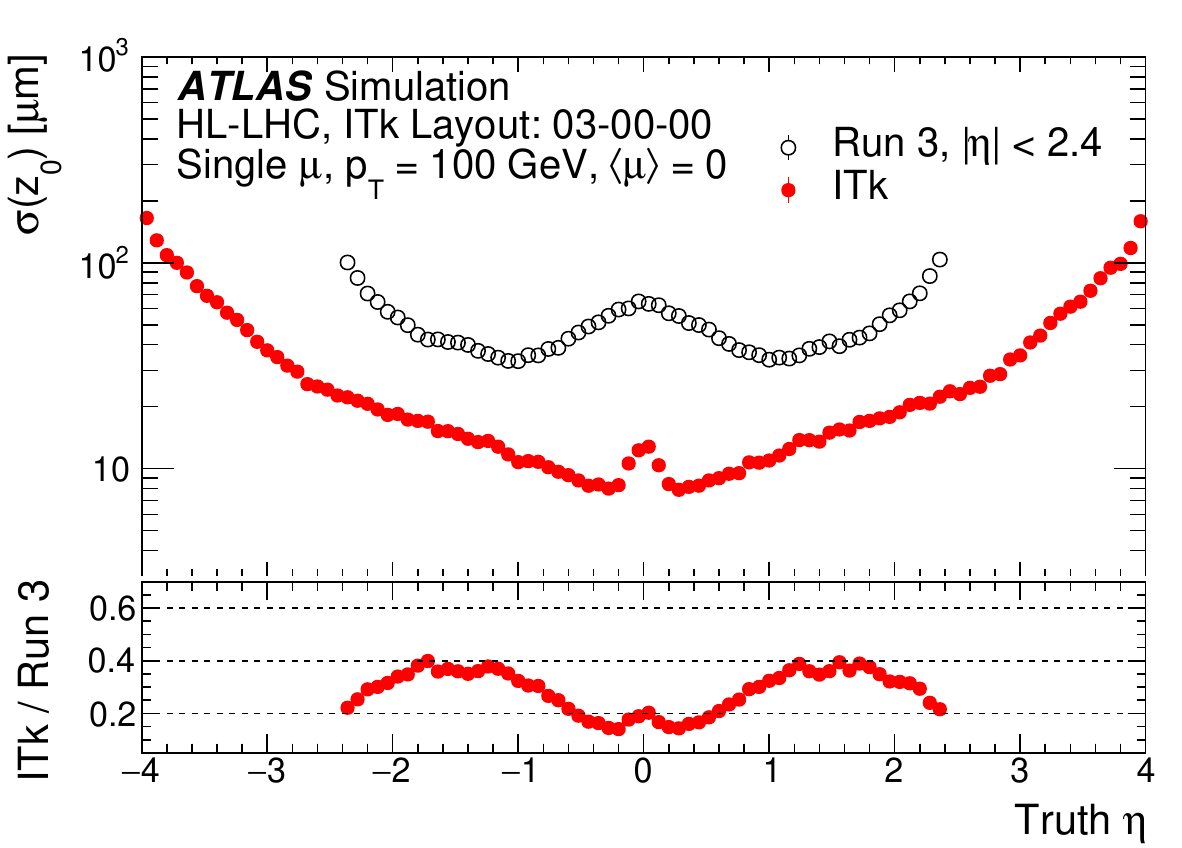}}\\
\caption{Longitudinal impact parameter ($z_0$) resolution as a function of $\eta$ for (a) 2~\GeV and (b) 100~\GeV muons without \pileup, compared between the ITk and the Run 3 detector.}
\label{fig:z0_res_vs_eta_mu}
\end{figure}

\begin{figure}[t]
\centering
\subfloat[]{\includegraphics[width=.48\linewidth]{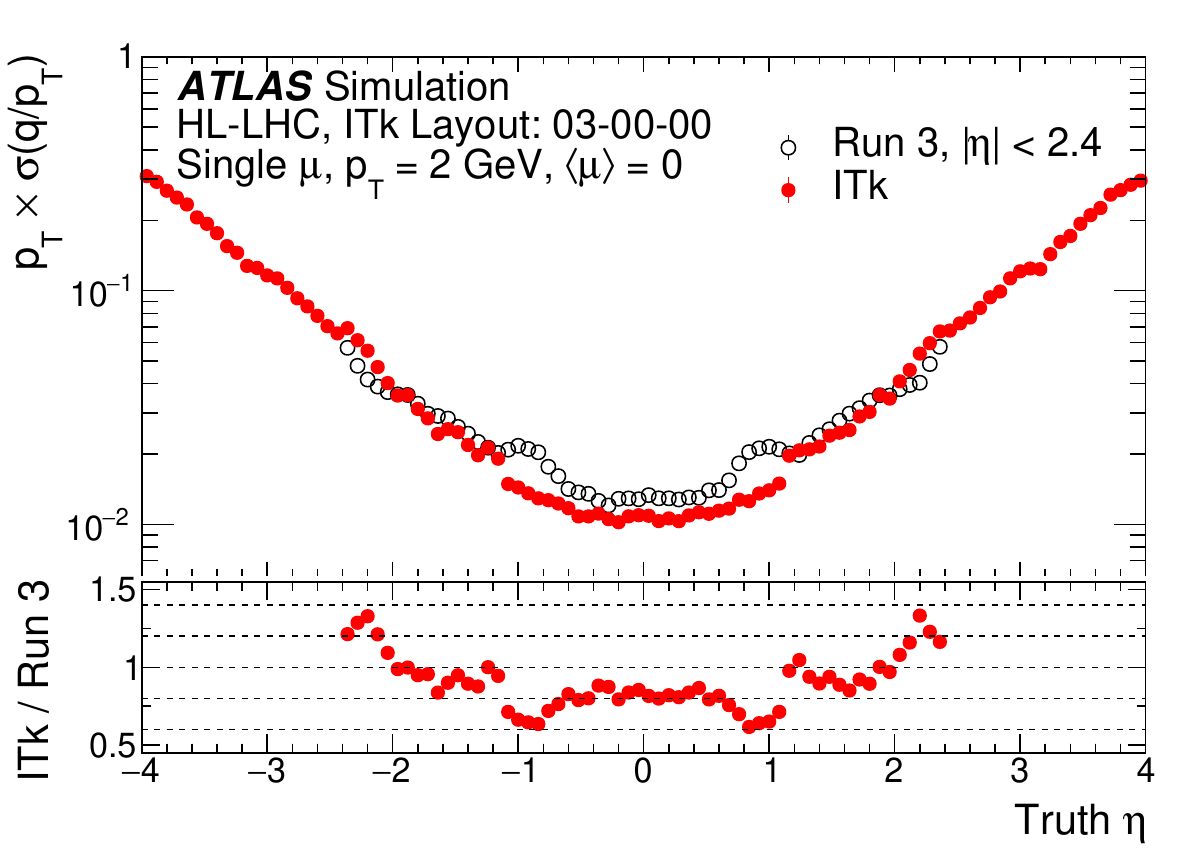}}
\subfloat[]{\includegraphics[width=.48\linewidth]{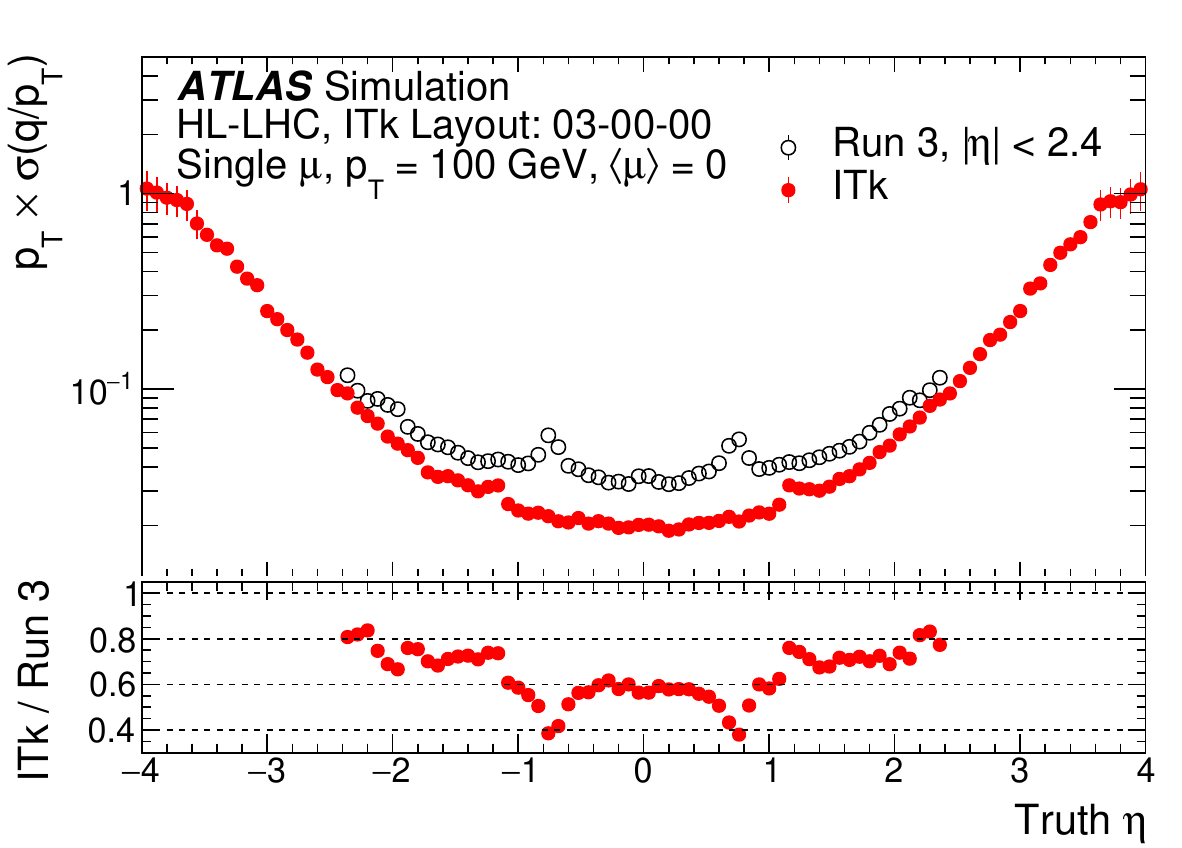}}\\
\caption{Relative transverse momentum resolution as a function of $\eta$ for (a) 2~\GeV and (b) 100~\GeV muons without \pileup, compared between the ITk and the Run 3 detector.}
\label{fig:ptqopt_res_vs_eta_mu}
\end{figure}

Figures~\ref{fig:d0_res_vs_eta_mu} to \ref{fig:ptqopt_res_vs_eta_mu} display the expected resolution on the transverse ($d_0$) and longitudinal ($z_0$) impact parameters and the transverse momentum for simulated muons with $\pT = 2$ and 100~\GeV, compared between the ITk and the Run 3 ATLAS detector.
The slight worsening of the $z_0$ resolution observed near $\eta=0$ in Figure~\ref{fig:z0_res_vs_eta_mu} is associated with the fact that position measurements from the pixel detector are calibrated taking into account the charge distribution within the cluster and the angle of incidence of the track.
As particles with $\eta \sim 0$ typically create one-pixel-wide clusters with normal incident angles, there is therefore no information to be exploited to improve the position resolution in such cases.
Due to the comparable radius of the innermost pixel layers and thanks to the smaller pixel pitch ($25\times100$ or $50\times50 \mu\mathrm{m}^2$ for ITk, $50\times250 \mu\mathrm{m}^2$ for the Run 3 detector~\cite{ATLAS-TDR-2010-19, PIX-2018-001}), the $d_0$ resolution is improved by up to 20\% and the $z_0$ resolution by up to a factor of two with ITk for 2~\GeV muons, mainly in the central part of the detector where the material budget is minimal.
This advantage is even more pronounced for 100~\GeV muons, which are less affected by multiple scatterings from material. In this case, the $d_0$ and $z_0$ resolutions are significantly improved, by a factor of up to two and four respectively.
The transverse momentum resolution with ITk is expected to surpass the Run~3 resolution due to the superior resolution in the bending plane provided by the silicon microstrip sensors in ITk compared with the straw tubes in the Run 3 detector.
While the transverse momentum resolution in the forward region appears relatively poor for 100~\GeV muons, the rate of such muons in that region of the detector is expected to be extremely low, since they correspond to forward muons with an energy of around 3~\TeV.

\FloatBarrier

\subsection{Primary vertex reconstruction and identification}

\begin{figure}[!htb]
\centering
\subfloat[]{\includegraphics[width=.48\linewidth]{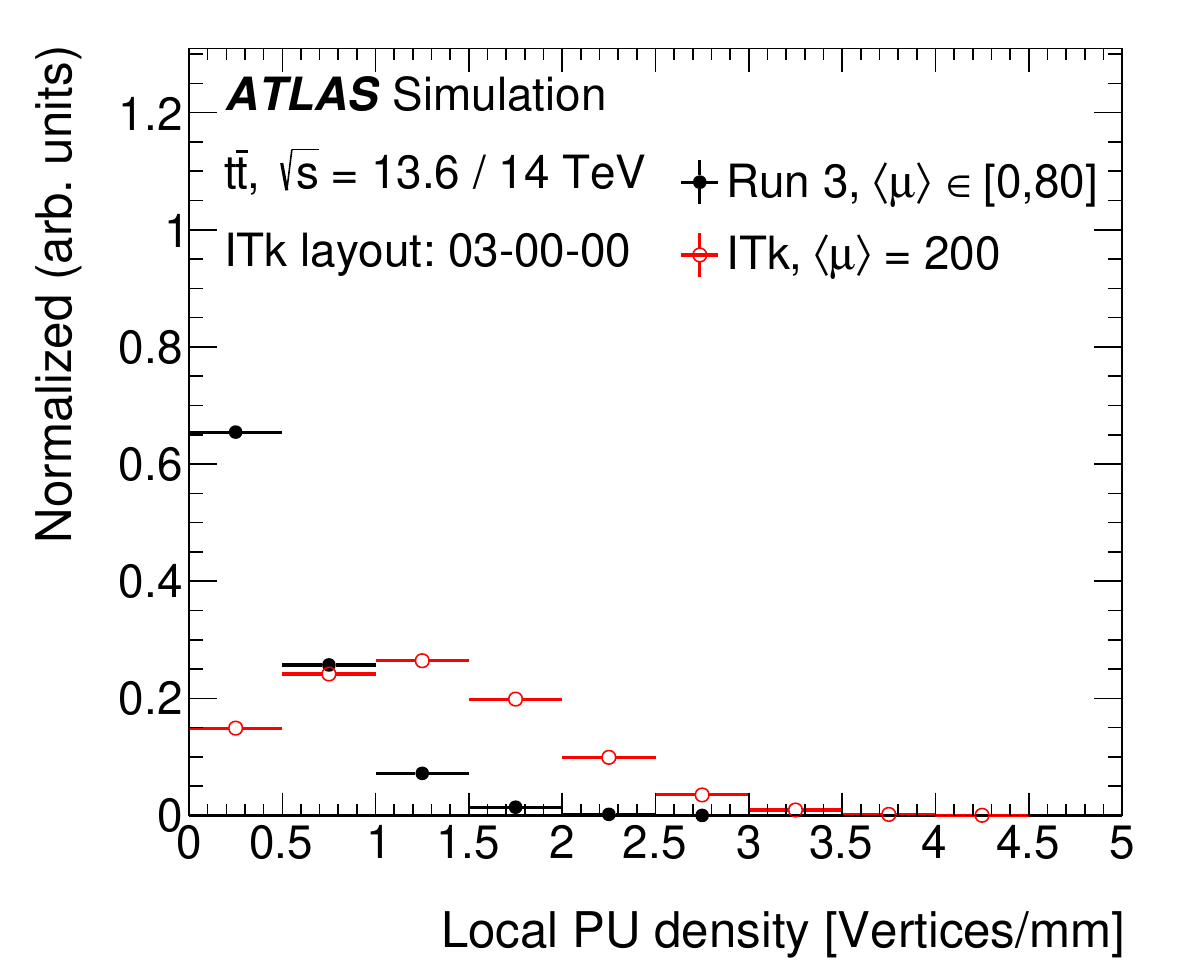} \label{fig:num_interactions_local_pu_density}}
\subfloat[]{\includegraphics[width=.48\linewidth]{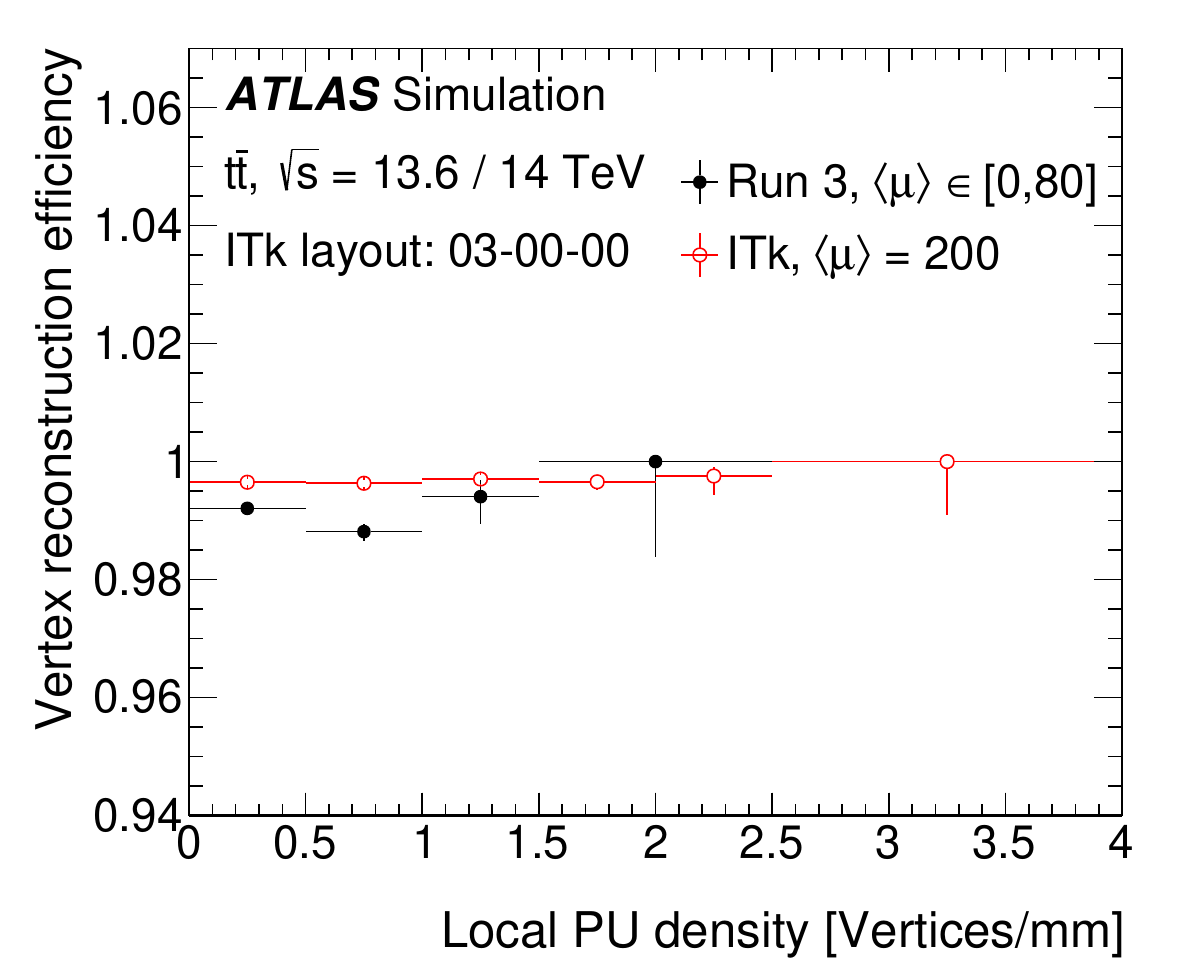} \label{fig:vtx_eff_vs_pudens}}\\
\caption{(a) Distribution of the number of local \pileup density around the hard-scatter vertex, evaluated in $t\bar{t}$ events with $\langle\mu\rangle=200$ in the ITk sample and with a uniform \pileup profile between 0~and~80 in the Run 3 sample. (b) Primary vertex reconstruction efficiency evaluated in $t\bar{t}$ events with $\langle\mu\rangle=200$ as a function of the local \pileup density around the hard-scatter vertex. For comparison, the performance obtained with the Run 3 ATLAS detector with a uniform \pileup profile between 0 and 80 is also shown.}
\end{figure}

The local \pileup density is defined as the number of proton--proton interactions within a 4~mm window around the true hard-scatter vertex, normalized by this distance.
In Figure~\ref{fig:num_interactions_local_pu_density}, the local \pileup density under HL-LHC conditions with $\langle\mu\rangle$ of 200 is shown, in comparison to the one used in the Run 3 sample.
Figure~\ref{fig:vtx_eff_vs_pudens} exhibits the efficiency of the hard-scatter vertex reconstruction, described in Section~\ref{sec:Vertex_reco}, denoting the fraction of events in which a vertex is reconstructed within 0.1 mm of the true hard-scatter position along the longitudinal direction. The figure underscores the resilient performance achieved even under the highest expected \pileup density in HL-LHC conditions.

\begin{figure}[!htb]
\centering
\subfloat[]{\includegraphics[width=.48\linewidth]{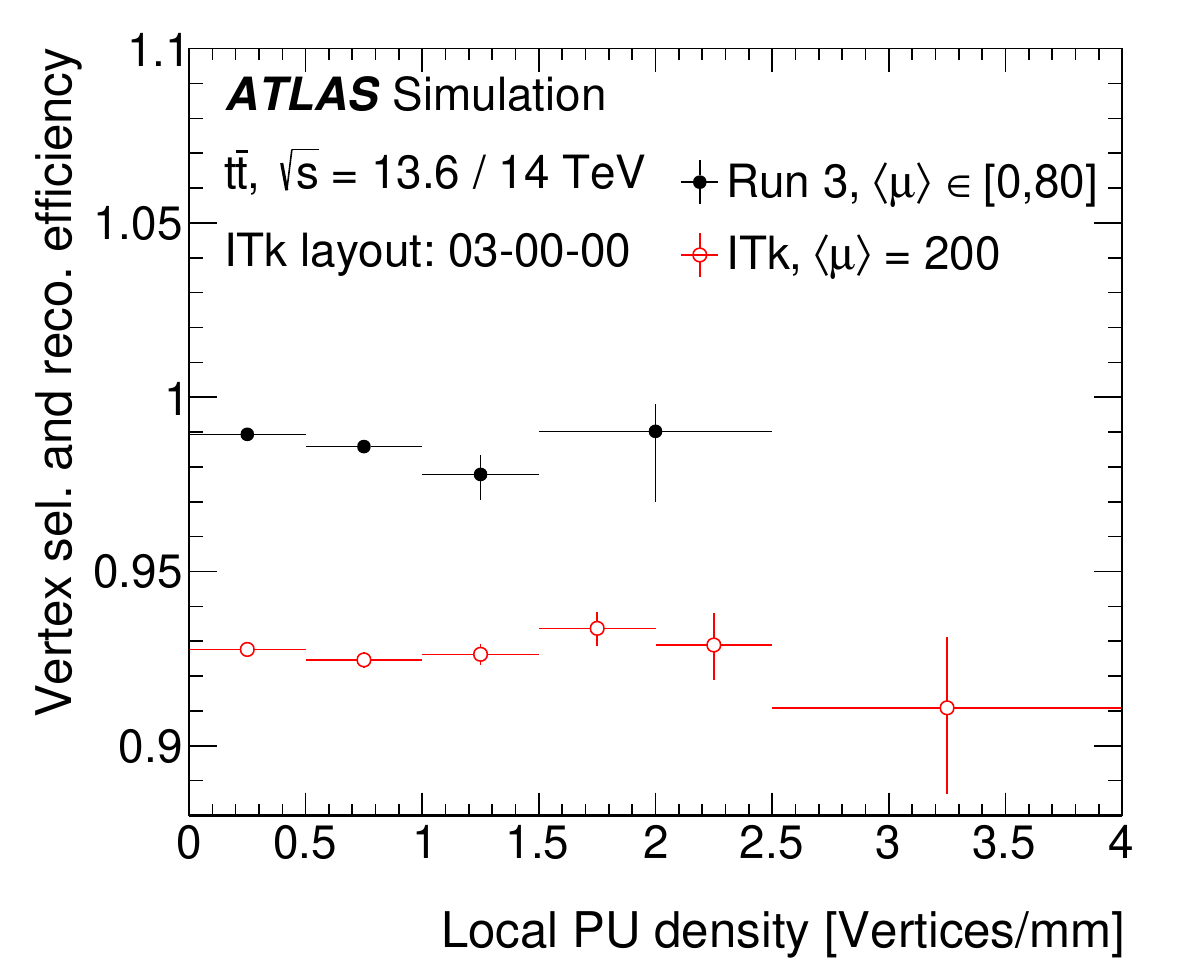} \label{fig:vtx_selreco_pudens}}
\subfloat[]{\includegraphics[width=.48\linewidth]{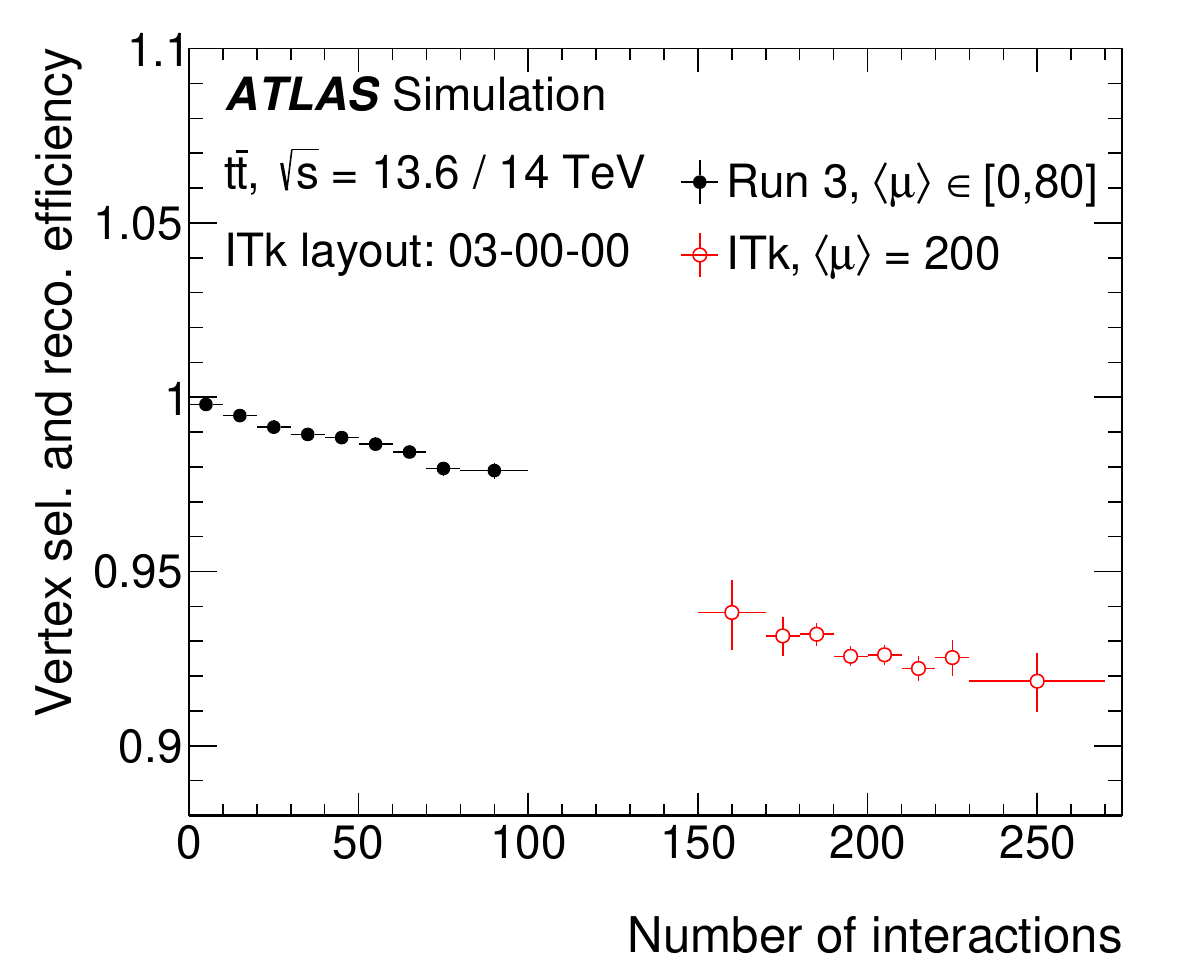} \label{fig:vtx_selreco_ntruth}}\\
\caption{Primary vertex combined reconstruction and selection efficiency evaluated in $t\bar{t}$ events with$\langle\mu\rangle=200$. The efficiency is presented as a function of (a) the local \pileup density around the hard-scatter vertex and (b) the number of interactions. For comparison, the performance obtained with the Run 3 ATLAS detector with a uniform \pileup profile between 0~and~80 is also shown.}
\end{figure}

Another relevant metric is the selection efficiency of the signal hard-scatter vertex, defined as the reconstructed primary vertex that has the largest number of tracks matched to true particles originating from the simulated hard-scatter interaction.
The combined reconstruction and selection efficiency, i.e., the fraction of events in which the hard-scatter vertex is reconstructed and selected as the highest $\Sigma \pT^{2}$ primary vertex, is expected to be influenced by various factors.
The highest-$\Sigma \pT^{2}$ vertex is chosen as the vertex of interest, but this choice may be compromised if the hard-scatter vertex is split into multiple reconstructed vertices due to nearby \pileup vertices.
This effect is expected to be primarily correlated with the local \pileup density, and the efficiency is demonstrated to be highly robust against this variable in Figure~\ref{fig:vtx_selreco_pudens}.

True \pileup vertices may also be merged into a single reconstructed vertex, increasing its associated $\Sigma \pT^{2}$.
Finally, there is a small probability for a \pileup interaction to genuinely yield a larger $\Sigma \pT^{2}$ than the simulated hard-scatter process, even with a perfect vertex reconstruction algorithm.
These effects are expected to be largely correlated with the number of interactions. Figure~\ref{fig:vtx_selreco_ntruth} displays the combined reconstruction and selection efficiency as a function of this variable.
The efficiency is expected to be reduced down to 92\% on average at a \pileup of 200.

\begin{figure}[h]
\centering
\includegraphics[width=0.7\textwidth]{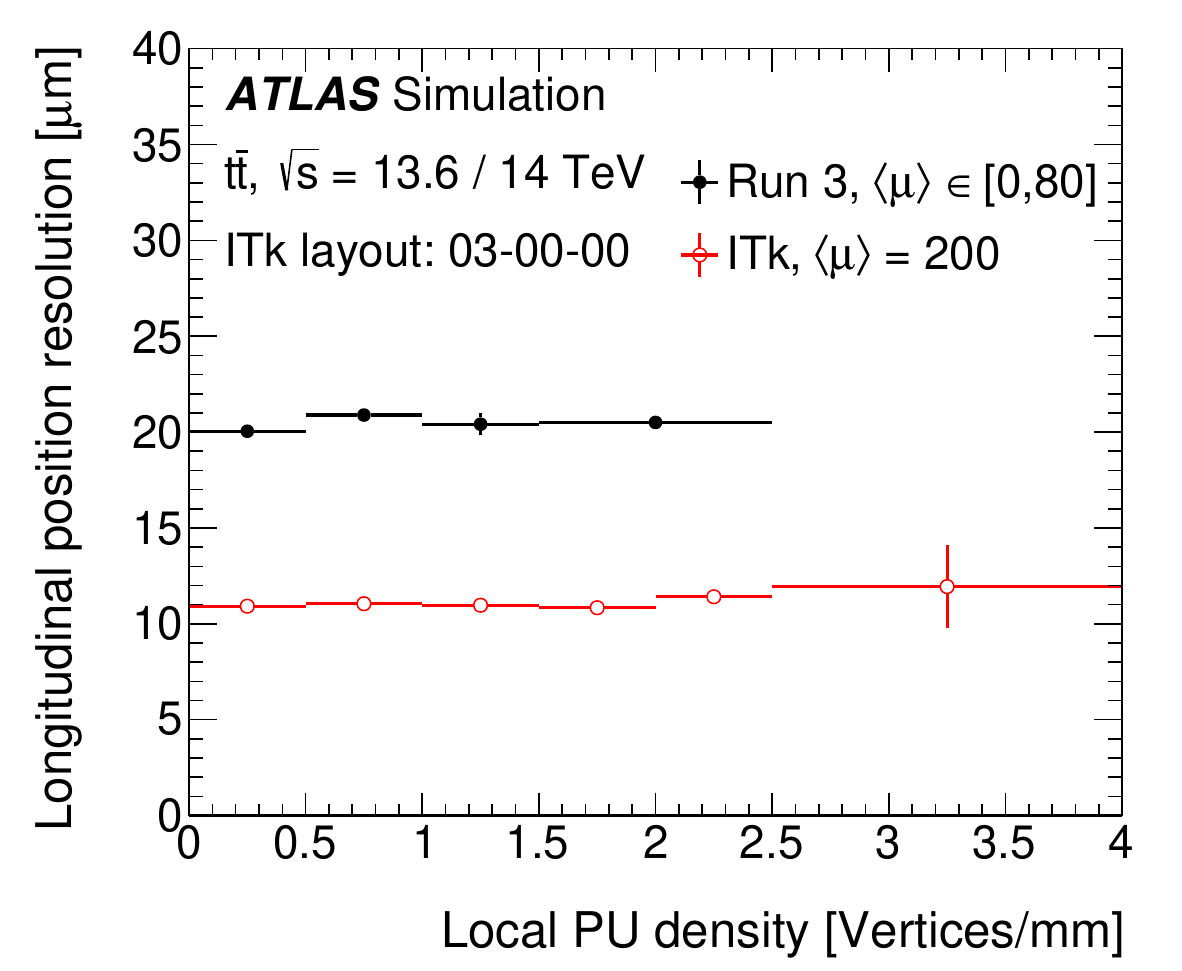}
\caption{Longitudinal position resolution of the reconstructed primary vertex, evaluated in $t\bar{t}$ events with $\langle\mu\rangle=200$. For comparison, the performance obtained with the Run 3 ATLAS detector with a uniform \pileup profile between 0~and~80 is also shown.}
\label{fig:vtx_z_res_vs_pudens}
\end{figure}

The longitudinal position resolution obtained with the ITk detector is presented in Figure~\ref{fig:vtx_z_res_vs_pudens} and exhibits a strong robustness against \pileup, being maintained at about 10 $\mu$m up to high \pileup density, improving by more than a factor of two the performance of the Run 3 detector. Strongly correlated with the improved track parameter resolution, this feature is thus expected to greatly benefit \pileup rejection in jets, hadronic $\tau$~reconstruction, lepton isolation, and flavor-tagging algorithms.

\FloatBarrier

\FloatBarrier


%
\section{Conclusion}
\label{sec:conclusion}
%

This paper presents the state-of-the-art tracking performance
expected with the ITk detector during the high-luminosity LHC operation phase.
The combination between the ITk design and the ATLAS tracking software is
expected to perform to a high standard in very challenging high
\pileup conditions.
The track seeding performance exploits in particular the large multiplicity
of high-precision silicon measurements to guarantee a high efficiency
and redundancy, offering a strong robustness against detector misalignment
or sensor defects.
The full tracking efficiency is found to reach
levels similar to Run~3, while the multiple high-precision silicon
measurements on each track enables a quasi-linear scaling of the track
multiplicity with \pileup, indicating much lower numbers of fake tracks expected than
in Run~3 despite the increased \pileup expected at high luminosity.
The resolution on track parameters, critical for the flavor-tagging and \pileup
rejection performances, are significantly improved relative to the Run~3
detector thanks to the smaller pixel pitch used in the ITk.
The tracking performance in the forward region with $2.4<|\eta|<4.0$, which will be newly covered by the ITk,
is also shown to reach very high standards, compatible with the requirements
expected from high-level object reconstruction and identification algorithms.
The vertex position resolution is also improved relative to the Run~3 performance,
while the vertex reconstruction and selection efficiency shows
a very strong robustness against \pileup.
Finally, the tracking performance is found to be degraded
in the core of high-$\pT$ jets due to the presence of merged charge
clusters, and mitigation techniques inspired by the ones used with the Run 3 detector
will have to be put in place for the ITk track reconstruction to guarantee
high tracking performance in this challenging environment, which is crucial for many beyond-the-Standard-Model searches.
The results presented in this paper represent a solid baseline for future
developments expected to happen in the coming years,
and will be used in particular as a reference for the expected
evolution of the ATLAS tracking software before the start of the Run 4
data-taking.


%
%
%
%
%
\section*{Acknowledgements}

%
%

%
%

We thank CERN for the very successful operation of the LHC and its injectors, as well as the support staff at
CERN and at our institutions worldwide without whom ATLAS could not be operated efficiently.

The crucial computing support from all WLCG partners is acknowledged gratefully, in particular from CERN, the ATLAS Tier-1 facilities at TRIUMF/SFU (Canada), NDGF (Denmark, Norway, Sweden), CC-IN2P3 (France), KIT/GridKA (Germany), INFN-CNAF (Italy), NL-T1 (Netherlands), PIC (Spain), RAL (UK) and BNL (USA), the Tier-2 facilities worldwide and large non-WLCG resource providers. Major contributors of computing resources are listed in Ref.~\cite{ATL-SOFT-PUB-2023-001}.

We gratefully acknowledge the support of ANPCyT, Argentina; YerPhI, Armenia; ARC, Australia; BMWFW and FWF, Austria; ANAS, Azerbaijan; CNPq and FAPESP, Brazil; NSERC, NRC and CFI, Canada; CERN; ANID, Chile; CAS, MOST and NSFC, China; Minciencias, Colombia; MEYS CR, Czech Republic; DNRF and DNSRC, Denmark; IN2P3-CNRS and CEA-DRF/IRFU, France; SRNSFG, Georgia; BMBF, HGF and MPG, Germany; GSRI, Greece; RGC and Hong Kong SAR, China; ICHEP and Academy of Sciences and Humanities, Israel; INFN, Italy; MEXT and JSPS, Japan; CNRST, Morocco; NWO, Netherlands; RCN, Norway; MNiSW, Poland; FCT, Portugal; MNE/IFA, Romania; MSTDI, Serbia; MSSR, Slovakia; ARIS and MVZI, Slovenia; DSI/NRF, South Africa; MICIU/AEI, Spain; SRC and Wallenberg Foundation, Sweden; SERI, SNSF and Cantons of Bern and Geneva, Switzerland; NSTC, Taipei; TENMAK, T\"urkiye; STFC/UKRI, United Kingdom; DOE and NSF, United States of America.

Individual groups and members have received support from BCKDF, CANARIE, CRC and DRAC, Canada; CERN-CZ, FORTE and PRIMUS, Czech Republic; COST, ERC, ERDF, Horizon 2020, ICSC-NextGenerationEU and Marie Sk{\l}odowska-Curie Actions, European Union; Investissements d'Avenir Labex, Investissements d'Avenir Idex and ANR, France; DFG and AvH Foundation, Germany; Herakleitos, Thales and Aristeia programmes co-financed by EU-ESF and the Greek NSRF, Greece; BSF-NSF and MINERVA, Israel; NCN and NAWA, Poland; La Caixa Banking Foundation, CERCA Programme Generalitat de Catalunya and PROMETEO and GenT Programmes Generalitat Valenciana, Spain; G\"{o}ran Gustafssons Stiftelse, Sweden; The Royal Society and Leverhulme Trust, United Kingdom.

In addition, individual members wish to acknowledge support from Armenia: Yerevan Physics Institute (FAPERJ); CERN: European Organization for Nuclear Research (CERN DOCT); Chile: Agencia Nacional de Investigaci\'on y Desarrollo (FONDECYT 1230812, FONDECYT 1230987, FONDECYT 1240864); China: Chinese Ministry of Science and Technology (MOST-2023YFA1605700, MOST-2023YFA1609300), National Natural Science Foundation of China (NSFC - 12175119, NSFC 12275265, NSFC-12075060); Czech Republic: Czech Science Foundation (GACR - 24-11373S), Ministry of Education Youth and Sports (FORTE CZ.02.01.01/00/22\_008/0004632), PRIMUS Research Programme (PRIMUS/21/SCI/017); EU: H2020 European Research Council (ERC - 101002463); European Union: European Research Council (ERC - 948254, ERC 101089007, ERC, BARD, 101116429), European Union, Future Artificial Intelligence Research (FAIR-NextGenerationEU PE00000013), Italian Center for High Performance Computing, Big Data and Quantum Computing (ICSC, NextGenerationEU); France: Agence Nationale de la Recherche (ANR-20-CE31-0013, ANR-21-CE31-0013, ANR-21-CE31-0022, ANR-22-EDIR-0002); Germany: Baden-Württemberg Stiftung (BW Stiftung-Postdoc Eliteprogramme), Deutsche Forschungsgemeinschaft (DFG - 469666862, DFG - CR 312/5-2); Italy: Istituto Nazionale di Fisica Nucleare (ICSC, NextGenerationEU), Ministero dell'Università e della Ricerca (PRIN - 20223N7F8K - PNRR M4.C2.1.1); Japan: Japan Society for the Promotion of Science (JSPS KAKENHI JP22H01227, JSPS KAKENHI JP22H04944, JSPS KAKENHI JP22KK0227, JSPS KAKENHI JP23KK0245); Norway: Research Council of Norway (RCN-314472); Poland: Ministry of Science and Higher Education (IDUB AGH, POB8, D4 no 9722), Polish National Agency for Academic Exchange (PPN/PPO/2020/1/00002/U/00001), Polish National Science Centre (NCN 2021/42/E/ST2/00350, NCN OPUS 2023/51/B/ST2/02507, NCN OPUS nr 2022/47/B/ST2/03059, NCN UMO-2019/34/E/ST2/00393, UMO-2020/37/B/ST2/01043, UMO-2021/40/C/ST2/00187, UMO-2022/47/O/ST2/00148, UMO-2023/49/B/ST2/04085, UMO-2023/51/B/ST2/00920); Spain: Generalitat Valenciana (Artemisa, FEDER, IDIFEDER/2018/048), Ministry of Science and Innovation (MCIN \& NextGenEU PCI2022-135018-2, MICIN \& FEDER PID2021-125273NB, RYC2019-028510-I, RYC2020-030254-I, RYC2021-031273-I, RYC2022-038164-I); Sweden: Carl Trygger Foundation (Carl Trygger Foundation CTS 22:2312), Swedish Research Council (Swedish Research Council 2023-04654, VR 2018-00482, VR 2022-03845, VR 2022-04683, VR 2023-03403, VR grant 2021-03651), Knut and Alice Wallenberg Foundation (KAW 2018.0458, KAW 2019.0447, KAW 2022.0358); Switzerland: Swiss National Science Foundation (SNSF - PCEFP2\_194658); United Kingdom: Leverhulme Trust (Leverhulme Trust RPG-2020-004), Royal Society (NIF-R1-231091); United States of America: U.S. Department of Energy (ECA DE-AC02-76SF00515), Neubauer Family Foundation.

%
%


%

%
%

%
%
%
%
%
%

%
%

%
%
%
\printbibliography

@article{Cornelissen:2008zzc,
    author = "Cornelissen, T. and Elsing, M. and Gavrilenko, I. and Liebig, W. and Moyse, E. and Salzburger, A.",
    title = "{The new ATLAS track reconstruction (NEWT)}",
    doi = "10.1088/1742-6596/119/3/032014",
    journal = "J. Phys. Conf. Ser.",
    volume = "119",
    pages = "032014",
    year = "2008"
}

@article{Fruhwirth:1987fm,
    author = "Fr{\"u}hwirth, R.",
    title = "{Application of Kalman filtering to track and vertex fitting}",
    reportNumber = "HEPHY-PUB-87-503",
    doi = "10.1016/0168-9002(87)90887-4",
    journal = "Nucl. Instrum. Meth. A",
    volume = "262",
    pages = "444--450",
    year = "1987"
}

@Booklet{CERN-LHCC-2012-022,
      author       = "{ATLAS Collaboration}",
      title        = "{Letter of Intent for the Phase-II Upgrade of the ATLAS Experiment}",
      howpublished = "{CERN-LHCC-2012-022, LHCC-I-023}",
      year          = "2012",
      url           = "https://cds.cern.ch/record/1502664",
}

@Booklet{CMS-TDR-014,
      author       = "{CMS Collaboration}",
      title        = "{The Phase-2 Upgrade of the CMS Tracker}",
      howpublished = "{CERN-LHCC-2017-009, CMS-TDR-014}",
      url           = "https://cds.cern.ch/record/2272264",
      year          = "2017",
}

@Booklet{CMS-TDR-15-02,
      author       = "{CMS Collaboration}",
      title        = "{Technical Proposal for the Phase-II Upgrade of the CMS Detector}",
      howpublished = "{CERN-LHCC-2015-010, LHCC-P-008, CMS-TDR-15-02}",
      url           = "https://cds.cern.ch/record/2020886",
      year          = "2015",
}

@Booklet{Aberle:2749422,
      author        = "Aberle, O. and Béjar Alonso, I and Brüning, O and
                       Fessia, P and Rossi, L and Tavian, L and Zerlauth, M and
                       Adorisio, C. and Adraktas, A. and Ady, M. and Albertone, J.
                       and Alberty, L. and Alcaide Leon, M. and Alekou, A. and
                       Alesini, D. and Ferreira, B. Almeida and Lopez, P. Alvarez
                       and Ambrosio, G. and Andreu Munoz, P. and Anerella, M. and
                       Angal-Kalinin, D. and Antoniou, F. and Apollinari, G. and
                       Apollonio, A. and Appleby, R. and Arduini, G. and Alonso,
                       B. Arias and Artoos, K. and Atieh, S. and Auchmann, B. and
                       Badin, V. and Baer, T. and Baffari, D. and Baglin, V. and
                       Bajko, M. and Ball, A. and Ballarino, A. and Bally, S. and
                       Bampton, T. and Banfi, D. and Barlow, R. and Barnes, M. and
                       Barranco, J. and Barthelemy, L. and Bartmann, W. and
                       Bartosik, H. and Barzi, E. and Battistin, M. and
                       Baudrenghien, P. and Alonso, I. Bejar and Belomestnykh, S.
                       and Benoit, A. and Ben-Zvi, I. and Bertarelli, A. and
                       Bertolasi, S. and Bertone, C. and Bertran, B. and Bestmann,
                       P. and Biancacci, N. and Bignami, A. and Bliss, N. and
                       Boccard, C. and Body, Y. and Borburgh, J. and Bordini, B.
                       and Borralho, F. and Bossert, R. and Bottura, L. and
                       Boucherie, A. and Bozzi, R. and Bracco, C. and Bravin, E.
                       and Bregliozzi, G. and Brett, D. and Broche, A. and
                       Brodzinski, K. and Broggi, F. and Bruce, R. and Brugger, M.
                       and Brüning, O. and Buffat, X. and Burkhardt, H. and
                       Burnet, J. and Burov, A. and Burt, G. and Cabezas, R. and
                       Cai, Y. and Calaga, R. and Calatroni, S. and Capatina, O.
                       and Capelli, T. and Cardon, P. and Carlier, E. and Carra,
                       F. and Carvalho, A. and Carver, L.R. and Caspers, F. and
                       Cattenoz, G. and Cerutti, F. and Chancé, A. and Rodrigues,
                       M. Chastre and Chemli, S. and Cheng, D. and Chiggiato, P.
                       and Chlachidze, G. and Claudet, S. and Coello De Portugal,
                       JM. and Collazos, C. and Corso, J. and Costa Machado, S.
                       and Costa Pinto, P. and Coulinge, E. and Crouch, M. and
                       Cruikshank, P. and Cruz Alaniz, E. and Czech, M. and
                       Dahlerup-Petersen, K. and Dalena, B. and Daniluk, G. and
                       Danzeca, S. and Day, H. and De Carvalho Saraiva, J. and De
                       Luca, D. and De Maria, R. and De Rijk, G. and De Silva, S.
                       and Dehning, B. and Delayen, J. and Deliege, Q. and
                       Delille, B. and Delsaux, F. and Denz, R. and Devred, A. and
                       Dexter, A. and Di Girolamo, B. and Dietderich, D. and
                       Dilly, J.W. and Doherty, A. and Dos Santos, N. and Drago,
                       A. and D.Drskovic and Ramos, D. Duarte and Ducimetière, L.
                       and Efthymiopoulos, I. and Einsweiler, K. and Esposito, L.
                       and Esteban Muller, J. and Evrard, S. and Fabbricatore, P.
                       and Farinon, S. and Fartoukh, S. and Faus-Golfe, A. and
                       Favre, G. and Felice, H. and Feral, B. and Ferlin, G. and
                       Ferracin, P. and Ferrari, A. and Ferreira, L. and Fessia,
                       P. and Ficcadenti, L. and Fiotakis, S. and Fiscarelli, L.
                       and Fitterer, M. and Fleiter, J. and Foffano, G. and Fol,
                       E. and Folch, R. and Foraz, K. and Foussat, A. and Frankl,
                       M. and Frasciello, O. and Fraser, M. and Menendez, P.
                       Freijedo and Fuchs, J-F. and Furuseth, S. and Gaddi, A. and
                       Gallilee, M. and Gallo, A. and Alia, R. Garcia and Gavela,
                       H. Garcia and Matos, J. Garcia and Garcia Morales, H. and
                       Valdivieso, A. Garcia-Tabares and Garino, C. and Garion, C.
                       and Gascon, J. and Gasnier, Ch. and Gentini, L. and
                       Gentsos, C. and Ghosh, A. and Giacomel, L. and Hernandez,
                       K. Gibran and Gibson, S. and Ginburg, C. and Giordano, F.
                       and Giovannozzi, M. and Goddard, B. and Gomes, P. and
                       Gonzalez De La Aleja Cabana, M. and Goudket, P. and
                       Gousiou, E. and Gradassi, P. and Costa, A. Granadeiro and
                       Grand-Clément, L. and Grillot, S. and Guillaume, JC. and
                       Guinchard, M. and Hagen, P. and Hakulinen, T. and Hall, B.
                       and Hansen, J. and Heredia Garcia, N. and Herr, W. and
                       Herty, A. and Hill, C. and Hofer, M. and Höfle, W. and
                       Holzer, B. and Hopkins, S. and Hrivnak, J. and Iadarola, G.
                       and Infantino, A. and Bermudez, S. Izquierdo and Jakobsen,
                       S. and Jebramcik, M.A. and Jenninger, B. and Jensen, E. and
                       Jones, M. and Jones, R. and Jones, T. and Jowett, J. and
                       Juchno, M. and Julie, C. and Junginger, T. and Kain, V. and
                       Kaltchev, D. and Karastathis, N. and Kardasopoulos, P. and
                       Karppinen, M. and Keintzel, J. and Kersevan, R. and
                       Killing, F. and Kirby, G. and Korostelev, M. and Kos, N.
                       and Kostoglou, S. and Kozsar, I. and Krasnov, A. and Krave,
                       S. and Krzempek, L. and Kuder, N. and Kurtulus, A. and
                       Kwee-Hinzmann, R. and Lackner, F. and Lamont, M. and
                       Lamure, A.L. and m, L. Lari and Lazzaroni, M. and Le
                       Garrec, M. and Lechner, A. and Lefevre, T. and Leuxe, R.
                       and Li, K. and Li, Z. and Lindner, R. and Lindstrom, B. and
                       Lingwood, C. and Löffler, C. and Lopez, C. and
                       Lopez-Hernandez, LA. and Losito, R. and Maciariello, F. and
                       Macintosh, P. and Maclean, E.H. and Macpherson, A. and
                       Maesen, P. and Magnier, C. and Durand, H. Mainaud and
                       Malina, L. and Manfredi, M. and Marcellini, F. and
                       Marchevsky, M. and Maridor, S. and Marinaro, G. and
                       Marinov, K. and Markiewicz, T. and Marsili, A. and Martinez
                       Urioz, P. and Martino, M. and Masi, A. and Mastoridis, T.
                       and Mattelaer, P. and May, A. and Mazet, J. and Mcilwraith,
                       S. and McIntosh, E. and Medina Medrano, L. and Mejica
                       Rodriguez, A. and Mendes, M. and Menendez, P. and Mensi, M.
                       and Mereghetti, A. and Mergelkuhl, D. and Mertens, T. and
                       Mether, L. and Métral, E. and Migliorati, M. and Milanese,
                       A. and Minginette, P. and Missiaen, D. and Mitsuhashi, T.
                       and Modena, M. and Mokhov, N. and Molson, J. and Monneret,
                       E. and Montesinos, E. and Moron-Ballester, R. and Morrone,
                       M. and Mostacci, A. and Mounet, N. and Moyret, P. and
                       Muffat, P. and Muratori, B. and Muttoni, Y. and Nakamoto,
                       T. and Navarro-Tapia, M. and Neupert, H. and Nevay, L. and
                       Nicol, T. and Nilsson, E. and Ninin, P. and Nobrega, A. and
                       Noels, C. and Nolan, E. and Nosochkov, Y. and Nuiry, FX.
                       and Oberli, L. and Ogitsu, T. and Ohmi, K. and Olave R. and
                       Oliveira, J. and Orlandi, Ph. and Ortega, P. and Osborne,
                       J. and Otto, T. and Palumbo, L. and Papadopoulou, S. and
                       Papaphilippou, Y. and Paraschou, K. and Parente, C. and
                       Paret, S. and Park, H. and Parma, V. and Pasquino, Ch. and
                       Patapenka, A. and Patnaik, L. and Pattalwar, S. and Payet,
                       J. and Pechaud, G. and Pellegrini, D. and Pepinster, P. and
                       Perez, J. and Espinos, J. Perez and Marcone, A. Perillo and
                       Perin, A. and Perini, P. and Persson, T.H.B. and Peterson,
                       T. and Pieloni, T. and Pigny, G. and Pinheiro de Sousa,
                       J.P. and Pirotte, O. and Plassard, F. and Pojer, M. and
                       Pontercorvo, L. and Poyet, A. and Prelipcean, D. and Prin,
                       H. and Principe, R. and Pugnat, T. and Qiang, J. and
                       Quaranta, E. and Rafique, H. and Rakhno, I. and Duarte, D.
                       Ramos and Ratti, A. and Ravaioli, E. and Raymond, M. and
                       Redaelli, S. and Renaglia, T. and Ricci, D. and Riddone, G.
                       and Rifflet, J. and Rigutto, E. and Rijoff, T. and
                       Rinaldesi, R. and Riu Martinez, O. and Rivkin, L. and
                       Rodriguez Mateos, F. and Roesler, S. and Romera Ramirez, I.
                       and Rossi, A. and Rossi, L. and Rude, V. and Rumolo, G. and
                       Rutkovksi, J. and Sabate Gilarte, M. and Sabbi, G. and
                       Sahner, T. and Salemme, R. and Salvant, B. and Galan, F.
                       Sanchez and Santamaria Garcia, A. and Santillana, I. and
                       Santini, C. and Santos, O. and Diaz, P. Santos and Sasaki,
                       K. and Savary, F. and Sbrizzi, A. and Schaumann, M. and
                       Scheuerlein, C. and Schmalzle, J. and Schmickler, H. and
                       Schmidt, R. and Schoerling, D. and Segreti, M. and Serluca,
                       M. and Serrano, J. and Sestak, J. and Shaposhnikova, E. and
                       Shatilov, D. and Siemko, A. and Sisti, M. and Sitko, M. and
                       Skarita, J. and Skordis, E. and Skoufaris, K. and Skripka,
                       G. and Smekens, D. and Sobiech, Z. and Sosin, M. and
                       Sorbio, M. and Soubelet, F. and Spataro, B. and Spiezia, G.
                       and Stancari, G. and Staterao, M. and Steckert, J. and
                       Steele, G. and Sterbini, G. and Struik, M. and Sugano, M.
                       and Szeberenyi, A. and Taborelli, M. and Tambasco, C. and
                       Rego, R. Tavares and Tavian, L. and Teissandier, B. and
                       Templeton, N. and Therasse, M. and Thiesen, H. and Thomas,
                       E. and Toader, A. and Todesco, E. and Tomás, R. and Toral,
                       F. and Torres-Sanchez, R. and Trad, G. and Triantafyllou,
                       N. and Tropin, I. and Tsinganis, A. and Tuckamantel, J. and
                       Uythoven, J. and Valishev, A. and Van Der Veken, F. and Van
                       Weelderen, R. and Vande Craen, A. and Vazquez De Prada, B.
                       and Velotti, F. and Verdu Andres, S. and Verweij, A. and
                       Shetty, N. Vittal and Vlachoudis, V. and Volpini, G. and
                       Wagner, U. and Wanderer, P. and Wang, M. and Wang, X. and
                       Wanzenberg, R. and Wegscheider, A. and Weisz, S. and
                       Welsch, C. and Wendt, M. and Wenninger, J. and Weterings,
                       W. and White, S. and Widuch, K. and Will, A. and Willering,
                       G. and Wollmann, D. and Wolski, A. and Wozniak, J. and Wu,
                       Q. and Xiao, B. and Xiao, L. and Xu, Q. and Yakovlev, Y.
                       and Yammine, S. and Yang, Y. and Yu, M. and Zacharov, I.
                       and Zagorodnova, O. and Zannini, C. and Zanoni, C. and
                       Zerlauth, M. and Zimmermann, F. and Zlobin, A. and Zobov,
                       M. and Zurbano Fernandez, I.",                                          
      title         = "{High-Luminosity Large Hadron Collider (HL-LHC): Technical
                       design report}",
      howpublished = "{CERN-2020-010}",               
      year          = "2020",
      url           = "https://cds.cern.ch/record/2749422",
}

@article{GeoModel,
	author = {Bandieramonte, Marilena and Bianchi, Riccardo Maria and Boudreau, Joseph and Dell'Acqua, Andrea and Tsulaia, Vakhtang},
	title = {The GeoModel tool suite for detector description},
	DOI= "10.1051/epjconf/202125103007",
	journal = {EPJ Web Conf.},
	year = 2021,
	volume = 251,
	pages = "03007",
}

@article{bib:bichsel,
  title = {Straggling in thin silicon detectors},
  author = {Bichsel, Hans},
  journal = {Rev. Mod. Phys.},
  volume = {60},
  pages = {663--699},
  numpages = {0},
  year = {1988},
  month = {Jul},
  publisher = {American Physical Society},
  doi = {10.1103/RevModPhys.60.663},
}

@article{CERN-EP-2001-032,
      author        = "Gorelov, I and Gorfine, G W and Hoeferkamp, M and Seidel,
                       S C and Ciocio, A and Einsweiler, Kevin F and Gilchriese, M
                       G D and Joshi, A and Kleinfelder, S A and Marchesini, R and
                       Milgrome, O and Palaio, N and Pengg, F X and Richardson, J
                       and Zizka, G and Ackers, M and Fischer, P and Keil, M and
                       Meuser, S and Stockmanns, T and Treis, J and Wermes, N and
                       Gössling, C and Hügging, F G and Wüstenfeld, J and
                       Wunstorf, R and Barberis, D and Beccherle, R and Cervetto,
                       M and Darbo, G and Gagliardi, G and Gemme, C and Morettini,
                       P and Netchaeva, P and Osculati, B and Parodi, F and Rossi,
                       L and Dao, K and Fasching, D and Blanquart, L and Breugnon,
                       P and Calvet, D and Clemens, J C and Delpierre, P A and
                       Hallewell, G D and Laugier, D and Mouthuy, T and Rozanov, A
                       N and Trouilleau, C and Valin, I and Aleppo, M and
                       Andreazza, A and Caccia, M and Lari, T and Meroni, C and
                       Ragusa, F and Troncon, C and Vegni, G and Rohe, T and Boyd,
                       G and Severini, H and Skubic, P L and Snow, J and Sícho, P
                       and Tomasek, L and Vrba, V and Holder, M and Lipka, D and
                       Ziolkowski, M and Cauz, D and D'Auria, S and del Papa, C
                       and Grassman, H and Santi, L and Becks, K H and Gerlach, P
                       and Grah,f C and Gregor, I and Harenberg, T and Linder, C",
      title         = "A Measurement of Lorentz Angle and Spatial Resolution of
                       Radiation Hard Silicon Pixel Sensors",
      journal       = "Nucl. Instrum. Meth. A",
      collaboration = "ATLAS Pixel Collaboration",
      reportNumber  = "CERN-EP-2001-032, 1-3",
      volume        = "481",
      pages         = "204-221. 26 p",
      month         = "Apr",
      year          = "2002",
      doi           = "10.1016/S0168-9002(01)01413-9",
}

@Booklet{ATL-SOFT-PUB-2007-007,
      author        = "Cornelissen, T and Elsing, M and Fleischmann, S and
                       Liebig, W and Moyse, E and Salzburger, A",
    title          = "{Concepts, Design and Implementation of the ATLAS New
                       Tracking (NEWT)}",
    howpublished   = "{ATL-SOFT-PUB-2007-007}",
    url            = "https://cds.cern.ch/record/1020106",
    year           = "2007",
}

@Report{CERN-LHCC-2020-015,
      author        = "{ATLAS Collaboration}",
      title         = "{ATLAS HL-LHC Computing Conceptual Design Report}",
      type          = "CERN-LHCC-2020-015, LHCC-G-178",
      year          = "2020",
      url           = "https://cds.cern.ch/record/2729668",
}

@article{Cornelissen:2008zza,
    author = "Cornelissen, Thijs and Elsing, M. and Gavrilenko, I. and Laporte, J. F. and Liebig, W. and Limper, M. and Nikolopoulos, K. and Poppleton, A. and Salzburger, A.",
    title = "{The global $\chi ^2$ track fitter in ATLAS}",
    doi = "10.1088/1742-6596/119/3/032013",
    journal = "J. Phys. Conf. Ser.",
    volume = "119",
    pages = "032013",
    year = "2008"
}

@article{Lund_2009,
	doi = {10.1088/1748-0221/4/04/p04001},
	year = 2009,
	month = {4},
	publisher = {{IOP} Publishing},
	volume = {4},
	number = {04},
	pages = {P04001--P04001},
	author = {E Lund and L Bugge and I Gavrilenko and A Strandlie},
	title = {Track parameter propagation through the application of a new adaptive Runge-Kutta-Nyström method in the {ATLAS} experiment},
	journal = {JINST}
}

@article{RD53,
title = {RD53 pixel chips for the ATLAS and CMS Phase-2 upgrades at HL-LHC},
journal = {Nucl. Instrum. Meth. A},
volume = {1067},
pages = {169682},
year = {2024},
doi = {10.1016/j.nima.2024.169682},
author = "{RD53 Collaboration}"
}

@Article{IDET-2017-10-fix,
    author         = "{ATLAS Collaboration}",
    title          = "{Modelling radiation damage to pixel sensors in the ATLAS detector}",
    journal        = "JINST",
    volume         = "14",
    year           = "2019",
    pages          = "P06012",
    doi            = "10.1088/1748-0221/14/06/P06012",
    reportNumber   = "CERN-EP-2019-061",
    eprint         = "1905.03739",
    archivePrefix  = "arXiv",
    primaryClass   = "physics.ins-det",
}

@Article{HION-2018-03-fix,
    author         = "{ATLAS Collaboration}",
    title          = "{Measurement of angular and momentum distributions of charged particles within and around jets in Pb+Pb and \(pp\) collisions at \(\sqrt{s_{\text{NN}}} = 5.02\,\text{TeV}\) with the ATLAS detector}",
    journal        = "Phys. Rev. C",
    volume         = "100",
    year           = "2019",
    pages          = "064901",
    doi            = "10.1103/PhysRevC.100.064901",
    reportNumber   = "CERN-EP-2019-158",
    eprint         = "1908.05264",
    archivePrefix  = "arXiv",
    primaryClass   = "nucl-ex",
    related        = "HION-2018-03-err",
    relatedstring  = "Erratum:",
}

@Article{IDTR-2022-04-fix,
    author         = "{ATLAS Collaboration}",
    title          = "{Software Performance of the ATLAS Track Reconstruction for LHC Run~3}",
    journal        = "Comput. Softw. Big Sci.",
    volume         = "8",
    year           = "2024",
    pages          = "9",
    doi            = "10.1007/s41781-023-00111-y",
    reportNumber   = "CERN-EP-2023-168",
    eprint         = "2308.09471",
    archivePrefix  = "arXiv",
    primaryClass   = "hep-ex",
}

@Booklet{ATL-PHYS-PUB-2019-006-fix,
    author         = "{ATLAS and CMS Collaborations}",
    title          = "{Report on the Physics at the HL-LHC and Perspectives for the HE-LHC}",
    eprint         = "1902.10229",
    archivePrefix  = "arXiv",
    primaryClass   = "hep-ex",
    year           = "2019",
}

@Booklet{ATL-PHYS-PUB-2021-023-fix,
    author         = "{ATLAS Collaboration}",
    title          = "{Expected performance of the ATLAS detector under different High-Luminosity LHC conditions}",
    howpublished   = "{ATL-PHYS-PUB-2021-023}",
    url            = "https://cds.cern.ch/record/2776650",
    year           = "2021",
}

@Booklet{ATL-PHYS-PUB-2022-018-fix,
    author         = "{ATLAS and CMS Collaborations}",
    title          = "{Snowmass White Paper Contribution: Physics with the Phase-2 ATLAS and CMS Detectors}",
    howpublished   = "{ATL-PHYS-PUB-2022-018, CMS PAS FTR-22-001}",
    url            = "https://cds.cern.ch/record/2805993",
    year           = "2022",
}

@Article{Evans:2008zzb,
      author         = "Evans, Lyndon and Bryant, Philip",
      title          = "{LHC Machine}",
      journal        = "JINST",
      volume         = "3",
      pages          = "S08001",
      doi            = "10.1088/1748-0221/3/08/S08001",
      year           = "2008",
      SLACcitation   = "%%CITATION = JINST,3,S08001;%%",
}

@Article{Cacciari:2008gp,
     author    = "Cacciari, Matteo and Salam, Gavin P. and Soyez, Gregory",
     title     = "{The anti-\(k_{t}\) jet clustering algorithm}",
     journal   = "JHEP",
     volume    = "04",
     year      = "2008",
     pages     = "063",
     eprint    = "0802.1189",
     archivePrefix = "arXiv",
     primaryClass  =  "hep-ph",
     doi       = "10.1088/1126-6708/2008/04/063",
     SLACcitation  = "%%CITATION = 0802.1189;%%"
}

@Article{Fastjet,
      author         = "Cacciari, Matteo and Salam, Gavin P. and Soyez, Gregory",
      title          = "{FastJet user manual}",
      journal        = "Eur. Phys. J. C",
      volume         = "72",
      year           = "2012",
      pages          = "1896",
      doi            = "10.1140/epjc/s10052-012-1896-2",
      eprint         = "1111.6097",
      archivePrefix  = "arXiv",
      primaryClass   = "hep-ph",
      reportNumber   = "CERN-PH-TH-2011-297",
      SLACcitation   = "%%CITATION = ARXIV:1111.6097;%%"
}

@article{Ball:2012cx,
      author         = "{NNPDF Collaboration} and Ball, Richard D. and others",
      title          = "{Parton distributions with LHC data}",
      journal        = "Nucl. Phys. B",
      volume         = "867",
      year           = "2013",
      pages          = "244",
      doi            = "10.1016/j.nuclphysb.2012.10.003",
      eprint         = "1207.1303",
      archivePrefix  = "arXiv",
      primaryClass   = "hep-ph",
      reportNumber   = "EDINBURGH-2012-08, IFUM-FT-997, FR-PHENO-2012-014,
                        RWTH-TTK-12-25, CERN-PH-TH-2012-037, SFB-CPP-12-47\,
                        --CERN-PH-TH-2012-037",
      SLACcitation   = "%%CITATION = ARXIV:1207.1303;%%"
}

@article{Ball:2014uwa,
      author         = "{NNPDF Collaboration} and Ball, Richard D. and others",
      title          = "{Parton distributions for the LHC run II}",
      journal        = "JHEP",
      volume         = "04",
      year           = "2015",
      pages          = "040",
      doi            = "10.1007/JHEP04(2015)040",
      eprint         = "1410.8849",
      archivePrefix  = "arXiv",
      primaryClass   = "hep-ph",
      reportNumber   = "EDINBURGH-2014-15, IFUM-1034-FT, CERN-PH-TH-2013-253,
                        OUTP-14-11P, CAVENDISH-HEP-14-11",
      SLACcitation   = "%%CITATION = ARXIV:1410.8849;%%"
}

@article{Sjostrand:2014zea,
      author         = "Sj{\"o}strand, Torbj{\"o}rn and Ask, Stefan and Christiansen,
                        Jesper R. and Corke, Richard and Desai, Nishita and Ilten,
                        Philip and Mrenna, Stephen and Prestel, Stefan and
                        Rasmussen, Christine O. and Skands, Peter Z.",
      title          = "{An introduction to PYTHIA 8.2}",
      journal        = "Comput. Phys. Commun.",
      volume         = "191",
      year           = "2015",
      pages          = "159",
      doi            = "10.1016/j.cpc.2015.01.024",
      eprint         = "1410.3012",
      archivePrefix  = "arXiv",
      primaryClass   = "hep-ph",
      reportNumber   = "LU-TP-14-36, MCNET-14-22, CERN-PH-TH-2014-190,
                        FERMILAB-PUB-14-316-CD, DESY-14-178, SLAC-PUB-16122,
                        --FERMILAB-PUB-14-316-CD",
      SLACcitation   = "%%CITATION = ARXIV:1410.3012;%%"
}

@Article{Bierlich:2022pfr,
    author        = {Bierlich, Christian and Chakraborty, Smita and Desai, Nishita and Gellersen, Leif and Helenius, Ilkka and Ilten, Philip and Lönnblad, Leif and Mrenna, Stephen and Prestel, Stefan and Preuss, Christian T. and Sjöstrand, Torbjörn and Skands, Peter and Utheim, Marius and Verheyen, Rob},
    title         = "{A comprehensive guide to the physics and usage of PYTHIA 8.3}",
    journal       = {SciPost Phys. Codeb.},
    pages         = {8},
    year          = {2022},
    publisher     = {SciPost},
    doi           = {10.21468/SciPostPhysCodeb.8},
    eprint        = "2203.11601",
    archivePrefix = "arXiv",
    primaryClass  = "hep-ph",
    reportNumber  = "LU-TP 22-16, MCNET-22-04, FERMILAB-PUB-22-227-SCD",
}

@Article{Lange:2001uf,
      author         = "Lange, D. J.",
      title          = "{The EvtGen particle decay simulation package}",
      booktitle      = "{Proceedings, 7th International Conference on B physics
                        at hadron machines (BEAUTY 2000)}",
      journal        = "Nucl. Instrum. Meth. A",
      volume         = "462",
      year           = "2001",
      pages          = "152",
      doi            = "10.1016/S0168-9002(01)00089-4",
      SLACcitation   = "%%CITATION = NUIMA,A462,152;%%"
}

@Article{Nason:2004rx,
      author         = "Nason, Paolo",
      title          = "{A new method for combining NLO QCD with shower Monte Carlo algorithms}",
      journal        = "JHEP",
      volume         = "11",
      pages          = "040",
      doi            = "10.1088/1126-6708/2004/11/040",
      year           = "2004",
      eprint         = "hep-ph/0409146",
      archivePrefix  = "arXiv",
}

@Article{Frixione:2007vw,
      author         = "Frixione, Stefano and Nason, Paolo and Oleari, Carlo",
      title          = "{Matching NLO QCD computations with parton shower
                        simulations: the POWHEG method}",
      journal        = "JHEP",
      volume         = "11",
      pages          = "070",
      doi            = "10.1088/1126-6708/2007/11/070",
      year           = "2007",
      eprint         = "0709.2092",
      archivePrefix  = "arXiv",
      primaryClass   = "hep-ph",
}

@Article{Alioli:2010xd,
      author         = "Alioli, Simone and Nason, Paolo and Oleari, Carlo and Re,
                        Emanuele",
      title          = "{A general framework for implementing NLO calculations in
                        shower Monte Carlo programs: the POWHEG BOX}",
      journal        = "JHEP",
      volume         = "06",
      pages          = "043",
      doi            = "10.1007/JHEP06(2010)043",
      year           = "2010",
      eprint         = "1002.2581",
      archivePrefix  = "arXiv",
      primaryClass   = "hep-ph",
}

@article{Frixione:2007nw,
      author         = "Frixione, Stefano and Ridolfi, Giovanni and Nason, Paolo",
      title          = "{A positive-weight next-to-leading-order Monte Carlo for
                        heavy flavour hadroproduction}",
      journal        = "JHEP",
      volume         = "09",
      pages          = "126",
      doi            = "10.1088/1126-6708/2007/09/126",
      year           = "2007",
      eprint         = "0707.3088",
      archivePrefix  = "arXiv",
      primaryClass   = "hep-ph",
      SLACcitation   = "%%CITATION = ARXIV:0707.3088;%%",
}

@article{Agostinelli:2002hh,
      author         = "Agostinelli, S. and others",
      title          = "{\textsc{Geant4} -- a simulation toolkit}",
      journal        = "Nucl. Instrum. Meth. A",
      volume         = "506",
      year           = "2003",
      pages          = "250",
      doi            = "10.1016/S0168-9002(03)01368-8",
      reportNumber   = "SLAC-PUB-9350, FERMILAB-PUB-03-339",
      SLACcitation   = "%%CITATION = NUIMA,A506,250;%%"
}

@Article{PERF-2007-01,
    author         = "{ATLAS Collaboration}",
    title          = "{The ATLAS Experiment at the CERN Large Hadron Collider}",
    journal        = "JINST",
    volume         = "3",
    year           = "2008",
    pages          = "S08003",
    doi            = "10.1088/1748-0221/3/08/S08003",
    primaryClass   = "hep-ex",
}

@Article{SOFT-2010-01,
    author         = "{ATLAS Collaboration}",
    title          = "{The ATLAS Simulation Infrastructure}",
    journal        = "Eur. Phys. J. C",
    volume         = "70",
    year           = "2010",
    pages          = "823",
    doi            = "10.1140/epjc/s10052-010-1429-9",
    eprint         = "1005.4568",
    archivePrefix  = "arXiv",
    primaryClass   = "physics.ins-det",
}

@Article{PERF-2012-05,
    author         = "{ATLAS Collaboration}",
    title          = "{A neural network clustering algorithm for the ATLAS silicon pixel detector}",
    journal        = "JINST",
    volume         = "9",
    year           = "2014",
    pages          = "P09009",
    doi            = "10.1088/1748-0221/9/09/P09009",
    reportNumber   = "CERN-PH-EP-2014-120",
    eprint         = "1406.7690",
    archivePrefix  = "arXiv",
    primaryClass   = "hep-ex",
}

@Article{PERF-2015-08,
    author         = "{ATLAS Collaboration}",
    title          = "{Performance of the ATLAS track reconstruction algorithms in dense environments in LHC Run~2}",
    journal        = "Eur. Phys. J. C",
    volume         = "77",
    year           = "2017",
    pages          = "673",
    doi            = "10.1140/epjc/s10052-017-5225-7",
    reportNumber   = "CERN-EP-2017-045",
    eprint         = "1704.07983",
    archivePrefix  = "arXiv",
    primaryClass   = "hep-ex",
}

@Article{PERF-2015-09,
    author         = "{ATLAS Collaboration}",
    title          = "{Jet reconstruction and performance using particle flow with the ATLAS Detector}",
    journal        = "Eur. Phys. J. C",
    volume         = "77",
    year           = "2017",
    pages          = "466",
    doi            = "10.1140/epjc/s10052-017-5031-2",
    reportNumber   = "CERN-EP-2017-024",
    eprint         = "1703.10485",
    archivePrefix  = "arXiv",
    primaryClass   = "hep-ex",
}

@Article{MUON-2018-03,
    author         = "{ATLAS Collaboration}",
    title          = "{Muon reconstruction and identification efficiency in ATLAS using the full Run~2 \(pp\) collision data set at \(\sqrt{s} = 13\,\text{TeV}\)}",
    journal        = "Eur. Phys. J. C",
    volume         = "81",
    year           = "2021",
    pages          = "578",
    doi            = "10.1140/epjc/s10052-021-09233-2",
    reportNumber   = "CERN-EP-2020-199",
    eprint         = "2012.00578",
    archivePrefix  = "arXiv",
    primaryClass   = "hep-ex",
}

@Article{PIX-2018-001,
    author         = "Abbott, B. and others",
    title          = "{Production and integration of the ATLAS Insertable B-Layer}",
    journal        = "JINST",
    volume         = "13",
    year           = "2018",
    pages          = "T05008",
    doi            = "10.1088/1748-0221/13/05/T05008",
    eprint         = "1803.00844",
    archivePrefix  = "arXiv",
    primaryClass   = "physics.ins-det",
}

@Article{SUSY-2018-19,
    author         = "{ATLAS Collaboration}",
    title          = "{Search for long-lived charginos based on a disappearing-track signature using \(136\,\text{fb}^{-1}\) of \(pp\) collisions at \(\sqrt{s} = 13\,\text{TeV}\) with the ATLAS detector}",
    journal        = "Eur. Phys. J. C",
    volume         = "82",
    year           = "2022",
    pages          = "606",
    doi            = "10.1140/epjc/s10052-022-10489-5",
    reportNumber   = "CERN-EP-2021-209",
    eprint         = "2201.02472",
    archivePrefix  = "arXiv",
    primaryClass   = "hep-ex",
}

@Article{FTAG-2019-07,
    author         = "{ATLAS Collaboration}",
    title          = "{ATLAS flavour-tagging algorithms for the LHC Run~2 \(pp\) collision dataset}",
    journal        = "Eur. Phys. J. C",
    volume         = "83",
    year           = "2023",
    pages          = "681",
    doi            = "10.1140/epjc/s10052-023-11699-1",
    reportNumber   = "CERN-EP-2022-226",
    eprint         = "2211.16345",
    archivePrefix  = "arXiv",
    primaryClass   = "physics.data-an",
}

@Article{STDM-2019-22,
    author         = "{ATLAS Collaboration}",
    title          = "{Measurement of the production of a \(W\) boson in association with a charmed hadron in \(pp\) collisions at \(\sqrt{s} = 13\,\text{TeV}\) with the ATLAS detector}",
    journal        = "Phys. Rev. D",
    volume         = "108",
    year           = "2023",
    pages          = "032012",
    doi            = "10.1103/PhysRevD.108.032012",
    reportNumber   = "CERN-EP-2022-291",
    eprint         = "2302.00336",
    archivePrefix  = "arXiv",
    primaryClass   = "hep-ex",
}

@Article{IDET-2020-01,
    author         = "{ATLAS Collaboration}",
    title          = "{Measurements of sensor radiation damage in the ATLAS inner detector using leakage currents}",
    journal        = "JINST",
    volume         = "16",
    year           = "2021",
    pages          = "P08025",
    doi            = "10.1088/1748-0221/16/08/P08025",
    reportNumber   = "CERN-EP-2021-055",
    eprint         = "2106.09287",
    archivePrefix  = "arXiv",
    primaryClass   = "hep-ex",
}

@Article{EGAM-2021-01,
    author         = "{ATLAS Collaboration}",
    title          = "{Electron and photon efficiencies in LHC Run~2 with the ATLAS experiment}",
    journal        = "JHEP",
    volume         = "05",
    year           = "2024",
    pages          = "162",
    doi            = "10.1007/JHEP05(2024)162",
    reportNumber   = "CERN-EP-2023-182",
    eprint         = "2308.13362",
    archivePrefix  = "arXiv",
    primaryClass   = "hep-ex",
}

@Article{SOFT-2022-02,
    author         = "{ATLAS Collaboration}",
    title          = "{Software and computing for Run~3 of the ATLAS experiment at the LHC}",
    year           = "2024",
    reportNumber   = "CERN-EP-2024-100",
    eprint         = "2404.06335",
    archivePrefix  = "arXiv",
    primaryClass   = "hep-ex",
}

@Booklet{ATL-SOFT-PUB-2023-001,
    author         = "{ATLAS Collaboration}",
    title          = "{ATLAS Computing Acknowledgements}",
    howpublished   = "{ATL-SOFT-PUB-2023-001}",
    url            = "https://cds.cern.ch/record/2869272",
    year           = "2023",
}

@Report{ATLAS-TDR-04,
    author         = "{ATLAS Collaboration}",
    title          = "{ATLAS Inner Detector: Technical Design Report, Volume 1}",
    type           = "ATLAS-TDR-4; CERN-LHCC-97-016",
    year           = "1997",
    url            = "https://cds.cern.ch/record/331063",
}

@Report{ATLAS-TDR-05,
    author         = "{ATLAS Collaboration}",
    title          = "{ATLAS Inner Detector: Technical Design Report, Volume 2}",
    type           = "ATLAS-TDR-5, CERN-LHCC-97-017",
    year           = "1997",
    url            = "https://cds.cern.ch/record/331064",
}

@Report{ATLAS-TDR-2010-19,
    author         = "{ATLAS Collaboration}",
    title          = "{ATLAS Insertable B-Layer Technical Design Report}",
    type           = "ATLAS-TDR-19; CERN-LHCC-2010-013",
    year           = "2010",
    url            = "https://cds.cern.ch/record/1291633",
    related        = "ATLAS-TDR-2010-19-addm",
    relatedstring  = "Addendum:",
}

@Report{ATLAS-TDR-25,
    author         = "{ATLAS Collaboration}",
    title          = "{ATLAS Inner Tracker Strip Detector: Technical Design Report}",
    type           = "ATLAS-TDR-025; CERN-LHCC-2017-005",
    year           = "2017",
    url            = "https://cds.cern.ch/record/2257755",
}

@Report{ATLAS-TDR-30,
    author         = "{ATLAS Collaboration}",
    title          = "{ATLAS Inner Tracker Pixel Detector: Technical Design Report}",
    type           = "ATLAS-TDR-030; CERN-LHCC-2017-021",
    year           = "2017",
    url            = "https://cds.cern.ch/record/2285585",
}

@Report{ATLAS-TDR-31,
    author         = "{ATLAS Collaboration}",
    title          = "{A High-Granularity Timing Detector for the ATLAS Phase-II Upgrade: Technical Design Report}",
    type           = "ATLAS-TDR-031; CERN-LHCC-2020-007",
    year           = "2020",
    url            = "https://cds.cern.ch/record/2719855",
}

@Report{ATLAS-TDR-PhaseII,
    author         = "{ATLAS Collaboration}",
    title          = "{ATLAS Phase-II Upgrade Scoping Document}",
    type           = "CERN-LHCC-2015-020, LHCC-G-166",
    year           = "2015",
    url            = "https://cds.cern.ch/record/2055248",
}

@Article{CMS-CMS-00-001,
    author         = "{CMS Collaboration}",
    title          = "{The CMS Experiment at the CERN LHC}",
    journal        = "JINST",
    volume         = "3",
    year           = "2008",
    pages          = "S08004",
    doi            = "10.1088/1748-0221/3/08/S08004",
}

@Booklet{ATL-PHYS-PUB-2014-021,
    author         = "{ATLAS Collaboration}",
    title          = "{ATLAS Pythia~8 tunes to \(7~\text{TeV}\) data}",
    howpublished   = "{ATL-PHYS-PUB-2014-021}",
    url            = "https://cds.cern.ch/record/1966419",
    year           = "2014",
}

@Booklet{ATL-PHYS-PUB-2016-020,
    author         = "{ATLAS Collaboration}",
    title          = "{Studies on top-quark Monte Carlo modelling for Top2016}",
    howpublished   = "{ATL-PHYS-PUB-2016-020}",
    url            = "https://cds.cern.ch/record/2216168",
    year           = "2016",
}

@Booklet{ATL-PHYS-PUB-2019-005,
    author         = "{ATLAS Collaboration}",
    title          = "{Expected performance of the ATLAS detector at the High-Luminosity LHC}",
    howpublished   = "{ATL-PHYS-PUB-2019-005}",
    url            = "https://cds.cern.ch/record/2655304",
    year           = "2019",
}

@Booklet{ATL-PHYS-PUB-2019-015,
    author         = "{ATLAS Collaboration}",
    title          = "{Development of ATLAS Primary Vertex Reconstruction for LHC Run 3}",
    howpublished   = "{ATL-PHYS-PUB-2019-015}",
    url            = "https://cds.cern.ch/record/2670380",
    year           = "2019",
}

@Booklet{ATL-PHYS-PUB-2020-005,
    author         = "{ATLAS Collaboration}",
    title          = "{Expected \(b\)-tagging Performance with the upgraded ATLAS Inner Tracker Detector at the High-Luminosity LHC}",
    howpublished   = "{ATL-PHYS-PUB-2020-005}",
    url            = "https://cds.cern.ch/record/2713377",
    year           = "2020",
}

@Booklet{ATL-PHYS-PUB-2022-027,
    author         = "{ATLAS Collaboration}",
    title          = "{Graph Neural Network Jet Flavour Tagging with the ATLAS Detector}",
    howpublished   = "{ATL-PHYS-PUB-2022-027}",
    url            = "https://cds.cern.ch/record/2811135",
    year           = "2022",
}

@Booklet{ATL-PHYS-PUB-2022-044,
    author         = "{ATLAS Collaboration}",
    title          = "{Reconstruction, Identification, and Calibration of hadronically decaying tau leptons with the ATLAS detector for the LHC Run~3 and reprocessed Run~2 data}",
    howpublished   = "{ATL-PHYS-PUB-2022-044}",
    url            = "https://cds.cern.ch/record/2827111",
    year           = "2022",
}

@Booklet{ATL-PHYS-PUB-2022-047,
    author         = "{ATLAS Collaboration}",
    title          = "{Neural Network Jet Flavour Tagging with the Upgraded ATLAS Inner Tracker Detector at the High-Luminosity LHC}",
    howpublished   = "{ATL-PHYS-PUB-2022-047}",
    url            = "https://cds.cern.ch/record/2839913",
    year           = "2022",
}

@Booklet{ATL-PHYS-PUB-2023-034,
    author         = "{ATLAS Collaboration}",
    title          = "{ATLAS Run~3 charged particle track seed finding performance}",
    howpublished   = "{ATL-PHYS-PUB-2023-034}",
    url            = "https://cds.cern.ch/record/2882156",
    year           = "2023",
}
\clearpage
 
\begin{flushleft}
\hypersetup{urlcolor=black}
{\Large The ATLAS Collaboration}

\bigskip

\AtlasOrcid[0000-0002-6665-4934]{G.~Aad}$^\textrm{\scriptsize 105}$,
\AtlasOrcid[0000-0001-7616-1554]{E.~Aakvaag}$^\textrm{\scriptsize 17}$,
\AtlasOrcid[0000-0002-5888-2734]{B.~Abbott}$^\textrm{\scriptsize 124}$,
\AtlasOrcid[0000-0002-0287-5869]{S.~Abdelhameed}$^\textrm{\scriptsize 120a}$,
\AtlasOrcid[0000-0002-1002-1652]{K.~Abeling}$^\textrm{\scriptsize 56}$,
\AtlasOrcid[0000-0001-5763-2760]{N.J.~Abicht}$^\textrm{\scriptsize 50}$,
\AtlasOrcid[0000-0002-8496-9294]{S.H.~Abidi}$^\textrm{\scriptsize 30}$,
\AtlasOrcid[0009-0003-6578-220X]{M.~Aboelela}$^\textrm{\scriptsize 46}$,
\AtlasOrcid[0000-0002-9987-2292]{A.~Aboulhorma}$^\textrm{\scriptsize 36e}$,
\AtlasOrcid[0000-0001-5329-6640]{H.~Abramowicz}$^\textrm{\scriptsize 158}$
% ...
\bigskip
\\

% ...
$^{17}$Department for Physics and Technology, University of Bergen, Bergen; Norway.\\
% ...
$^{30}$Physics Department, Brookhaven National Laboratory, Upton NY; United States of America.\\
% ...
$^{36}$$^{(a)}$Facult\'e des Sciences Ain Chock, Universit\'e Hassan II de Casablanca;$^{(b)}$Facult\'{e} des Sciences, Universit\'{e} Ibn-Tofail, K\'{e}nitra;$^{(c)}$Facult\'e des Sciences Semlalia, Universit\'e Cadi Ayyad, LPHEA-Marrakech;$^{(d)}$LPMR, Facult\'e des Sciences, Universit\'e Mohamed Premier, Oujda;$^{(e)}$Facult\'e des sciences, Universit\'e Mohammed V, Rabat;$^{(f)}$Institute of Applied Physics, Mohammed VI Polytechnic University, Ben Guerir; Morocco.\\
% ...
$^{46}$Physics Department, Southern Methodist University, Dallas TX; United States of America.\\
% ...
$^{50}$Fakult\"{a}t Physik , Technische Universit{\"a}t Dortmund, Dortmund; Germany.\\
% ...
$^{56}$II. Physikalisches Institut, Georg-August-Universit\"{a}t G\"ottingen, G\"ottingen; Germany.\\
% ...
$^{105}$CPPM, Aix-Marseille Universit\'e, CNRS/IN2P3, Marseille; France.\\
% ...
$^{120}$$^{(a)}$New York University Abu Dhabi, Abu Dhabi;$^{(b)}$United Arab Emirates University, Al Ain; United Arab Emirates.\\
% ...
$^{124}$Homer L. Dodge Department of Physics and Astronomy, University of Oklahoma, Norman OK; United States of America.\\
% ...
$^{158}$Raymond and Beverly Sackler School of Physics and Astronomy, Tel Aviv University, Tel Aviv; Israel.\\
% ...
$^{a}$ Also Affiliated with an institute covered by a cooperation agreement with CERN.\\
$^{b}$ Also at An-Najah National University, Nablus; Palestine.\\
$^{c}$ Also at Borough of Manhattan Community College, City University of New York, New York NY; United States of America.\\
$^{d}$ Also at Center for High Energy Physics, Peking University; China.\\
$^{e}$ Also at Center for Interdisciplinary Research and Innovation (CIRI-AUTH), Thessaloniki; Greece.\\
$^{f}$ Also at CERN, Geneva; Switzerland.\\
% ...

\end{flushleft}


%

%
%

%

%
%
%
%

\end{document}